\documentclass[superscriptaddress,amsmath,amssymb,aps,prl,longbibliography,twocolumn]{revtex4-2} %twocolumn
\usepackage{lipsum}
\usepackage{graphicx}% Include figure files
\usepackage{dcolumn}% Align table columns on decimal point
\usepackage{bm}% bold math
\usepackage{amsmath}
\usepackage{amssymb}
\usepackage{natbib}
\usepackage{url}
\usepackage{upgreek}
\usepackage[breaklinks=true]{hyperref}% add hypertext capabilities
\usepackage{mathtools}
\usepackage{bookmark}

\newcommand{\comment}[1]{}

\begin{document}
\title{Magnetic-Field Universality of the Kondo Effect Revealed by Thermocurrent Spectroscopy}
\author{Chunwei Hsu}
\affiliation {\small \textit Kavli Institute of Nanoscience, Delft University of Technology, Lorentzweg 1, Delft 2628 CJ, The Netherlands}
\author{Theo A. Costi}
\affiliation {\small \textit Peter Grünberg Institut, Forschungszentrum Jülich, 52425 Jülich, Germany}
\affiliation {\small \textit Institute for Advanced Simulation, Forschungszentrum Jülich, 52425 Jülich, Germany}
\author{David Vogel} 
\affiliation {\small \textit Department of Chemistry, University of Basel, St. Johanns‐Ring 19, 4056 Basel, Switzerland}
\author{Christina Wegeberg}
\affiliation{\small \textit Department of Chemistry, University of Basel, St. Johanns‐Ring 19, 4056 Basel, Switzerland}
\author{Marcel Mayor}
\affiliation {\small \textit Department of Chemistry, University of Basel, St. Johanns‐Ring 19, 4056 Basel, Switzerland}
\affiliation {\small \textit Institute for Nanotechnology (INT), Karlsruhe Institute of Technology (KIT), P.O. Box 3640, 76021 Karlsruhe, Germany}
\affiliation {\small \textit Lehn Institute of Functional Materials (LIFM), School of Chemistry, Sun Yat-Sen University (SYSU), 510275 Guangzhou, China}
\author{Herre S.J. van der Zant}
\affiliation {\small \textit Kavli Institute of Nanoscience, Delft University of Technology, Lorentzweg 1, Delft 2628 CJ, The Netherlands}
\author{Pascal Gehring}
\email{Email: pascal.gehring@uclouvain.be\\}
\affiliation {\small \textit Kavli Institute of Nanoscience, Delft University of Technology, Lorentzweg 1, Delft 2628 CJ, The Netherlands}
\affiliation {\small \textit IMCN/NAPS, Université Catholique de Louvain (UCLouvain), 1348 Louvain-la-Neuve, Belgium}

\date{\today}
\begin{abstract}
Probing the universal low-temperature magnetic-field scaling of Kondo-correlated quantum dots via electrical conductance has proved to be experimentally challenging. Here, we show how to probe this in nonlinear thermocurrent spectroscopy applied to a molecular quantum dot in the Kondo regime. Our results demonstrate that the bias-dependent thermocurrent is a sensitive probe of universal Kondo physics, directly measures the splitting of the Kondo resonance in a magnetic field, and opens up possibilities for investigating nanosystems far from thermal and electrical equilibrium. 
\end{abstract}
\maketitle
{\it Introduction.---}
The Kondo effect, originally describing the anomalous increase with decreasing temperature in the resistivity of nonmagnetic metals containing a small concentration of magnetic impurities \cite{deHaas1934,Kondo1964,Hewson1997}, is now a ubiquitous phenomenon in physics, forming the starting point for understanding the Mott transition \cite{Georges1996}, heavy fermions \cite{vonLoehneysen2007}, and transport through correlated nanostructures, such as quantum dots~\cite{Goldhaber-Gordon1998a,Cronenwett1998}, molecules~\cite{Park2002} and adatoms on surfaces~\cite{Madhavan1998}. In the so-called "QCD Kondo effect" \cite{Ozaki2016}, it also constitutes one of the first known examples of asymptotic freedom~\cite{Gross1973,Politzer1974}, a property of the strong interaction in particle physics. 

A key feature of the Kondo effect is its universality  \cite{Anderson1970a,Wilson1975,Andrei1983}. For example, the temperature dependence of the linear conductance $G(T)$ of a spin-$1/2$ quantum dot is described by a unique universal scaling function $G(T)/G(0)=g(T/T_{\rm K})$ of $T/T_{\rm K}$, where $T$ is the temperature and $T_{\rm K}$ is the Kondo scale, and is used as a hallmark for establishing a spin-$1/2$ Kondo effect in quantum dot systems \cite{Goldhaber-Gordon1998b}. The same holds for exotic realizations of the Kondo effect \cite{Gonzalez-Buxton1998,Roch2009,Parks2010,Iftikhar2018}, with each having its own characteristic set of universal scaling functions. Thus, universality in Kondo systems provides hallmarks for identifying the particular Kondo effect in a given experiment \cite{Roch2009,Costi2009,Parks2010}.

\begin{figure*}
    \centering
    \includegraphics{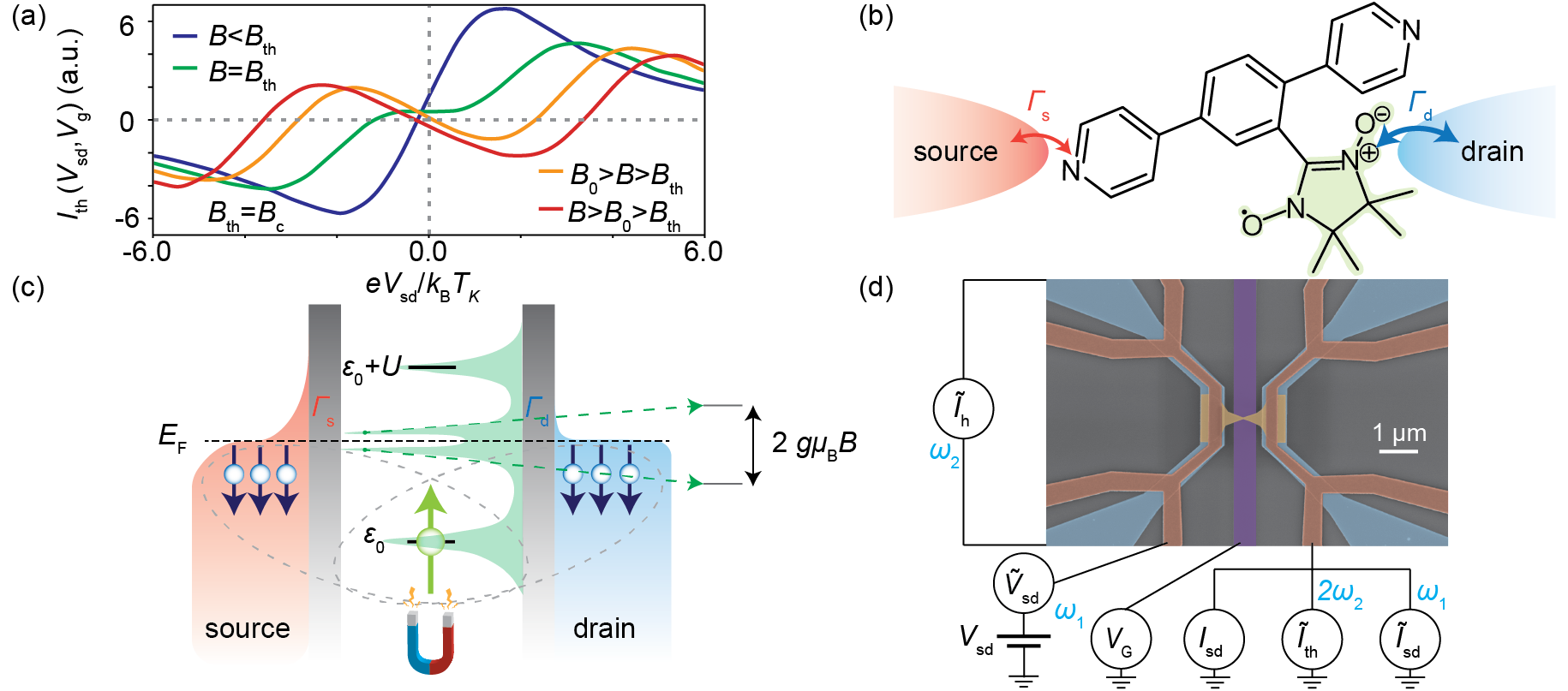}
    \caption{
    (a) Sketch of $I_{\rm th}$ vs $V_{\rm sd}$ in the Kondo regime for several $B$.
    $\left(\partial I_{\rm th}/\partial V_{\rm sd}\right)\rvert_{V_{\rm sd}=0}$ is a universal function of $B/T_{\rm K}$ and changes sign ("kink") at a universal field $B=B_{\mathrm{th}}=B_\mathrm{c}$ [Fig.~\ref{fig:fig4}(a)], while $I_{\rm th}(V_{\rm sd}=0)$  \cite{Svilans2018} changes sign at $B=B_0>B_\mathrm{c}$ and is nonuniversal [Fig.~\ref{fig:fig4}(a), Ref.~\onlinecite{Costi2019a} and Sec.~SM.3.5.5 \cite{Note1}].
    (b) Molecular junction of a NNR molecule anchored to source and drain leads.
    (c) Anderson model of (b) in a magnetic field $B$. A singly occupied level $\varepsilon_0$ with Coulomb repulsion $U$ and gate voltage $V_{\rm g}=(\varepsilon_0+U/2)/\mathit{\Gamma}$ coupled to hot and cold source and drain leads with strength $\rm \mathit{\Gamma}=\mathit{\Gamma}_s+\mathit{\Gamma}_d$ gives rise to a spin-$1/2$ Kondo effect for $V_{\rm g}\approx 0$ resulting in a Kondo resonance at the Fermi energy, $E_\mathrm{F}$. The field $B$ splits the up and down levels at $\varepsilon_0$ by $g\mu_{\rm B}B$ and the Kondo resonance in d$I/$d$V_{\rm sd}$ by $2g\mu_{\rm B}B$ \cite{Wingreen1994,Moore2000} (for $g\mu_{\rm B}B\gg k_{\rm B}T_{\rm K}$). A thermal bias $\Delta T>0$ causes a thermocurrent $I_\mathrm{th}$ to flow between source and drain, measured as described in (d). 
    (d) False-coloured scanning electron microscopy image of the thermoelectric device. 
    Bias and thermal voltages are generated by a DC+AC bias voltage source, $V_\mathrm{sd}$+$\tilde{V}_\mathrm{sd}(\omega_\mathrm{1})$, and a AC heater current source, $\tilde{I}_\mathrm{h}(\omega_\mathrm{2})$, on the hot left lead. The resulting DC, AC electrical currents and AC thermocurrent, $I_\mathrm{sd}$, $\tilde{I}_\mathrm{sd}(\omega_\mathrm{1})$ and $\tilde{I}_\mathrm{th}(2\omega_\mathrm{2})$, are measured simultaneously on the cold right lead. }
    \label{fig:fig1}
\end{figure*}

In this Letter, we address another aspect of universality of Kondo-correlated quantum dots, namely the universal magnetic-field scaling in the low-temperature ($T\ll T_{\rm K}$) Fermi-liquid regime of quantum dots. While our interest is in the thermocurrent, we first specify what we mean by low-temperature magnetic field scaling in the context of the more familiar differential conductance 
 $G(T,V_{\rm sd})=\textrm{d}I/\textrm{d}V_{\rm sd}$ (derivative of the electrical current with respect to source-drain voltage).   Specifically, for the asymmetrically coupled quantum dot device investigated in this Letter [Fig.~\ref{fig:fig1}(b)], described within the Anderson impurity model [Fig.~\ref{fig:fig1}(c)], $G(T\ll T_{\rm K},V_{\rm sd}\ll T_{\rm K})$ is given, for {\it arbitrary} magnetic fields $B$, within higher-order Fermi liquid theory \cite{Oguri2018b,Oguri2018a,Filippone2018} as
\begin{equation}\label{eq:dIdV-FL}
\frac{\mathrm{d}I}{\mathrm{d}V_{\rm sd}}\propto a_0-c_{T}\left(\frac{\pi T}{T_{\rm K}}\right)^2-c\frac{V_{\rm sd}}{T_{\rm K}} -c_{V}\left(\frac{V_{\rm sd}}{T_{\rm K}}\right)^2, 
\end{equation}
with field-dependent coefficients $a_0, c_T, c$ and $c_V$ \footnote{See Supplemental Material at [URL] for derivation of Eqs.~(\ref{eq:dIdV-FL}) and (\ref{eq:Ith-FL}), expressions for $a_0(B), c_T(B), c(B), c_V(B)$ and $s_0(B), S_1(B)$, evaluations, further experimental details, fitting procedures, synthesis and characterization, results at finite temperature/thermal-bias and including Refs.~\cite{Park1999,ONeill2007,Gehring2017,Costi1994,KWW1980a,Bulla2008,Lavagna2015,Hofstetter2000,Peters2006,Weichselbaum2007,Bulla1998,Campo2005,Rosch2003b,Merker2013,Hershfield1992,Meir1992,Jauho1994,Mora2015,Aligia2015,Amasha2005,Houck2005,Liu2009,Yoshida2009,Costi2010,Lacroix1981,Eckern2020,software_packages,Pyurbeeva2021,turnbull_phosphonofluoresceins_2021,hirel_nitronyl_2001,wilmarth_application_1955,romanenko_spin-labeled_2020,wang_temperature-dependent_2018,zhivetyeva_interaction_2018,stone_spin-labeled_1965,chechik_spin-labelled_2004,Tsutsumi2021,Sela2009,Aligia2011,Sierra2017}}. The low-temperature magnetic- field scaling that we refer to is reflected in the universal dependence of the curvature coefficients $c_V(B)\propto -\partial^2G/\partial V_\mathrm{sd}^2$ and $c_T(B)\propto -\partial^2G/\partial T^2$ on $B/T_{\rm K}$ in the Kondo regime \cite{Oguri2018b,Oguri2018a,Filippone2018} \footnote{$a_0(B)$ and $c(B)$ in (\ref{eq:dIdV-FL}) depend strongly on particle-hole and/or lead coupling asymmetry and are nonuniversal functions of $B/T_{\rm K}$.}. Surprisingly, the exact dependence of $c_V$ and $c_T$ on magnetic field has only recently been calculated via a generalization of Nozi\`eres Fermi-liquid theory \cite{Nozieres1974} to nonequilibrium and particle-hole asymmetric situations \cite{Oguri2018a,Oguri2018b,Filippone2018}.  The results show that $c_V$ and $c_T$ are universal functions of magnetic field which change sign at a universal crossover field $B=B_c$ describing the onset of the splitting of the Kondo resonance in $\textrm{d}I/\textrm{d}V_{\rm sd}$, in agreement with predictions for $B_c$ for the Kondo model \cite{Costi2000}. Nevertheless, establishing this universality and the splitting of the Kondo resonance in $\textrm{d}I/\textrm{d}V_{\rm sd}$ is intrinsically difficult \cite{Kogan2004,Quay2007,Kretinin2011,Hata2021}. Yet, both serve as useful experimental hallmarks of the Kondo effect in quantum dots.

Here, we propose a different approach to address magnetic-field scaling in the strong-coupling Kondo regime of quantum dots by employing the recently developed thermocurrent spectroscopy~\cite{Gehring2021}. We experimentally show that the thermocurrent, $I_{\rm th}$, of a molecular quantum dot in the Kondo regime exhibits a clear feature as a function of magnetic field, in the form of a zero-bias ($V_{\rm sd}=0$) kink appearing for fields $B$ larger than a certain value, which we denote by $B_{\rm th}$. We explain this behavior within higher-order Fermi-liquid theory \cite{Oguri2018b,Filippone2018} for $V_{\rm sd} \ll T_{\rm K}$, and an
approximate nonequilibrium Green function approach \cite{VanRoermund2010}
for $V_{\rm sd} \gtrsim T_{\rm K}$. Within the former, to leading order
in $V_\mathrm{sd}, T$ and $\Delta T$, where $\Delta T$ is the applied thermal-bias, we find in the low-temperature strong-coupling regime $\Delta T \ll T\ll T_{\rm K}$, 
\begin{equation}\label{eq:Ith-FL}
I_{\rm th} (T, V_{\rm sd}) = \gamma \frac{\pi^2 k^2_{\rm B}}{3} T\Delta T[s_0(B)+s_1(B)V_{\rm sd}],
\end{equation}
with constant $\gamma$ and coefficients $s_0(B)$ and $s_1(B)$ \cite{Note1}.
Remarkably, we show that, (i), $s_1(B)/s_1(0)$ and $c_{V}(B)/c_V(0)$ are described by essentially the same universal scaling function in the Kondo regime, showing that $\frac{\mathrm{d}I_\mathrm{th}}{\mathrm{d}V_\mathrm{sd}} \big|_{\mathrm{V_\mathrm{sd}=0}}$ [$\propto s_1(B)$] probes magnetic field universality, and, (ii), $B_{\rm th}$ coincides with $B_{\rm c}$, thus demonstrating that thermocurrent spectroscopy provides a new route to directly probe the splitting of the Kondo resonance \cite{Costi2000} and extract the universal field $B_{\rm c}=B_{\rm th}$.
Our findings are concisely summarized in the sketch in Fig.~\ref{fig:fig1}(a). We note, that in contrast to the zero-bias thermocurrent slope,  the zero-bias thermocurrent, $I_{\rm th}(T,V_{\rm sd}=0)$ [$\propto s_0(B)$], measured in Ref.~\onlinecite{Svilans2018} as a function of gate-voltage ($V_{\rm g}$) and magnetic field and found to change sign at a certain field $B_0$, is nonuniversal [Fig.~\ref{fig:fig4}(a), Ref.~\onlinecite{Costi2019a} and Sec.~SM.3.5.5 \cite{Note1}]. Thus $I_{\rm th}(T,V_{\rm sd}=0)$ does not provide a hallmark for the splitting of the Kondo resonance and cannot be used to extract $B_c$, in contrast to the thermocurrent spectroscopy proposed in this Letter. 
\begin{figure*}[t]
    \centering
    \includegraphics{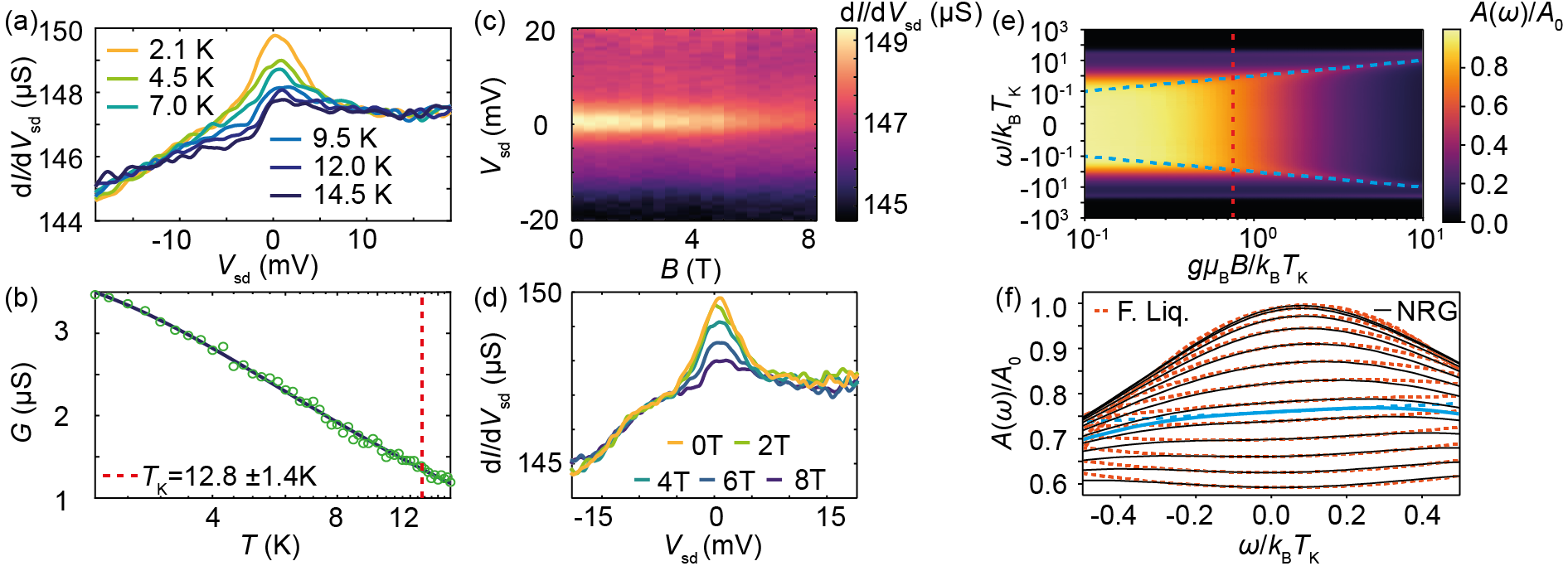}
    \caption{
    (a) d$I/$d$V_\mathrm{sd}$, of the molecular quantum dot vs $V_{\rm sd}$ at different temperatures. 
    (b) Peak conductance of Kondo resonance vs temperature and fit to a spin-$1/2$ Kondo model yielding $T_\mathrm{K}\mathrm{\approx12.8}~$K, see Secs.~SM.1.3 and SM.2.3.
    (c)-(d) d$I$/d$V_\mathrm{sd}$ vs $V_\mathrm{sd}$ at different magnetic fields. (e) NRG Spectral function $A(\omega)/A_0$ with $A_0=1/2\pi\Gamma$ of the Anderson model vs energy ($\omega$) and magnetic field for $V_\mathrm{g}=(\varepsilon_0+U/2)/\mathit{\Gamma}=-1$ and $U/\mathit{\Gamma}=8$. Vertical dashed line: splitting field $B_\mathrm{c}=0.75k_{\rm B}T_{\rm K}/g\mu_{\rm B}$. Blue dashed lines: positions [$\omega(B)=\pm g\mu_{\mathrm B}B$] of the split Kondo peaks in $A(\omega)$ for $B\gg B_{\rm c}$ \cite{Wingreen1994,Moore2000}. 
    (f) $A(\omega)/A_0$ from line cuts in (e) at $g\mu_{\mathrm B}B/k_{\mathrm B}T_\mathrm{K}=0,0.1,\dots,0.7,0.75,0.8,\dots,1.2$ (solid lines), compared to
    $A(\omega)/A_0$ from Fermi-liquid theory (dashed lines). Blue lines: 
 $B=B_\mathrm{c}$.
A g factor of $g=2$ is used, as measured by electron paramagnetic resonance  (Sec.~SM.4.2 \cite{Note1}).} 
    \label{fig:fig2}
\end{figure*}

{\it Experiment and results.---}
The experiment is carried out on a molecular quantum dot consisting of an organic radical molecule (nitronyl nitroxide radical, NNR) made up of a backbone and a nitronyl-nitroxide side group where an unpaired electron resides as shown in Fig.~\ref{fig:fig1}(b). Such free radical molecules are model systems to study the spin-$1/2$ Kondo effect [Fig.~\ref{fig:fig1}(c)] \cite{Zhang2013,Frisenda2015,Gaudenzi2017}. Furthermore, their asymmetric structure and the additional pyridine anchoring sites allow us to achieve asymmetric and strong coupling between the source/drain leads and the molecule (quantified by couplings $\it\Gamma_\mathrm{s}$ and $\it\Gamma_\mathrm{d}$, Fig~\ref{fig:fig1}(b)). We form a NNR-molecule quantum dot in the thermoelectric device shown in Fig.~\ref{fig:fig1}(d) by immersing electromigrated nanogaps in the molecular solution~\cite{Joeri2019}. The thermoelectric device incorporates a local backgate and two microheaters in direct thermal contact with the source/drain leads [see Fig.~\ref{fig:fig1}(d) and Sec.~SM.1.1 \cite{Note1}].

Evidence for a Kondo effect is shown by the strong suppression of the zero-bias peak in the measured $\textrm{d}I/\textrm{d}V_{\rm sd}$, both as a function of increasing $T$ [Figs.~\ref{fig:fig2}(a)-\ref{fig:fig2}(b)] and $B$ [Figs.~\ref{fig:fig2}(c)-\ref{fig:fig2}(d)]. The $T$-dependence of the zero-bias peak height [Fig.~\ref{fig:fig2}(b)] is well described by the numerical renormalization group (NRG) conductance of a spin-$1/2$ Kondo model and yields $T_\mathrm{K}=12.8$~K (Sec.~SM.1.3 \cite{Note1}). Based on the structure of the molecule, an asymmetric coupling is expected. Assuming, $\mathit{\Gamma}_\mathrm{d} \gg \mathit{\Gamma}_\mathrm{s}$ (see Sec.~SM.3.5.3 \cite{Note1} for $\mathit{\Gamma}_\mathrm{d} \ll \mathit{\Gamma}_\mathrm{s}$), we find  $\mathit{\Gamma}_\mathrm{s}/\mathit{\Gamma}_\mathrm{d}\approx0.017$. An underscreened Kondo effect \cite{Roch2009,Parks2010}, requiring a larger molecular spin ($S>1/2$), is excluded, since such an effect results in a split Kondo resonance in d$I/$d$V_\mathrm{sd}$ starting already at zero field, which is not observed in Figs.~\ref{fig:fig2}(c)-\ref{fig:fig2}(d). Thus, a single-level Anderson model describing a $S=1/2$ Kondo effect [Fig.~\ref{fig:fig1}(c)] is justified by the data. In the remainder of this Letter, the base temperature is kept at $T\approx2$~K$\ll T_\mathrm{K}$ while the thermocurrent is  measured for a small thermal bias $\Delta T\approx 0.6 K\ll T\ll T_\mathrm{K}$ so that we probe the strongly-coupled Kondo regime (see Secs.~SM.2.4-5 and Secs.~SM.3.6.4-5 \cite{Note1} for thermal bias and temperature effects).

A closer look at the field dependence of d$I$/d$V_\mathrm{sd}$ in Figs.~\ref{fig:fig2}(c)-\ref{fig:fig2}(d), indicates that the expected splitting of the Kondo peak at  $B_\textrm{c}\approx7.15$~T \footnote{Estimated using $B_\mathrm{c} \approx 0.5k_\mathrm{B} T_\mathrm{K}^\mathrm{HWHM}/g \mu_\mathrm{B}=0.75k_\mathrm{B} T_\mathrm{K}/g \mu_\mathrm{B}$ ~\cite{Costi2000,Oguri2018b,Filippone2018}, with $T_\mathrm{K}^\mathrm{HWHM}\approx 1.5 \mathrm{T_K}$ from Sec.~SM.3.2 \cite{Note1}, and using $T_\mathrm{K}$ in Fig.~\ref{fig:fig2}(b).} is not observed. This is in part due to a large non-Kondo (field and temperature independent) contribution in Figs.~\ref{fig:fig2}(a) and \ref{fig:fig2}(d) which may mask the appearance of a splitting at zero bias. In addition, the largest field used, $B=8 $~T, was only marginally larger than $B_{\rm c}$. For a device where higher fields relative to $B_{\rm c}$ could be applied, such a splitting is observed (Sec.~SM.2.1 \cite{Note1}). Despite these complications in extracting $B_{\rm c}$ from d$I/$d$V_\mathrm{sd}$ for the device studied, there is also a general problem in doing so, which can be appreciated by attempting this from exact theoretical results. This is illustrated in Figs.~\ref{fig:fig2}(e)-\ref{fig:fig2}(f) which show the spectral function  $A(\omega= eV_{\mathrm{sd}})\sim $~d$I/$d$V_\mathrm{sd}$ within the NRG method and within Fermi-liquid theory. While the precise value of $B_{\rm c}$ is impossible to determine visually in Fig.~\ref{fig:fig2}(e) (vertical dashed line), it can be deduced from the line cuts in Fig.~\ref{fig:fig2}(f) as the field where the curvature of $A(\omega)$ vanishes. However, such accuracy in second derivatives of $A(\omega= eV_{\mathrm{sd}})\sim $~d$I/$d$V_\mathrm{sd}$  is difficult to attain from experimental data with finite error bars.

\begin{figure}
    \centering
    \includegraphics{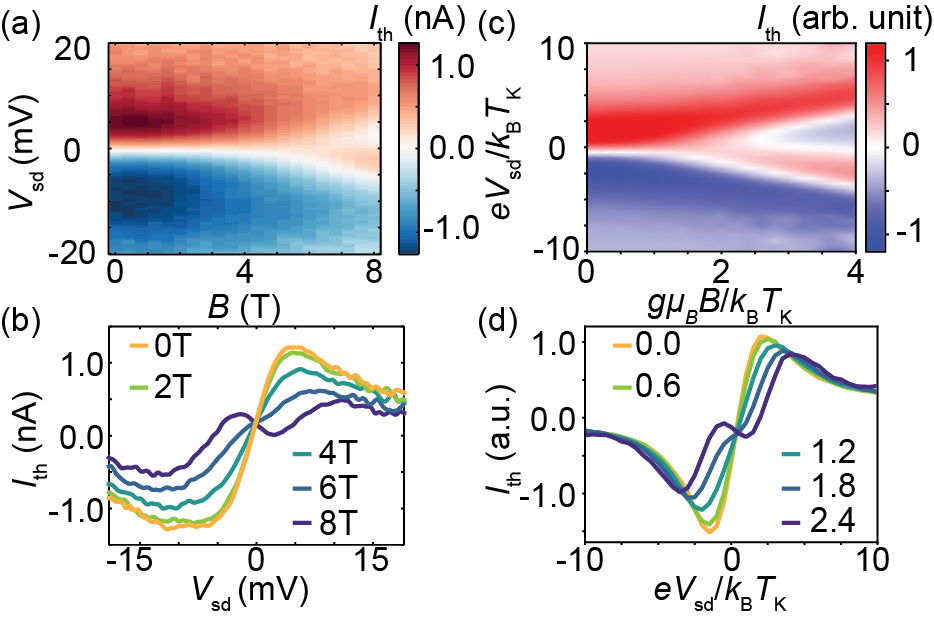}
    \caption{
    (a)-(b) Measured $I_\mathrm{th}$, vs $V_\mathrm{sd}$ at different magnetic fields.
    (c)-(d) Calculated $I_\mathrm{th}$ vs $V_\mathrm{sd}$ at different magnetic fields $g\mu_\mathrm{B}B/k_\mathrm{B}T_\mathrm{K}$ for the Anderson model in Fig.~\ref{fig:fig1}(c) with $V_{\rm g}=-2.5$, $U/\mathit{\Gamma}=8$, $\Delta T/T_\mathrm{K}=0.2$ and $T/T_\mathrm{K}=0.5$.
    } 

    \label{fig:fig3}
\end{figure}

\begin{figure}[t!]
    \centering
    \includegraphics{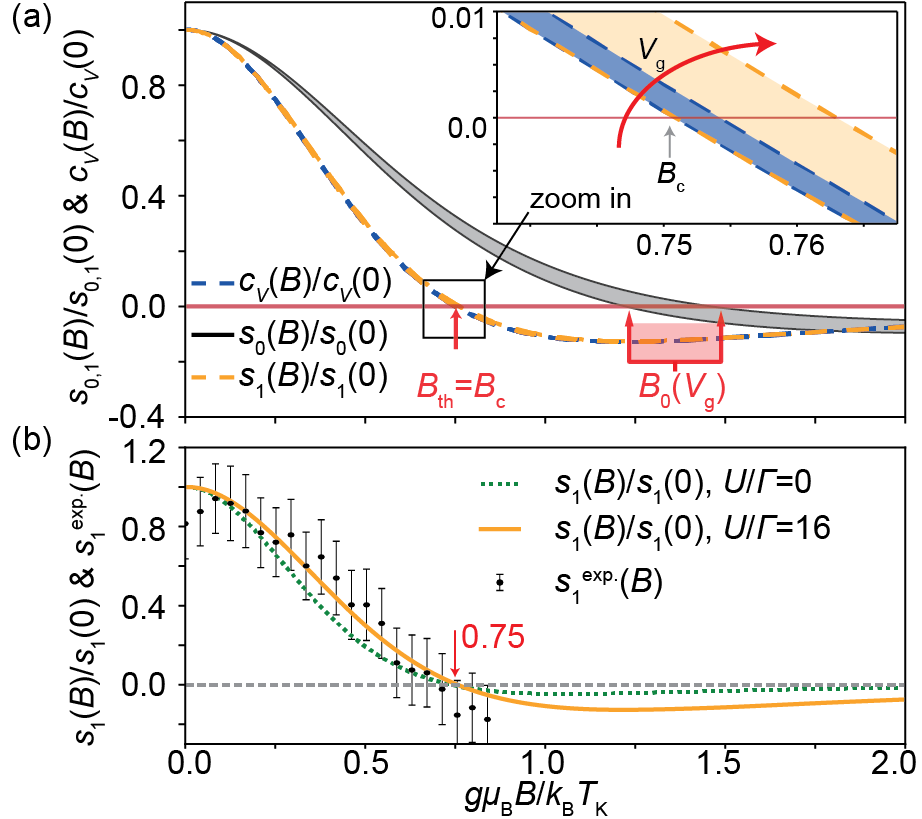}
    \caption{(a) Normalized zero-bias thermocurrent slope $s_1(B)/s_1(0)$ and curvature coefficient $c_\mathrm{V}(B)/c_{\rm V}(0)$ vs $g\mu_{\rm B}B/k_{\rm B}T_{\rm K}$, for gate voltages $1\leq V_{\rm g}=(\varepsilon_0+U/2)/\mathit{\Gamma}\leq 5$ in the Kondo regime exhibiting scaling collapse [Anderson model in Fig.~\ref{fig:fig1}(c) with $U/\mathit{\Gamma}=16$]. Also shown is the nonuniversal normalized zero-bias thermocurrent $s_0(B)/s_0(0)$ with a sign change at a strongly $V_{\rm g}$-dependent $B_0$. Inset: $c_V(B)/c_V(0)$ and $s_1(B)/s_1(0)$ in the region around $B=B_\mathrm{th}=B_{\rm c}\approx 0.75k_{\rm B}T_{\rm K}/g\mu_{\rm B}$, and their $V_{\rm g}$-dependence. (b) Least-squares fit of the experimental zero-bias thermocurrent slope $s_{1}^{\mathrm{exp.}}(B)$ to the universal curve for $s_\mathrm{1}(B)/s_1(0)$. Error bars denote 1$\sigma$ confidence intervals (see Sec.SM.1.4 \cite{Note1}). A fit to the noninteracting case with $U=0$ and $\varepsilon_0=-0.1\mathit{\Gamma}$ (green dotted line) yielded an RMS deviation that was 194\% larger than for the Kondo scaling curve. The estimated experimental $B_{\rm th}\approx 0.69k_{\rm B}T_{\rm K}/g\mu_{\rm B}$ is close to theory ($0.75k_{\rm B}T_{\rm K}/g\mu_{\rm B}$).
    }
    \label{fig:fig4}
\end{figure}

Thermocurrent spectroscopy resolves the above difficulty. Figures~\ref{fig:fig3}(a)-\ref{fig:fig3}(b) show the measured thermocurrent versus bias voltage and magnetic field, while Figs.~\ref{fig:fig3}(c)-\ref{fig:fig3}(d) show analogous theory results within an approximate nonequilibrium Green function approach (Sec.~SM.3.6 \cite{Note1}). First, notice that the large non-Kondo contribution to the differential conductance [Figs.~\ref{fig:fig2}(a) and \ref{fig:fig2}(d)], is absent in the thermocurrent Fig.~\ref{fig:fig3}(b), with the latter being largely symmetric in magnitude around zero bias, in agreement with theory [Fig.~\ref{fig:fig3}(d)]. This is because the thermocurrent effectively measures a difference of electrical currents (in the presence or absence of thermal bias), and thus filters out the non-Kondo contributions.
By the same token the thermocurrent therefore probes universal aspects of Kondo physics more precisely than the differential conductance. Second, we now see a clear feature, in the form of a zero-bias kink with a negative slope of the thermocurrent, appearing in $I_{\rm th}$ at a field $B=B_{\rm th}$. This is qualitatively captured, together with the behavior at $V_{\rm sd}\gtrsim T_{\rm K}$,  by the approximate approach [Figs.~\ref{fig:fig3}(c) and \ref{fig:fig3}(d)]. However, the precise field at which this feature occurs and its connection to $B_{\rm c}$ requires a more exact theory, which is provided by the higher-order Fermi-liquid theory [Eq.~(\ref{eq:Ith-FL}) and Sec.~SM.3.5 \cite{Note1}].
Preempting the result of this theory, we note that analyzing the experimental data in Figs.~\ref{fig:fig3}(a) and \ref{fig:fig3}(b) for the slope of the thermocurrent d$I_{\rm th}$/d$V_{\rm sd}(V_\mathrm{sd}=0)$ at zero-bias voltage as a function of magnetic field, we find that this slope vanishes (i.e., the kink appears) at $B_{\rm th}\approx 6.6 $~T. This value is within 10\% of the expected $B_{\rm c}\approx 7.15$~T and already suggests that $B_{\rm th}=B_c$, and, hence, that 
the splitting of the Kondo resonance can be directly measured 
in the bias voltage dependence of $I_{\rm th}(V_{\rm sd})$.

Equation~(\ref{eq:Ith-FL}), with $s_{i}(B),i=0,1$ evaluated exactly for all $B$ within the NRG (Sec.~SM.3.5 \cite{Note1}), allows us to address the experimentally observed sign change of   $\left(\partial I_{\rm th}(V)/\partial V_\mathrm{sd} \right)_{V=0} \propto  s_1(B) $ upon increasing $B$ above $B_{\rm th}$ (the "kink") and to extract $B_{\rm th}$. 
Figure~\ref{fig:fig4}(a) shows the normalized zero-bias thermocurrent slope $\propto s_1(B)/s_1(0)$, the normalized zero-bias thermocurrent $\propto s_0(B)/s_0(0)$ and the normalized curvature coefficient $\propto c_V(B)/c_V(0)$
as a function of $B$ and for a range of $V_{\rm g}$ in the Kondo regime. First, notice that both $c_V(B)/c_V(0)$ and $s_1(B)/s_1(0)$ [in contrast to $s_0(B)/s_0(0)$] are universal scaling functions of $g\mu_{\rm B} B/k_{\rm B}T_{\rm K}$ with only a weak dependence on $V_{\rm g}$ [inset Fig.~\ref{fig:fig4}(a)], and  while distinct, they lie within about 1\% of each other [inset Fig.~\ref{fig:fig4}(a) and Fig.~S13 \cite{Note1}]. Thus, measuring the field dependence of $s_1(B)$ via thermocurrent spectroscopy, requiring only a first derivative with respect to  bias voltage, equivalently probes the magnetic-field universality from an electrical conductance measurement, which, however, requires a second derivative with respect to  bias voltage and is consequently less accurate.  Furthermore, since both $s_1(B)$  and $c_{V}(B)$ change sign at  the same magnetic field, i.e., $B_{\rm th}= B_{c}\approx 0.75k_\mathrm{B}T_\mathrm{K}/\mu_\mathrm{B}$, thermocurrent measurements of Kondo correlated quantum dots at finite bias voltage provide a new way to determine the splitting of the Kondo resonance via a sign change in the slope of the thermocurrent with respect to  bias voltage.

In Fig.~\ref{fig:fig4}(b) we show a direct comparison between theory and experiment for the slope of the zero-bias thermocurrent as a function of magnetic field. The experimental data follows well the universal curve for $s_1(B)/s_1(0)$, and the aforementioned value extracted from this data for $B_{\rm th}\approx 6.6 $~T ($g\mu_{\rm B}B_{\rm th}/k_{\rm B}T_{\rm K}=0.69$), is consistent with the expected splitting field of $B_{\rm th}\approx 7.15 $~T ($g\mu_{\rm B}B_{\rm th}/k_{\rm B}T_{\rm K}\approx 0.75$). The largest available field, $8$~T, did not allow accessing the minimum of the $s_1(B)$ vs B curve or the slow increase of $s_1(B)$ to zero at $B\gg T_{\rm K}$. The agreement between theory and experiment at the largest fields measured $B>B_{\rm th}$ is reduced, but still within the error bounds of the experimental data. The extracted $B_{\rm c}=B_{\rm th}$ from the thermocurrent validates the theory prediction with higher accuracy than has so far been reported (see Sec.~SM.3.5.2 \cite{Note1}). 
The large energy level separation in a molecular quantum dot grants the observed good agreement between the theory and experiment, even under a simple single-level assumption in the transport window [Fig.~\ref{fig:fig1}(c)]. 

{\it Conclusion.---}
In summary, we have studied the effect of a magnetic field on a Kondo-correlated molecular quantum dot via nonlinear thermocurrent spectroscopy. We demonstrated theoretically and confirmed experimentally, that the nonequilibrium thermocurrent, via its zero-bias slope $s_1(B)$, exhibits universal Fermi-liquid magnetic-field scaling, and that the vanishing of $s_1(B)$ at $B=B_{\mathrm{th}}$ with $B_{\mathrm{th}}= B_{\rm c}$, directly probes the splitting of the Kondo resonance. Since the thermocurrent is largely robust against parasitic conductive phenomena, it provides a more clear cut signature of this hallmark than is available from conductance measurements only. The ability to tune thermal and voltage bias, as well as temperature and magnetic field, opens up possibilities for using thermocurrent spectroscopy to yield insights into Kondo physics of nanoscale systems and may prompt theoretical investigations to address the largely unexplored area of nanosystems far  from thermal and electrical equilibrium.
\begin{acknowledgments}
We thank J. de Bruijckere for his support in analysis software and M. van der Star for his help in sample fabrication. This work is part of The Netherlands Organization for Scientific Research (NWO). P.G. acknowledges financial support from the F.R.S.-FNRS of Belgium and a Marie Skłodowska-Curie Individual Fellowship under Grant TherSpinMol (ID: 748642) from the EU’s Horizon 2020 research and innovation programme. H.S.J.v.d.Z, C.H., M.M. and D.V. acknowledge funding by the EU (FET-767187-QuIET). Computing time granted through JARA on the supercomputer JURECA at Forschungszentrum Jülich is gratefully acknowledged (T.A.C.). M.M. acknowledges support from the Swiss National Science Foundation (SNF grant numbers 200020-178808) and the 111 project (90002-18011002). C.W. thanks the Independent Research Fund Denmark for an international postdoctoral grant (9059-00003B). 
\end{acknowledgments}

%\bibliography{Main.bib}
\providecommand{\noopsort}[1]{}\providecommand{\singleletter}[1]{#1}%
%

%\bibliography{ref}
% Produces the bibliography via BibTeX.
\end{document}

\title{Supplemental Material for\\Magnetic Field Universality of the Kondo Effect Revealed by Thermocurrent Spectroscopy}

\author{Chunwei Hsu}
\affiliation {\small \textit Kavli Institute of Nanoscience, Delft University of Technology, Lorentzweg 1, Delft 2628 CJ, The Netherlands}
\author{Theo A. Costi}
\affiliation {\small \textit Peter Grünberg Institut, Forschungszentrum Jülich, 52425 Jülich, Germany}
\affiliation {\small \textit Institute for Advanced Simulation, Forschungszentrum Jülich, 52425 Jülich, Germany}
\author{David Vogel} 
\affiliation {\small \textit Department of Chemistry, University of Basel, St. Johanns‐Ring 19, 4056 Basel, Switzerland}
\author{Christina Wegeberg}
\affiliation{\small \textit Department of Chemistry, University of Basel, St. Johanns‐Ring 19, 4056 Basel, Switzerland}
\author{Marcel Mayor}
\affiliation {\small \textit Department of Chemistry, University of Basel, St. Johanns‐Ring 19, 4056 Basel, Switzerland}
\affiliation {\small \textit Institute for Nanotechnology (INT), Karlsruhe Institute of Technology (KIT), P.O. Box 3640, 76021 Karlsruhe, Germany}
\affiliation {\small \textit Lehn Institute of Functional Materials (LIFM), School of Chemistry, Sun Yat-Sen University (SYSU), 510275 Guangzhou, China}
\author{Herre van der Zant}
\affiliation {\small \textit Kavli Institute of Nanoscience, Delft University of Technology, Lorentzweg 1, Delft 2628 CJ, The Netherlands}
\author{Pascal Gehring}
\email{Email: pascal.gehring@uclouvain.be\\}
\affiliation {\small \textit Kavli Institute of Nanoscience, Delft University of Technology, Lorentzweg 1, Delft 2628 CJ, The Netherlands}
\affiliation {\small \textit IMCN/NAPS, Université Catholique de Louvain (UCLouvain), 1348 Louvain-la-Neuve, Belgium}
\date{\today}
\maketitle

\tableofcontents
\renewcommand\theequation{S\arabic{equation}}
\renewcommand\thesection{SM.\arabic{section}}
\renewcommand\thesubsection{\thesection.\arabic{subsection}}
\renewcommand\thesubsubsection{\thesubsection.\arabic{subsubsection}}
\renewcommand\thefigure{S\arabic{figure}}
\section{Methods}
\subsection{Device fabrication.} 
\label{subsec:device-fabrication}
The theromelectric device is fabricated by standard electron-beam lithography with a double-layer PMMA 495K/950K resist and electron-beam evaporation deposition technique. First, the local back gate (dark purple pattern in Fig.~1d) with Ti+Pd thickness of 1+7 nm is created on a SiO$_2$/Si wafer with an oxide thickness of 817 nm. The heaters (light blue pattern) are fabricated with Ti+Pd at a thickness of 3+27 nm subsequently. Afterwards, a insulating layer of $\rm Al_2O_3$ of 10 nm is deposited with an atomic layer deposition technique at 300 $^\circ{}$C. The Au bridge (golden pattern) is then made with a thickness of 12 nm without an adhesive layer. Finally the contact electrodes (orange pattern) of Ti+Au thickness of 5+65 nm are made to connect the Au bridge made in the previous step. 

After the fabrication steps, the device is wire-bonded, mounted onto a variable temperature insert and pumped down. A high current with a feedback control is used to initiate the electromigration of the Au bridge in vacuum~\cite{Park1999}. A nm-sized gap is formed by electromigration and self-breaking~\cite{ONeill2007}. After the successful electromigration, the nanogap was then immersed in a molecular solution of 1 mM dissolved in dichloromethane. The sample was immediately pumped and cooled down to 2.0 K to prevent further opening of the electromigrated nanogap and potential degradation of the radical molecule. \\
\\
\
\subsection{Lock-in modulation technique.}
    After device formation and cool down, the electrical and thermoelectric measurements are carried out. In order to measure the electrical and thermoelectric signals at the same time, we employ a lock-in technique \cite{Gehring2021}, where the AC thermal bias $\Delta \Tilde{T}$ and electrical bias $\Tilde{V}_\mathrm{sd}$ were applied at two different frequencies. The differential conductance (d$I$/d$V_\mathrm{sd}$) of the quantum dot was directly obtained with a lock-in amplifier. In a typical experiment $\Tilde{V}_\mathrm{sd} = 100 \upmu$V at $\omega_1 = 13$~Hz is used. A thermal bias at a frequency $\omega_2=3$~Hz is applied by the microheater and the resulting thermocurrent ($I_{\mathrm{th}}$) through the quantum dot is measured at the second harmonic 2$\omega_2$ with respect to the excitation~\cite{Gehring2017}.
   \newline
   \newline
\subsection{Extraction of the Kondo temperature.}
\label{subsec:TK-extraction}
The temperature dependence of the zero-bias peak for a spin-$1/2$ Kondo effect, calculated within NRG \cite{Costi1994}, is well described by the empirical formula \cite{Goldhaber-Gordon1998}.
\begin{equation}\label{eqn:peakfit}
    G(T)=G_l\left(\frac{T^2}{T_\mathrm{K}^2}(2^{1/s}-1)+1\right)^{-s}+G_c,
\end{equation}
where $G_l=\frac{2e^2}{h} \frac{4\mathit{\Gamma}_\mathrm{s} \mathit{\Gamma}_\mathrm{d}}{(\mathit{\Gamma}_\mathrm{s}+\mathit{\Gamma}_\mathrm{d})^2}$ is the maximum conductance peak reached at $T\ll T_\mathrm{K}$, $s=0.22$ for a spin-$1/2$ Kondo effect, $G_c$ is the background conductance which is additional to the Kondo resonance due to a direct tunneling/parallel channel in the quantum dot. Equation (\ref{eqn:peakfit}) defines the Kondo temperature $T_\mathrm{K}$ as $G(T=T_\mathrm{K})-G_c=(G(T=0)-G_c)/2$. The fitting of Eq.~(\ref{eqn:peakfit}) is shown Fig.~2b as a function of temperature after subtracting the background conductance, resulting in $T_\mathrm{K}=12.8$~K.
\newline
\newline
\subsection{Fitting \texorpdfstring{$s_{1}^{\rm exp.}(B)$}{} to the universal curve \texorpdfstring{$s_1(B)/s_1(0)$}{}.}
\label{subsec:fitting-to-s1}
In the Kondo regime of a spin-$1/2$ quantum dot with asymmetric lead couplings $\mathit{\Gamma}_d\gg \mathit{\Gamma}_s$, the universal curve for $s_1(B)/s_1(0)=f_1(b)$ with $b=g\mu_{\rm B}B/k_{\rm B}T_{\rm K}$ can be calculated within NRG for all $b$ (\ref{subsubsec:probe-splitting-Ith}). An approximate interpolation formula for this function, valid for a large range of $b$ from $b\ll 1$ to $b\gg 1$, is given by (\ref{subsubsec:universal-scaling-functions})
\begin{equation}\label{eqn:slopefit1}
    f_1(b)= \frac{(1-\alpha_0 b^2)}{(1+\beta_0 b^2)^{z}},
\end{equation}
where $\alpha_0\approx 1.77, \beta_0\approx 1.12$ and $z\approx 2.6$. 
We used Eq.~(\ref{eqn:slopefit1}) to carry out a least-squares fit of $r\times s_{1}^{\rm exp.}(B)$ to $f_1(b)$, yielding $r=2.18\times 10^{6}\,\mathrm{mVnA^{-1}}$.
\pagebreak
\section{Additional experimental results}% and comparison to calculations}
\subsection{Experimental data on molecular quantum dot QD2}
\begin{figure}[htb!]
    \centering
    \includegraphics{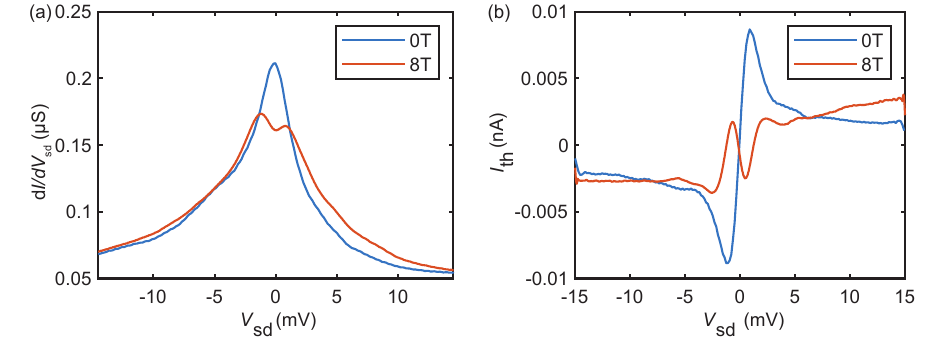}
    \caption{(a) Measured  d$I$/d$V_{\rm sd}$ of QD2 at magnetic fields of 0 and 8~T. (b) Measured  $I_{\rm th}$ of QD2 at magnetic fields of 0 and 8~T.}
    \label{fig:Qdot2}
\end{figure}

An additional set of data on a different molecular quantum dot device, QD2, is presented in Fig.~\ref{fig:Qdot2}. The experimental d$I$/d$V_{\rm sd}$ and $I_\mathrm{th}$ of QD2 are shown in Fig.~\ref{fig:Qdot2}a,b respectively. Worth-noting, the coupling between the molecule and the electrodes is smaller than that of the device shown in the main text (we denote the quantum dot in the main text by QD1 in the following). This is seen readily by the smaller HWHM of the zero-bias peak and the smaller base-line d$I$/d$V_{\rm sd}$. Due to this weaker coupling, meaning a lower $T_{\rm K}$ and $B_{\rm c}$, a splitting of d$I$/d$V_{\rm sd}$ at zero bias is observed at $B=8\rm ~T$, different from the case of QD1. The $I_{\rm th}$ of QD2  presented in Fig.~\ref{fig:Qdot2}b  shows the same trend as in Fig.~3b. At $B=0 \rm~T$, from $V_{\rm sd}<0$ to $V_{\rm sd}>0$, $I_{\rm th}$ first goes to a large negative value, crossing the origin and then to a large positive value, in a S-like shape. At $B=8~\rm T$, a "kink" appears as the slope of $I_{\rm th}$ changes its sign at $V_{\rm sd}=0$, similar to the case in Fig.~3b of QD1. This additional set of measurement provides an example where the splitting of d$I$/d$V_{\rm sd}$ is comparable to the kink of $I_{\rm th}$, consistent with the calculations shown in the main text. However, further magnetic and temperature dependence of QD2 was unable to be obtained as the sample was damaged in a measurement with high heater current.

\pagebreak
\clearpage
\subsection{\texorpdfstring{Proportionality between $-{\rm d}^2I/{\rm d}V_{\rm sd}^2$ and  $I_{\rm th} $}{}}
\label{subsec:exp-proportionality}
\begin{figure}[htb!]
    \centering
    \includegraphics{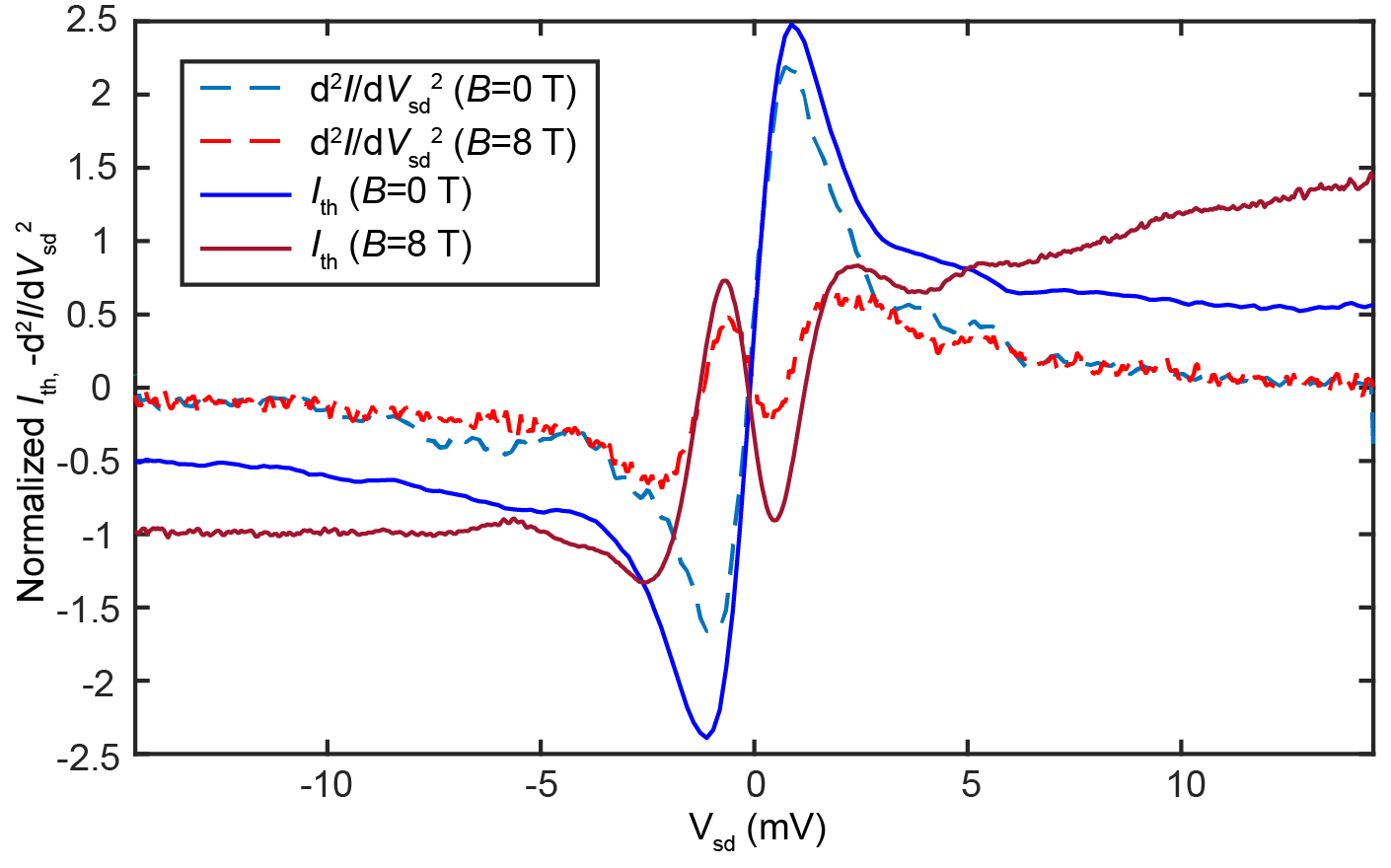}
    \caption{Normalized -d$^2I$/d$V_{\rm sd}^2 $ and $I_\mathrm{th}$ of QD2.}
    \label{fig:dIdVIth}
\end{figure}
In the weaker coupling case of QD2, an interesting feature is observed by taking the derivative of d$I$/d$V_{\rm sd}$ in Fig~\ref{fig:Qdot2}. Here, by plotting the normalized value of -d$^2I$/d$V_{\rm sd}^2$ on top of $I_\mathrm{th}$, a proportionality between the two quantities is observed as shown in Fig.~\ref{fig:dIdVIth}. This experimental observation is further discussed theoretically in the later section of \ref{sec:proportionality}.

\subsection{Gaussian filter for electronic cross-talk}
\label{subsec:crosstalk}
\begin{figure}[htb!]
    \centering
    \includegraphics{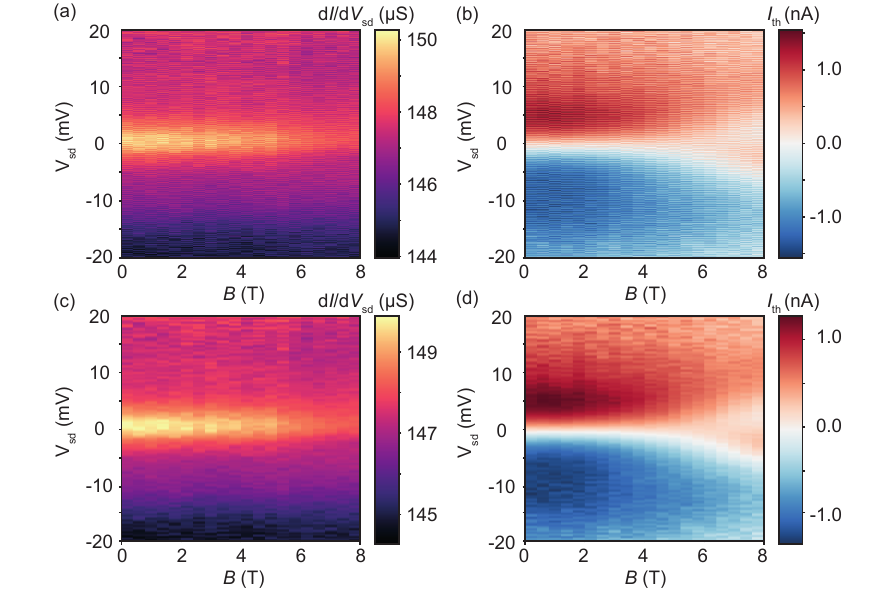}
    \caption{(a)-(b) Unfiltered data of the electronic/thermoelectric measurement of the quantum dot shown in Fig.~2,3 (QD1). (c)-(d) Fast-Fourier-Transform (FFT) Filtering of data in (a)-(b). The data with electronic cross-talk is rectified with a FFT filter, where frequency components smaller than 1/3 of maximum frequency in the Fourier domain are kept. }
    \label{fig:FFT}
\end{figure}

In the simultaneous stimulation of d$I$/d$V_{\rm sd}$ and $I_{\rm th}$ with lock-in amplifiers, oscillations in the measured signals are observed in the measurements as a result of electronics cross-talk. Therefore, a Gaussian filter is applied to the high frequency oscillations in the measurements to restore the measurement signals. The implementation of Gaussian filter is basically weighted moving average function and can be found in the \textit{smoothdata} function in the proprietary software Matlab with version of 2017b and newer. 
\newline Figure~\ref{fig:FFT}a,b show an example of unfiltered data of Fig. 2 of QD1. An periodic interference in both d$I$/d$V_{\rm sd}$ and $I_{\rm th}$ can be observed as a result of lock-in cross-talk. With the Gaussian filter described above, we obtain the restored signal in Fig.~2,3 in the main text. The error bars are estimated from the moving standard deviation on the raw data, using \textit{movstd} in Matlab. This gives error bars around 0.5~$\upmu$S for Fig. 2a,c and d, which are not plotted in the main text figure for data visibility.   To better demonstrate the effect of Gaussian filter applied in the main text, we also show another commonly used Fast-Fourier transform filter as a comparison in Fig.~\ref{fig:FFT}c,d. In this case we rectify the data by cutting off Fourier components with frequency that are larger than 1/3 of maximum frequency in the frequency domain. Essentially, both the Gaussian and FFT filters have a similar effect on the data and filter out the periodic interference. By comparing the two methods, it is clear that the underlying features can be restored as long as the interference signal is periodic.

\pagebreak
\clearpage
\subsection{Temperature dependence of QD1}
\label{subsec:exp-temp-dependence-thermocurrent}
\begin{figure}[htb!]
    \centering
    \includegraphics{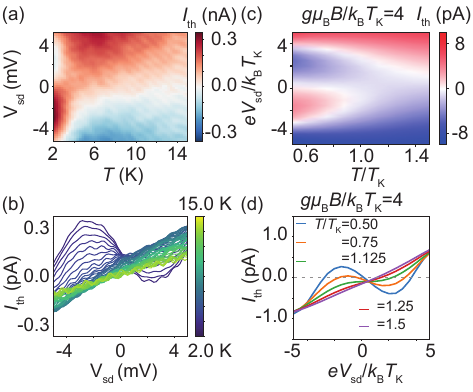}
    \caption{(a)-(b) Temperature dependent thermocurrent measurement of QD1 at a magnetic field 8~T above $B_{\rm th}$. (c)-(d) Calculated temperature dependence of thermocurrent at a magnetic field $B> B_{\rm th}$ using the approximate nonequilbrium equation of motion method of  Sec.~\ref{subsec:EOM-method}.}
    \label{fig:T_sweep_8T}
\end{figure}
 In the Letter, we presented measurements of the magnetic field dependence of the thermocurrent in the low temperature strong coupling regime ($T\ll T_{\rm K}$). We showed that the
 zero-bias thermocurrent slope changes sign from positive to negative (i.e., a kink develops at zero-bias) upon increasing the magnetic field above $B_{\rm th}$ with $B_{\rm th}=6.6$~T. Here, we investigate how this kink in $I_{\rm th}$, present for $B>B_{\rm th}$ and $T\ll T_{\rm K}$, behaves with increasing temperature. Figs.~\ref{fig:T_sweep_8T}(a) and (b) show results for a specific field $B=8$~T ($>B_{\rm th}$). One sees, that the kink vanishes above a certain temperature.  For the case shown (B=8~T), this temperature is approximately 5~K. The temperature above which the kink in $I_{\rm th}$ vanishes is not universal (e.g., it will depend on
 the value of $B>B_{\rm th}$) and is not simply related to $T_{\rm K}$. Thus one cannot extract $T_{\rm K}$ from the vanishing of the kink with increasing temperature. Calculations at finite
 temperatures (comparable to $T_{\rm K}$) lie outside the regime of validity of the asymptotically exact Fermi-liquid theory used to explain the $B$-dependence of the low temperature thermocurrent in the strong coupling regime ($\Delta T\ll T \ll T_{\rm K}$) in the Letter. In order to explain the observations, we therefore to resort to an approximate nonequilbrium equation of motion approach (Sec.~\ref{subsec:EOM-method}). The results, shown in Figs.~\ref{fig:T_sweep_8T}(c) and (d) for $B>B_{\rm th}$, capture the observed experimental trends, including the vanishing of the kink in the thermocurrent above a certain nonuniversal temperature ($T\approx 1.125 T_{\rm K}$ for the case shown).
%qualitatively understand the experimental observation, we calculate the thermocurrent at 
%$B>B_{\rm th}$ as a function of temperature with the equation of motion method (see \ref{subsec:EOM-method}), as shown in Fig.~\ref{fig:T_sweep_8T}c and d. Both experimental and theoretical results show a thermocurrent inversion at a temperature around 4~K and 0.75~$T_{\rm K}$ (9.6~K)  respectively. A good qualitative agreement is observed. This demonstrates that the change of slope in $I_{\rm th}$ around zero bias is robust up to a temperature $T\approx 0.3~T_{\rm K}$ experimentally. On the other hand, the experimentally observed splitting of d$I$/d$V_{\rm sd}$ depends on the temperature and is close to the theoretical splitting field only at $T\ll T_{\rm K}$~\cite{Kretinin2011}. 
\pagebreak
\clearpage
\subsection{Thermal bias dependent measurement of QD1}
\begin{figure}[htb!]
    \centering
    \includegraphics{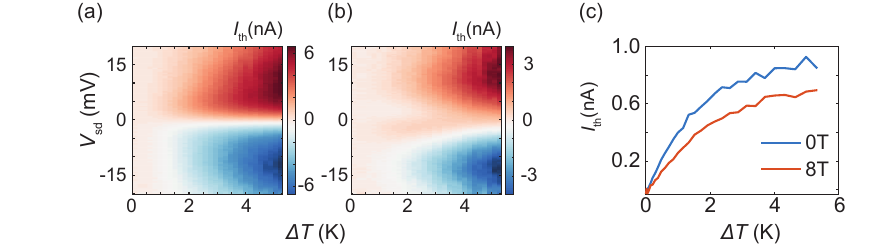}
    \caption{(a)-(b), Thermal bias dependence of $I_{\rm th}$ at $B=0~\rm T$ and $B=8~\rm T$. (c), Zero-bias $I_{\rm th}$ as a function of thermal bias, $\Delta T$.}
    \label{fig:Ith_vs_dT}
\end{figure}

 To further investigate the nonlinear effect caused by the thermal gradient ($\Delta T$), we study the thermocurrent response of the device as a function of thermal bias. As shown in Fig.~\ref{fig:Ith_vs_dT}(a), at zero magnetic field and finite bias, $I_\mathrm{th}$ increases as  $\Delta T$ rises. At first sight, this increase can be understood as a result of the change in thermal bias, which can be seen directly from the definition, $I_{\mathrm{th}}=-L\Delta T$, in the linear response regime. Nonetheless, as we consider $I_\mathrm{th}$ as a function of $\Delta T$ at a fixed bias voltage, a clear non-linear relation is observed (see Fig.~\ref{fig:Ith_vs_dT}(c) for the case at zero-bias), indicating the non-linear regime. 
 
 The Kondo feature, in the form of thermocurrent, changes as $\Delta T$ increases in this non-linear regime. Figure~\ref{fig:Ith_vs_dT}(b) shows $I_\mathrm{th}$ as a function of $\Delta T$ at 8~T, where the same kink presented in Fig.~3a,b is observed at low $\Delta T$. This inversion is suppressed when $\Delta T$ increases above a value of about 2~K. There are two contributions to this disappearance of the inversion in $I_\mathrm{th}$ at zero-bias. First, this can come from the global heating of the sample which raises the base temperature of the junction. By comparing the Kondo-peak height at different $\Delta T$ and $T$ (see \ref{subsec:T calibration}), we can roughly estimate that for a $\Delta T\approx5$~K, the junction temperature is increased to $T\approx 5$~K rather than 2~K. This is consistent with the observation in Fig.~\ref{fig:T_sweep_8T}(a), where thermocurrent inversion around zero-bias disappears also around 5~K. Another possible contribution for this absence of inversion at 8 T with $\Delta T>2$~K can also originate from the population of off-resonant levels by the thermal bias. This results the mixing for the Zeeman-split levels and smears out the inversion observed in $I_\mathrm{th}$.

 The above observations shows that our thermal bias dependent measurement successfully probes the non-linear thermoelectric regime, in combination with the Kondo effect. Particularly, the increasing of zero-bias $I_\mathrm{th}$ with a split spectral function upon the increase of $\Delta T$, demonstrates the breaking of particle-hole symmetry, or a large shift in the off-resonance levels upon the splitting of the Kondo spectral function at $\Delta T$. Such non-linear regime is addressed theoretically in Sec. \ref{sec:theory-intro}.

\pagebreak
\clearpage
\subsection{Temperature calibration by Kondo peak}\label{subsec:T calibration}
\begin{figure}[htb!]
    \centering
    \includegraphics{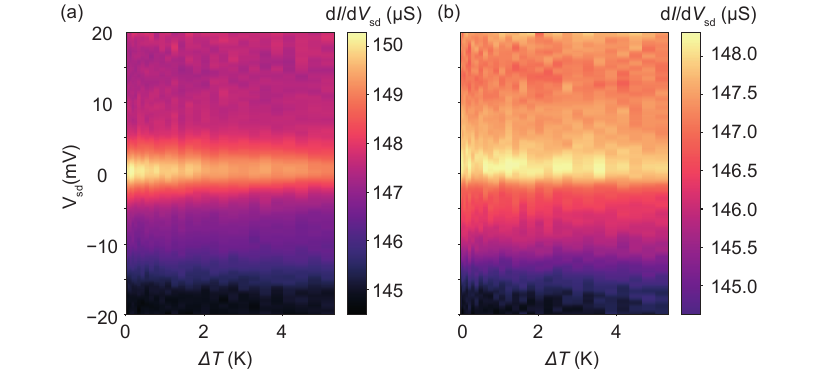}
    \caption{(a)-(b) Differential conductance of QD1 as a function of thermal bias $\Delta T$ at (a) B=0~T, and, (b) B=8~T.}
    \label{fig:dT_dIdV}
\end{figure}
\begin{figure}[hbt!]

    \includegraphics{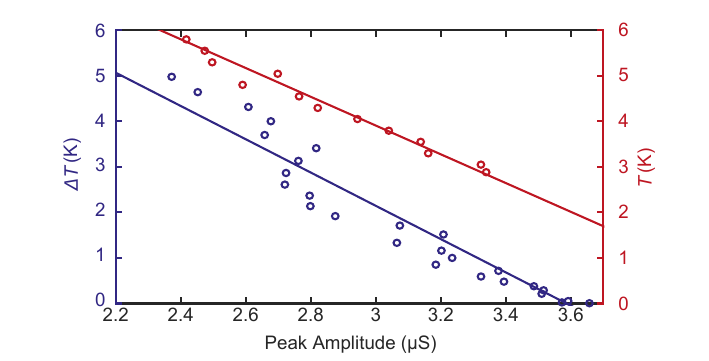}
    \caption{Estimated base temperature $T$(K), at different Kondo peak amplitudes and $\Delta T$.}
    \label{fig:EstimatedT}

\end{figure}

The thermal bias generated by the heater can also heat up the base temperature at the molecular quantum dot. To get an estimate of the local base temperature at this position, we use the Kondo peak amplitude at $B=0$~T as a conversion from thermal bias $\Delta T$ to the local base temperature $T$. The Kondo peak amplitudes at different $\Delta T$ and $T$ are extracted from Fig.~\ref{fig:dT_dIdV}(a) and Fig.~\ref{fig:T_sweep_8T}(a) respectively. In the case of Fig.~\ref{fig:dT_dIdV}, the global sample space was kept at 2~K while the local thermal bias was swept to 5.32~K; in the case of Fig.~\ref{fig:T_sweep_8T}, the local thermal bias was kept at 0.59~K while the global sample space temperature is swept from 2 to 15~K. We first plot $\Delta T$ as a function of peak amplitude, which reaches 0 at peak amplitude above 3.6 $\upmu \mathrm{S}$, indicating negligible heating at the molecular quantum dot (blue curve). We then plot $T$ as a function of peak amplitude (red curve). Therefore, for a fixed $\Delta T$ we can find the corresponding peak amplitude, which also corresponds to a fixed $T$ (Fig.~\ref{fig:EstimatedT}). For example, for $\Delta T =3$~K, we find a peak amplitude close to 2.8 $\upmu $S which translates to a base temperature of about 4.5 K.  Note that the red curve is offset vertically by 0.8~K such that the global sample temperature $T=2$~K corresponds to an amplitude of 3.6 $\upmu$S, where only negligible thermal bias is generated.

\pagebreak
\clearpage
\subsection{Stability diagram of the QD1}
\begin{figure}[htb!]
    \centering
    \includegraphics{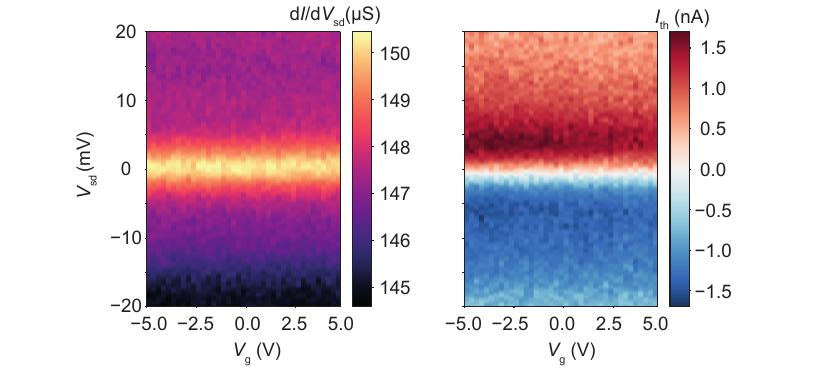}
    \caption{Stability diagram of QD1.}
    \label{fig:stability}
\end{figure}
The experimental results of the molecular quantum dot shown in the main text are measured without a gate. This is due to the small gate coupling as can be observed in Fig.~\ref{fig:stability}. The high differential conductance (2~$G_0$) here also indicates that the coupling between the source/drain leads to the dot is high as compared to the coupling to the gate.

\newpage
\clearpage
\section{Nonlinear thermocurrent of Kondo-correlated quantum dots in a magnetic field: Theory}
\label{sec:theory-intro}
This section provides a theory of the nonlinear thermocurrent $I_{\rm th}(V,T,B,\Delta T)$ of Kondo-correlated quantum dots in the presence of a magnetic field. For $\Delta T \ll T \ll T_{\rm K}$ and $V\ll T_{\rm K}$, we use an asymptotically exact higher-order Fermi-liquid theory \cite{Oguri2018a,Oguri2018b} combined with numerical renormalization group (NRG) calculations \cite{KWW1980a,Bulla2008} to evaluate $I_{\rm th}$ to lowest order in $\Delta T$ and $V$, while for $\Delta T, T$ and $V$ comparable to, or larger than $T_{\rm K}$ we use an approximate nonequilibrium Green function equation of motion method \cite{Roermund2010,Lavagna2015} to map out the trends in the voltage, temperature and thermal bias dependence of the thermocurrent at different magnetic fields (Sec.~\ref{subsec:EOM-method}). Within the former, we show that the zero-bias slope of the thermocurrent $s_1(B)=\left(\partial I_{\rm th}(V)/\partial V\right)_{V=0}$ changes sign at a certain field $B_{\rm th}$ as observed in the experiment. Moreover, we show that $B_{\rm th}$ coincides essentially exactly with the field $B_c$ at which the Kondo resonance in the differential conductance splits (Sec.~\ref{subsubsec:probe-splitting-Ith}) and also that $s_1(B)$ is a universal function of $B/T_{\rm K}$ (Sec.~\ref{subsubsec:universal-scaling-functions}). 
Significant deviations from these results only arise for weak correlations or outside the Kondo regime (e.g. on approaching the mixed valence regime). All NRG calculations used the full density matrix approach to spectral functions \cite{Hofstetter2000,Peters2006,Weichselbaum2007} and the self-energy method \cite{Bulla1998} and the Campo discretization scheme \cite{Campo2005} with discretization parameter $\Lambda=4$ and z-averaging with $N_z=4$.

\subsection{Model}
\label{subsec:model} 
Our starting point is the  two-lead single-level Anderson impurity model,
\begin{align}
  H= &H_{\rm dot}+H_{\rm leads}+ H_{\rm tunneling}.\label{eq:ham}
\end{align}
The first term,
  $H_{\rm dot}=\sum_{\sigma}\varepsilon_{0\sigma}n_{0\sigma} +Un_{0\uparrow}n_{0\downarrow}$,
describes the dot Hamiltonian, where $\varepsilon_{0\sigma}$ is the level energy with spin $\sigma$, $n_{0\sigma}=d^{\dagger}_{\sigma}d_{\sigma}$ is the occupation number for spin $\sigma={\uparrow,\downarrow}$ electrons on the dot, and $U=2E_{C}$ is the local Coulomb repulsion 
on the dot with $E_C=e^2/2C$ the charging energy of the dot with capacitance $C$ and $e$ is the elementary charge. The dot level energy $\varepsilon_{0\sigma}=\varepsilon_{0}-g\mu_BB\sigma/2$ includes the Zeeman shift
$\pm g\mu_BB/2$ of the level $\varepsilon_0$ where $g$ is the g-factor, $\mu_B$ is the Bohr magneton and $B$ is the local magnetic field.
We shall work at a fixed (dimensionless) gate voltage $V_{g}\equiv(\varepsilon_0+U/2)/\Gamma$ in the Kondo regime (as in the experiments) with $V_{g}=0$ corresponding to the particle-hole symmetric (mid-valley) point.
The second term, 
 $ H_{\rm leads}=\sum_{k\alpha\sigma}\epsilon_{k\alpha}c_{k\alpha\sigma}^{\dagger}c_{k\alpha\sigma}$,
 describes the leads, where $\alpha={L,R}$ labels the left (source) and right (drain) leads, and $\epsilon_{k\alpha\sigma}=\epsilon_{k}-\mu_{\alpha}$ is the kinetic energy of the lead electrons, each measured relative to their respective chemical potential $\mu_{\alpha=L,R}$. The bias voltage $V$ across the dot is defined in terms of the difference of chemical potentials between left and right leads via $-eV=\mu_L - \mu_R$. Furthermore, we assume that the voltage drops symmetrically across the leads, i.e., $\mu_L=-eV/2$ and $\mu_R=+eV/2$. In the notation of Ref.~\onlinecite{Oguri2018a}, where the voltage drop across the leads is written as $\mu_L=\alpha_L eV$ and $\mu_R=-\alpha_R eV$, our
choice corresponds to $\alpha_L=\alpha_R=-1/2$ with $\alpha_L+\alpha_R=-1$. The Fermi distributions of the leads are given by
$f_{\alpha}^{T_{\alpha}}(\omega)=f((\omega-\mu_{\alpha})/k_BT_{\alpha})$ where $f(x)=1/(1+e^x)$ is the Fermi function and $T_{\alpha=L,R}$ are
the temperatures of the left and right lead electrons. As in the experiment, we take the left lead to be the hot electrode with temperature $T_{L}=T_{R}+\Delta T$ where $\Delta T$ is the finite temperature difference across the dot. In the experiment, the thermal temperature difference is determined by the heater power used to heat the left electrode above the base temperature $T$, which is also the temperature of the right electrode, i.e., $T_R=T$. Finally, the last term in the Hamiltonian,
  $H_{\rm tunneling}=\sum_{k\alpha\sigma}t_{\alpha}(c_{k\alpha\sigma}^{\dagger}d_{\sigma} +d_{\sigma}^{\dagger}c_{k\alpha\sigma})$, 
describes the tunneling of electrons from the leads onto and off the dot with tunneling amplitudes $t_{\alpha}$. The corresponding tunneling rates are given by $\Gamma_{\alpha}= \pi N_{\rm F}t_\alpha^2$, where $N_{\rm F}=1/(2D)$ is the energy idependent lead electron density of states and $D=1$ is the half bandwidth of the leads. The total tunneling rate on and off the dot is denoted
  by $\Gamma=\Gamma_{L}+\Gamma_{R}$  and the asymmetry parameter of the lead couplings is defined by $\lambda=\Gamma_L/\Gamma_R$. Note that $\Gamma$ corresponds to the half-width of a
  Coulomb Blockade peak, the usual notation for theoretical work on the Anderson model \cite{Hewson1997}. In the literature on quantum dots, the full width $\tilde{\Gamma}=2\Gamma$ is often used.

\begin{figure}[htb!]
  \centering 
\includegraphics[width=0.98\textwidth]{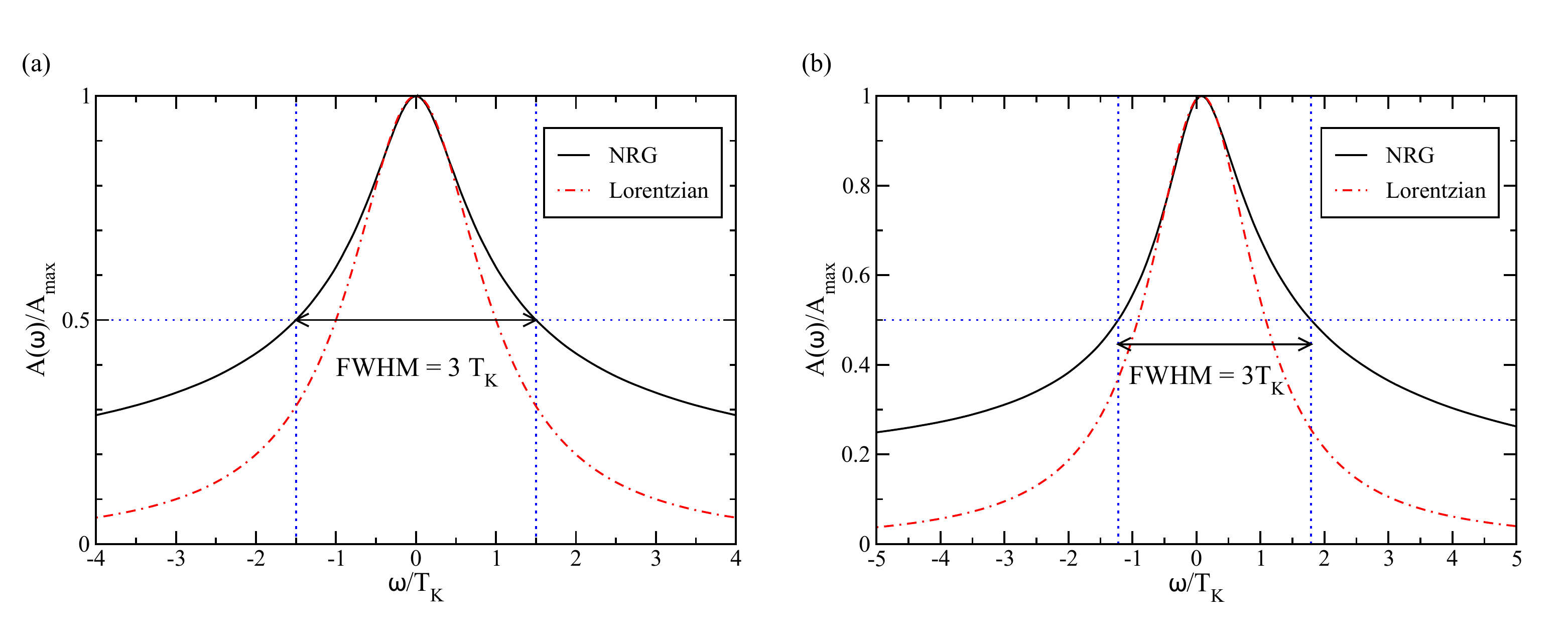}
\caption 
{Numerical renormalization group calculation of the low energy Kondo resonance in the $T=V=B=0$ normalized spectral function $A(\omega)/A_{\text{max}}$ of the Anderson impurity model for $U=8\Gamma$. $A_{\text{max}}$ is the maximum value of $A(\omega)$.
  The spectral function is plotted versus $\omega/T_{\rm K}$, where $T_{\rm K}$ is the  Kondo scale from the susceptibility in Eq.~(\ref{eq:tk-spin}).
  (a) Symmetric case ($\varepsilon_0=-U/2 = -4\Gamma$). (b) asymmetric case  ($\varepsilon_0=-3\Gamma$).
  In both cases, the FWHM of the Kondo resonance is very close to $3T_{\rm K}$, showing that $T_{\rm K}^{\mathrm{HWHM}}\approx 1.5 T_{\rm K}$ \cite{Costi2019a}. Note also, that a noninteracting resonant level of width $T_{\rm K}$, described by a Lorentzian (dashed-dotted lines), while approximately capturing the behavior of the Kondo resonance at $\omega\ll T_{\rm K}$, inevitably fails to describe the correct lineshape of the Kondo resonance at higher energies. For $\omega\gg T_{\rm K}$, the latter is better described by logarithmic tails \cite{Rosch2003b}.}
\label{fig:KRwidth}
\end{figure}
\subsection{Kondo scales}
\label{subsec:Kondo-scales}
For strong correlations $U/\Gamma \gg 1$ and local level position $\varepsilon_0/\Gamma \ll -1$, and $(\varepsilon_0+U)/\Gamma \gg +1$ the quantum dot level is approximately singly occupied with $n_{0}=\sum_{\sigma}\langle n_{0\sigma}\rangle\approx 1$ and has a spin $1/2$, resulting in the well known Kondo effect\cite{Hewson1997}, in which physical properties undergo a crossover from those of a weakly coupled system at $g\mu_BB,|e|V,k_{B}T \gg k_{B}T_{\rm K}$ to those of a strongly coupled system at $g\mu_BB,|e|V,k_{B}T \ll k_{B}T_{\rm K}$, where the low energy crossover scale $k_{B}T_{\rm K}$ is termed the Kondo scale, and $T_{\rm K}$ is the Kondo temperature. In applying theory predictions that use a certain definition of $T_{\rm K}$ to interpret an experiment that uses another definition of $T_{\rm K}$, one needs to know the relation between the different Kondo scales. We therefore briefly outline the the Kondo scales relevant to the present study and how they are related.

The first, $T_{\rm K}^{\mathrm{G}}$, is extracted from temperature dependence of the universal Kondo conductance curve, $G_{K}(T)$, via
$G(T_{\rm K}^{\mathrm{G}})=G(0)/2$. The second is defined below in Eq.~(\ref{eq:tk-spin}) via the $T=0$ spin susceptibility, and will be denoted as $T_{\rm K}$. It is the natural scale to use for the higher-order Fermi-liquid theory calculations below. Since $T_{\rm K}$ and
$T_{\rm K}^{\mathrm{G}}$ have been shown to lie within a few $\%$ of each other in the Kondo regime \cite{Merker2013}, we shall equate them
$T_{\rm K}\approx T_{\rm K}^{\mathrm{G}}$. Another commonly used Kondo scale is Haldane's expression from perturbative scaling on the asymmetic Anderson impurity model, $T_{\rm K}^{\mathrm{H}}=\sqrt{U\Gamma/2} e^{\pi \varepsilon_0 (\varepsilon_0+U)/2\Gamma U}$\cite{Hewson1997}. This shows explicitly the (approximate) dependence of the Kondo scale on the model parameters  $\Gamma, \varepsilon_0$ and $U$ in the Kondo limit and for $U\gg \Gamma$. We shall not use it in this paper. We  simply mention that it is of order $T_{\rm K}$ (e.g., for typical parameters $\varepsilon_0=-6\Gamma$ and $U=8\Gamma$ in the asymmetric Kondo regime we find $T_{\rm K}^{\mathrm{H}}\approx 0.84 T_{\rm K}$). Finally, since the splitting of the Kondo resonance in a magnetic field was first discussed in terms
of the scale $T_{\rm K}^{\mathrm{HWHM}}$, defined as the half-width at half-maximum of the $T=0$ Kondo resonance \cite{Costi2000}, we make
reference to this scale also. In particular, using the relation $T_{\rm K}^{\mathrm{HWHM}}\approx 1.5 T_{\rm K}$ [see Fig.~\ref{fig:KRwidth}] allows one to express the original estimate of the splitting field $g\mu_BB_c\approx 0.5 k_BT_{\rm K}^{\mathrm{HWHM}}$ \cite{Costi2000} in terms of recent estimates of this within higher-order Fermi-liquid theory which use instead $T_{\rm K}$. Thus, in terms of $T_{\rm K}$, we have $g\mu_BB_c\approx 0.75 k_BT_{\rm K}$, which is indeed the result found in higher-order Fermi-liquid theory \cite{Oguri2018b,Filippone2018}. The slower decay of the Kondo resonance away from the Fermi level, as compared to a
Lorentzian for a non-interacting level, accounts for the $50\%$ larger linewidth of the former. This is also expected due to the onset of logarithmic tails at higher energies $|\omega|\gg T_{\rm K}$ \cite{Rosch2003b}.

In a few places below, we shall, for the sake of brevity, omit physical units in expressions such as $B/T_{\rm K}$ or $V/T_{\rm K}$. In such cases, these should be understood as meaning $g\mu_BB/k_BT_{\rm K}$ and $eV/k_BT_{\rm K}$, respectively.

\subsection{Nonlinear transport quantities}
\label{subsec:currents}
The electrical current $I=I(V,T,B,\Delta T)$ is defined as the current in the left lead, i.e., $I\equiv I_L=-e\langle \dot{N}_{L}\rangle$, where $N_{L}=\sum_{k\sigma}\langle c_{kL\sigma}^{\dagger}c_{kL\sigma}\rangle$ is the number of electrons in the left lead. Likewise, the current in the right lead is defined by $I_R=-e\langle \dot{N}_{R}\rangle$, where $N_{R}=\sum_{k\sigma}\langle c_{kR\sigma}^{\dagger}c_{kR\sigma}\rangle$ is the number of electrons in the right lead. By current conservation, we have that $I_L+I_R=0$, or $I_R=-I_L = -I$. An explicit expression for the
current can be derived in terms of the lesser, retarded and advanced Green functions of the dot \cite{Hershfield1992,Meir1992}. However, since we use proportionate couplings $\Gamma_{L}=\lambda \Gamma_R$, the dependence of the current on the lesser Green function can be eliminated \cite{Jauho1994}, resulting in an expression for the current in terms of the difference of retarded and advanced Green functions, i.e., in terms of solely
the nonequilibrium spectral function $A_{\sigma}(\omega,V,T,B,\Delta T)$\footnote{The dependence of the dot spectral function on the thermal bias $\Delta T$ is implicit through the coupling of the dot to left and right reservoirs with temperature difference $T_L-T_R=\Delta T$} of the dot\cite{Hershfield1992,Meir1992,Jauho1994}, 
\begin{align}
  I(V,T,B,\Delta T) = & -\frac{2e}{\hbar}\sum_{\sigma}\int d\omega\frac{\Gamma_L\Gamma_R}{\Gamma}\left[f_{L}^{T+\Delta T}(\omega)-f_{R}^{T}(\omega)\right]A_{\sigma}(\omega,V,T,B,\Delta T),
        \label{eq:electrical-current}\\
   = & \gamma \int d\omega \left[f_{L}^{T+\Delta T}(\omega)-f_{R}^{T}(\omega)\right]A(\omega,V,T,B,\Delta T).
        \label{eq:electrical-current-1}
\end{align}
In the above, $\gamma=-2e \Gamma_L\Gamma_R/(\hbar\Gamma)$ and 
$A(\omega,V,T,B,\Delta T)=\sum_{\sigma}A_{\sigma}(\omega,V,T,B,\Delta T)$ is
the total spectral function of the dot. 
The thermocurrent, $I_{\rm th}$, is then defined as the difference of the electrical currents at finite  and zero thermal bias $\Delta T$,
\begin{align}
I_{\rm th}(V,T,B,\Delta T) = & I(V,T,B,\Delta T)-I(V,T,B,\Delta T=0).\label{eq:Ith-subtraction}
\end{align}
The differential conductance
\begin{align}
G(V,T,B)= &\frac{dI}{dV}\label{eq:dIdV},
\end{align}
is obtained from the current at zero thermal bias $\Delta T=0$. From Eq.~(\ref{eq:electrical-current-1}), the linear conductance $G_l=G(V=0,T,B)$ at $T=B=0$ is given by
\begin{align}
G_l= & \gamma e A(0,0,0,0) = \frac{2e^2}{h}\frac{4\Gamma_L\Gamma_R}{\Gamma^2}\sin^{2}(\pi n_0/2)\leq G_{\rm max},\label{eq:dIdV_max}
\end{align}
where $G_{\rm max}= G_0 4\Gamma_L\Gamma_R/\Gamma^2$, $G_0=2e^2/h$ is the conductance quantum and we used the Fermi-liquid result for the spectral function $A(0,0,0,0)=\frac{2}{\pi\Gamma}\sin^2(\pi n_0/2)\leq \frac{2}{\pi \Gamma}$ (with equality at mid-valley when $n_0=1$). We shall use $G_{\rm max}$ as a normalization
in showing $dI/dV$ results in Secs.~\ref{subsubsec:b-dependence}-\ref{subsubsec:prop-Ith}. Note that while $G_{\rm max}=G_0$ for symmetric coupling to the leads, for the molecular quantum dot in the experiment, the lead couplings are highly asymmetric $\Gamma_{L}=\lambda \Gamma_R \ll \Gamma_R$ and $G_{\rm max} \ll G_0$.

\subsection{Nonlinear thermocurrent for \texorpdfstring{$\Delta T\ll T$}{} and  asymmetric couplings}
\label{subsec:small-DeltaT-limit}
In the limit of a small thermal bias $\Delta T \ll T$ (and arbitrary voltage bias), the thermocurrent  in  Eq.~(\ref{eq:Ith-subtraction}) can be
approximated as
\begin{align}
  I_{\rm th} = & \gamma \int 
                         d\omega\left[f^{T+\Delta T}_L(\omega)-f^{T}_{R}(\omega)\right]A(\omega,V,T,B,\Delta T)\nonumber\\
   - & \gamma \int  d\omega\left[f^{T}_L(\omega)-f^{T}_{R}(\omega)\right]A(\omega,V,T,B, \Delta T=0)\label{eq:Ith-exact}\\
\approx & \gamma \int  d\omega\Delta T \left(\frac{\partial f_L^{T+\Delta T}(\omega)}{\partial \Delta T}\right)_{\Delta T=0} A(\omega,V,T,B, \Delta T=0)\nonumber\\
    + & \gamma \int  d\omega\Delta T \left[f^{T}_L(\omega)-f^{T}_{R}(\omega)\right]\left(\frac{\partial A(\omega,V,T,B, \Delta T)}{\partial\Delta T }\right)_{\Delta T =0},\label{eq:Ith-linearDeltaT}
\end{align}
  where we have omitted the arguments $V,T$ and $B$ of $I_{\rm th}$ for simplicity of notation and we have expanded the Fermi function $f_{L}^{T+\Delta T}(\omega)$ and the  spectral function $A(\omega,V,T,B,\Delta T)$ to order $\Delta T$. Using,
  \begin{align}
    \left(\frac{\partial f_L^{T+\Delta T}(\omega)}{\partial \Delta T}\right)_{\Delta T=0} = & \frac{\omega-\mu_L}{T}\left(-\frac{\partial f_L^{T+\Delta T}(\omega)}{\partial\omega}\right)_{\Delta T = 0} = \frac{\omega-\mu_L}{T}\left(-\frac{\partial f_L^{T}(\omega)}{\partial\omega}\right) ,\nonumber 
    \end{align}
  we have,
  \begin{align}
  I_{\rm th} =  & \frac{\gamma\Delta T}{T} \int  d\omega
  (\omega -\mu_L)\left(-\frac{\partial f_L^{T}(\omega)}{\partial\omega}\right) 
                         A(\omega,V,T,B,\Delta T=0)\nonumber\\
    + & \gamma \Delta T\int  d\omega \left[f^{T}_L(\omega)-f^{T}_{R}(\omega)\right]\left(\frac{\partial A(\omega,V,T,B, \Delta T)}{\partial\Delta T }\right)_{\Delta T =0},\label{eq:Ith-linearDeltaT-a}\\
    & \equiv I_{\rm th}^a + I_{\rm th}^b
\end{align}
 The above  equation is valid for $\Delta T \ll T$ and arbitrary $V$, however, we are mainly interested in explaining  the experimentally observed field induced sign change in the slope of the
    thermocurrent with respect to $V$ around $V=0$ at low temperatures $\Delta T \ll T\ll T_{\rm K}$ within the approach of Sec.~\ref{subsec:higher-order+NRG}. Further restricting to $\Delta T \ll T\ll T_{\rm K}$, a Sommerfeld expansion of the first term, $I_{\rm th}^a/\Delta T$, yields to lowest order in $T$,
    \begin{align}
      I_{\rm th}^{a} = &
      \frac{\gamma\Delta T}{T}\int d\omega (\omega -\mu_L)\left(-\frac{\partial f_L^{T}(\omega)}{\partial\omega}\right) 
                         A(\omega,V,T,B,\Delta T=0)\nonumber\\
      \approx & \frac{\gamma\Delta T}{T}\frac{\pi^2(k_{B}T)^2}{3}\left(\frac{\partial A(\omega,V,T=0,B,\Delta T=0)}{\partial\omega}\right)_{\omega=\mu_L}\label{eq:Sommerfeld1}\\
                                                               \equiv  & \gamma\frac{\pi^2k_{B}^2}{3} T\Delta T s(\omega=\mu_L,V,T=0,B),\label{eq:Sommerfeld2}
  \end{align}
  where $s(\omega=\mu_L,V,T=0,B)$ in (\ref{eq:Sommerfeld2}) denotes the derivative of the $T=0$ spectral function  at $\omega=\mu_L$ on the RHS of (\ref{eq:Sommerfeld1}). Expanding $s(\omega=\mu_L,V,T=0,B)$  in powers of $V$, i.e., 
  $s(\omega=\mu_L,V,T=0,B)=s_0(B)+ \tilde{s}_1(B)V +\dots$, gives
    \begin{align}
      I_{\rm th}^a \approx
      & \gamma\frac{\pi^2k_{B}^2}{3} T\Delta T \left[ s_0(B) + \tilde{s}_1^a(B)V + \dots\right].\label{eq:Sommerfeld3}
  \end{align}

The second term in Eq.~(\ref{eq:Ith-linearDeltaT}) also gives a contribution to the thermocurrent which is linear in $V, T$ and $\Delta T$ for $\Delta T\ll T\ll T_{\rm K}$,
\begin{align}
I_{\rm th}^b & = \gamma\Delta T \int  d\omega \left[f^{T}_L(\omega)-f^{T}_{R}(\omega)\right]\left(\frac{\partial A(\omega,V,T,B, \Delta T)}{\partial\Delta T }\right)_{\Delta T =0}\label{eq:Sommerfeld3a}\\ 
& \approx \gamma\Delta T\frac{\pi^2k_{B}^2}{3} [\tilde{s}_1^b(B)TV+\dots],\label{eq:Sommerfeld3b}
\end{align}
which defines $\tilde{s}_1^b(B)$. The total thermocurrent $I_{\rm th}=I_{\rm th}^a+I_{\rm th}^b$ is given by
   \begin{align}
      I_{\rm th}(\Delta T, V, T ,B) \approx
      & \gamma\frac{\pi^2k_{B}^2}{3} T \Delta T\left[ s_0(B) + s_1(B)V + \dots\right],\label{eq:Sommerfeld4}
  \end{align}
where $s_1(B)=\tilde{s}_1^a(B)+\tilde{s}_1^b(B)$. For $V\ll T_{\rm K}$ and $\Delta T \ll  T \ll T_{\rm K}$, explicit expressions for $s_0(B)$ and $s_1(B)$ will be obtained below within higher-order Fermi-liquid theory and evaluated numerically exactly within the NRG.
  
  Finally, we note that for $V=0$, the second term in Eq.~(\ref{eq:Ith-linearDeltaT}) vanishes, and the first term reduces to the expected linear response thermocurrent ($\Delta T\ll T$ finite),
\begin{align}
    I_{\rm th}(\Delta T,V=0,T,B) = G(T,B) S(T,B)\Delta T,\label{eq:linear-transport-thermocurrent}
\end{align} 
where $G(T,B)$ and $S(T,B)$ are the linear conductance and linear Seebeck coefficient, respectively.

\subsection{Nonlinear thermocurrent and dI/dV: higher-order Fermi-liquid theory and NRG calculations}
\label{subsec:higher-order+NRG}  
In order to evaluate the nonlinear thermocurrent in Eq.~(\ref{eq:Sommerfeld4}) to order $V$, and hence $\left(\partial I_{\rm th}(V)/\partial V\right)_{V=0}$, the spectral function $A(\omega,V,T,B)=\sum_{\sigma}A_{\sigma}(\omega,V,T,B)$ is required to order $\omega^2, V^2$ and $T^2$ [see Eq.~(\ref{eq:Sommerfeld1})].  Moreover, this is required for the case of asymmetric lead couplings $\Gamma_L\ll \Gamma_R$ , as is relevant to the molecular junction in the experiment (the case $\Gamma_L\gg \Gamma_R$ will also be considered). This can be obtained
from the expression for the local level retarded self-energy of the quantum dot, $\Sigma^{r}_{\sigma}(\omega,V,T,B)$, to order
$\omega^2,V^2$ and $T^2$ from Ref.~\onlinecite{Oguri2018a}. Substitution of this into the expression for the spectral function,
\begin{align}
  A_{\sigma}(\omega,V,T,B) = & -\frac{1}{\pi}{\rm Im} \left[\frac{1}{\omega - \varepsilon_{0\sigma}+i(\Gamma_L+\Gamma_R)-\Sigma_{\sigma}^{r}(\omega,V,T,B)}\right],\label{eq:Spec}
  \end{align}
yields, after some algebra, 
\begin{align}
  \pi \Gamma A_{\sigma}(\omega,V,T,B) = &  a_{0\sigma} + a_{1\sigma}\frac{\omega}{T_{\rm K}} + a_{2\sigma}\left(\frac{\omega}{T_{\rm K}}\right)^2\nonumber\\ + & a_{3\sigma}\left[ (\beta_0+3\alpha^2)\left(\frac{V}{T_{\rm K}}\right)^2 + \pi^2 \left(\frac{T}{T_{\rm K}}\right)^2\right]\nonumber\\
  + & a_{4\sigma}\alpha \frac{V}{T_{\rm K}} + 2 a_{5\sigma} \alpha \frac{V\omega}{T_{\rm K}^2},\label{eq:Spec-fliq}
\end{align}
where $T_{\rm K}$ is defined via the $T=0$ static spin susceptibility 
\begin{align}
\chi_s(T=0) = & (g \mu_B)^2/4k_B T_{\rm K},\label{eq:tk-spin} 
\end{align}
the (field-dependent) coefficients $a_{i\sigma}, i=0,\dots,5$ will be defined below, and the constants $\beta_0$ and $\alpha$
are related to the lead couplings and to how the bias voltage drops across the leads \cite{Oguri2018a},
\begin{align}
  \beta_0 = & \frac{3\Gamma_L \Gamma_R}{(\Gamma_L+\Gamma_R)^2},\\
  \alpha = & \frac{\alpha_L\Gamma_L-\alpha_R\Gamma_R}{\Gamma_L+\Gamma_R}.
\end{align}
The expression (\ref{eq:Spec-fliq}) generalizes that in Ref.~\onlinecite{Oguri2018a} [Eq.(5.4) therein] to arbitrary lead couplings and arbitrary voltage drops
across the leads, and reduces to that in the limit of both a symmetric coupling to the leads ($\Gamma_L=\Gamma_R$) and a symmetric voltage drop across the leads ($\alpha_L=\alpha_R=\pm 1/2$). For the usual situation encountered in quantum dots, in which the voltage drop across the leads is symmetric, i.e., for $\alpha_L=\alpha_R=\pm 1/2$, we have that $\beta_0+3\alpha^2=3/4$, independent of the value of the lead couplings $\Gamma_{\alpha=L,R}$. Hence, the prefactor of the $V^2$ term in the spectral function (\ref{eq:Spec-fliq}) also becomes independent of the lead couplings. The dependence on the
latter, then only enters explicitly in (\ref{eq:Spec-fliq})  via the terms proportional to $\alpha V$
and $\alpha V\omega$. For molecular quantum dots, such as those we address in this paper, the lead coupling asymmetry is large and $\alpha$ is therefore finite. In this case,  the terms proportional to $\alpha V$ and $\alpha V\omega$ in Eq.~(\ref{eq:Spec-fliq})  need to be taken into account in calculating $dI/dV$ and $I_{\rm th}$ up to the specified order in $V$ and $T$.

The explicit expressions for the (field-dependent) coefficients $a_{i\sigma}, i=0,\dots,5$ appearing in Eq.~(\ref{eq:Spec-fliq}), read
\begin{align}
  a_{0\sigma} = & \sin^2(\delta_\sigma), \label{eq:a0}\\
  a_{1\sigma} = & \pi T_{\rm K}\sin(2\delta_\sigma)\chi_{\sigma\sigma}, \label{eq:a1}\\
  a_{2\sigma} = & \left(\pi T_{\rm K}\right)^2\left[\cos(2\delta_\sigma)(\chi_{\sigma\sigma}^2+\frac{1}{2}\chi_{\uparrow\downarrow}^2)- \frac{\sin(2\delta_\sigma)}{2\pi}\frac{\partial\chi_{\sigma\sigma}}{\partial \varepsilon_{0\sigma}}\right]=a_{2\sigma}^{(2)}+a_{2\sigma}^{(3)}, \label{eq:a2}\\
  a_{3\sigma} = & \frac{\left(\pi T_{\rm K}\right)^2}{3}\left[\frac{3}{2}\cos(2\delta_\sigma)\chi_{\uparrow\downarrow}^2- \frac{\sin(2\delta_\sigma)}{2\pi}\frac{\partial\chi_{\uparrow\downarrow}}{\partial \varepsilon_{0,-\sigma}}\right]=a_{3\sigma}^{(2)}+a_{3\sigma}^{(3)}, \label{eq:a3}\\                   
  a_{4\sigma} = & 2\pi T_{\rm K}\sin(2\delta_\sigma)\chi_{\uparrow\downarrow}, \label{eq:a4}\\
  a_{5\sigma} = & \left(\pi T_{\rm K}\right)^2\left[\cos(2\delta_\sigma)(\chi_{\sigma\sigma}\chi_{\uparrow\downarrow}-\frac{1}{2}\chi_{\uparrow\downarrow}^2)- \frac{\sin(2\delta_\sigma)}{2\pi}\frac{\partial\chi_{\uparrow\downarrow}}{\partial \varepsilon_{0\sigma}}\right]=a_{5\sigma}^{(2)}+a_{5\sigma}^{(3)}, \label{eq:a5}
\end{align}
where the static susceptibilities $\chi_{\sigma\sigma'}(T)=\int_{0}^{1/T}d\tau \langle\delta n_{0\sigma}(\tau) \delta n_{0\sigma'}(0) \rangle$, with $\delta n_{0\sigma}(\tau)=n_{0\sigma}(\tau) -\langle n_{0\sigma}(\tau)\rangle$, their derivatives, as well as the phase shifts, $\delta_\sigma=\pi n_{0\sigma}$, are evaluated at $T=0$. The latter can all be calculated, essentially exactly, within the NRG approach \cite{Bulla2008}. 
For later convenience, we also define $a_{i}=\sum_{\sigma}a_{i\sigma}$.

The meaning of the different terms in  (\ref{eq:a0})-(\ref{eq:a5}) is as follows: terms involving static susceptibilities $\chi_{\sigma_2,\sigma_3}$ describe two-body fluctutations (denoted as $a_{i\sigma}^{(2)}$), whereas those involving the nonlinear static susceptibilities $\chi_{\sigma_1\sigma_2\sigma_3}^{3}=\partial \chi_{\sigma_2\sigma_3}/\partial \varepsilon_{0\sigma_1}$ describe three-body fluctuations (denoted as $a_{i\sigma}^{(3)}$) \cite{Oguri2018a}. The competition between the two- and  three-body terms ultimately drives the splitting of the Kondo resonance in $dI/dV$ at the field $B_c$ (next section), as recently demonstrated for carbon nanotube quantum dots in the $SU(2)$ Kondo regime\cite{Hata2021}. As will become evident below, the same competition accounts for the observed sign change of the zero-bias thermocurrent slope at $B_{\rm th}\approx B_c$ for the molecular quantumm dot in the main text.

\subsubsection{Splitting of the Kondo resonance in \texorpdfstring{$dI/dV$}{} and \it{A}}
\label{subsec:splitting-dIdV}
Before presenting results for the thermocurrent, it is useful to summarize the understanding of the splitting of the Kondo resonance in $dI/dV$ and in the spectral function within the above higher-order (microscopic) Fermi-liquid theory \cite{Oguri2018a}. Similar results  can be obtained within an extension of the phenomenological Fermi-liquid theory for the Kondo model \cite{Nozieres1974} to the Anderson model away from particle-hole symmetry \cite{Mora2015,Filippone2018}.

Consider first the differential conductance $G=dI/dV$.  Substituting the spectral function from Eq.~(\ref{eq:Spec-fliq}) into the expression for the current (at $\Delta T =0$) in Eq.~(\ref{eq:electrical-current}), and evaluating  $G=dI/dV$, yields to order $V^2$ and $T^2$\footnote{This expression, for asymmetric couplings, has recently also been obtained in Ref.~\cite{Tsutsumi2021}. The general form of this expression is also known, but with approximate coefficients, from previous work \cite{Sela2009,Aligia2011}.} 
\begin{align}
G = \frac{dI}{dV} = & g_0\left[a_0 - c_{T}\pi^2\left(\frac{T}{T_{\rm K}}\right)^2 - c \frac{V}{T_{\rm K}} - c_{V}\left(\frac{V}{T_{\rm K}}\right)^2\right],\label{eq:dIdV-fliq}
\end{align}
where $g_0$ is given by 
$$
g_0 = G_0 \frac{2\Gamma_L\Gamma_R}{(\Gamma_L+\Gamma_R)^2},
$$
and the curvature coefficients $c_{T}$ and $c_{V}$ and the coefficient $c$ of the linear voltage term are given by, 
\begin{align}
  c_{T} = & -\frac{1}{3}(a_{2} +3 a_{3}),\label{eq:ct}\\
  c_{V} = & -\left[\frac{1}{4}a_{2}+3(\beta_0+3\alpha^2) a_{3}\right] \to  -\frac{1}{4}(a_{2}+9 a_{3}),\label{eq:cv}\\  
  c  = & -2\alpha a_{4}.\label{eq:c}
\end{align} 
The second expression for $c_{V}$ in (\ref{eq:cv}) is that for a symmetric voltage drop across the leads ($\alpha_L=\alpha_R=\pm 1/2$). For both a symmetric voltage drop across the leads ($\alpha_L=\alpha_R=\pm 1/2$) and
equal lead couplings ($\Gamma_L=\Gamma_R$), i.e., for $\alpha=0$, the results of Ref.~\onlinecite{Oguri2018a} are recovered (note that due to slightly different definitions, our $c_{T,V}$ differ from those in Eq.~(5.12) of  Ref.~\onlinecite{Oguri2018a} by a factor of $2$).

The splitting of the Kondo resonance in $dI/dV$ with increasing magnetic field can be formulated precisely in the case where the
resonance lies at $V=0$, i.e., when the coefficient, $c=-2\alpha a_4$, of the linear voltage term in (\ref{eq:c}) vanishes.
This is the case, either at particle-hole symmetry (mid-valley gate voltage), when $a_4=0$, or when the
voltage drop across the leads is symmetric ($\alpha_L=\alpha_R=\pm 1/2$) and the lead couplings are equal ($\Gamma_L=\Gamma_R$).
In these cases, a splitting in $G(V,T=0,B)$ versus $V$ upon increasing $B$ occurs when a 
local maximum in this function for $B=0$ at $V=V_0=0$ turns into a local minimum upon increasing $B$, i.e., 
when the curvature coefficient $c_V$ changes sign at some field $B=B_V$. 
Similarly, a splitting in $G(V=0,T,B)$ versus $T$ upon increasing $B$ occurs when a 
local maximum in this function for $B=0$ at $T=T_0=0$ turns into a local minimum upon increasing $B$, i.e., when 
the curvature coefficient $c_T$ changes sign at some field $B=B_T$.

In all other situations, the ``zero-bias'' peak in $dI/dV$, just like the Kondo resonance in the spectral function, will not lie exactly 
at $V=0$ (or $\omega=0$, in the case of the spectral function) \cite{Aligia2015}.
For these cases we shall follow convention and simply define the splitting to occur
as above, namely at the fields where the corresponding curvature coefficients change sign. We expect that this definition is
reasonable deep in the Kondo regime where the Kondo resonance is pinned to within  a fraction of $T_{\rm K}$
around $V=0$ in $dI/dV$ (or a fraction of $T_{\rm K}$ around $\omega=0$ in the spectral function).
\begin{figure}[thb!]
  \centering 
\includegraphics[width=0.8\textwidth]{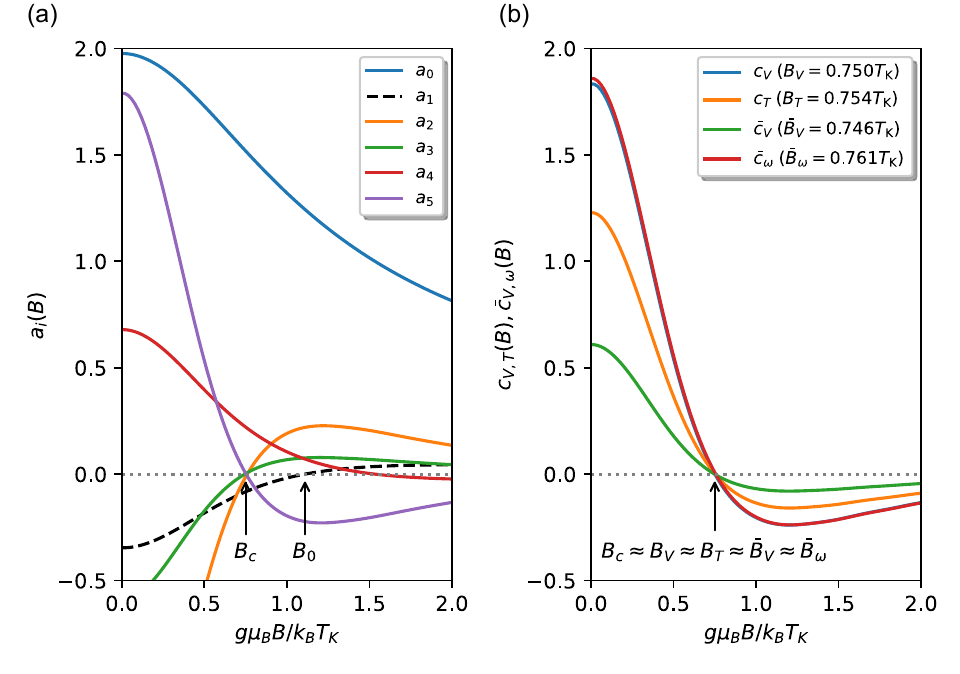}
\caption 
{ NRG calculation of (a) the $B$-dependence of the coefficients $a_{i}=\sum_{\sigma}a_{i\sigma}$ appearing in the spectral function (\ref{eq:Spec-fliq}), and, (b), the $B$-dependence of the curvature coefficients $c_V,c_T,\bar{c}_V,\bar{c}_{\omega}$ appearing 
  in the spectral function ($\bar{c}_V,\bar{c}_{\omega}$) and in $dI/dV$ ($c_V,c_T$). The coefficients $a_2$ and $a_3$ (and also $a_5$) enter the curvatures for the spectral function and $dI/dV$ and change sign at $B_c\approx 0.75T_{\rm K}$. 
  All curvatures in (b) change sign at approximately the same field $B_c\approx 0.75 T_{\rm K}$. The coefficient 
  $a_1$ (dashed line) determines the sign of the $V=0$ thermocurrent and changes sign at a gate-voltage dependent field $B_0(V_g)> B_c$ (here, for $V_g=-1$, $B_0(V_g=-1)\approx 1.11 T_{\rm K}$), 
  see  Sec.~\ref{subsubsec:contrast-offset+slope} and Refs.~\onlinecite{Costi2019a}. The parameters used ($U=8\Gamma$, $\varepsilon_0=-5\Gamma, V_g=-1$)   correspond to typical values for the Kondo regime. The NRG calculations used a discretization parameter $\Lambda=4$ and z-averaging with $N_z$=4 \cite{Campo2005}.
}
\label{fig:acoeff+ccoeff}
\end{figure}

The splitting of the Kondo resonance in the spectral function can be described in an analogous manner to that above for $dI/dV$.
For this purpose, we write the  total spectral function, $A(\omega,V,T,B)=\sum_{\sigma}A_{\sigma}(\omega,V,T,B)$, in a similar way to that in (\ref{eq:dIdV-fliq}) for $dI/dV$:
 \begin{align}
\pi\Gamma A(\omega,V,T,B) = & a_0 + a_1\frac{\omega}{T_{\rm K}}  -\bar{c}_{\omega}\left(\frac{\omega}{T_{\rm K}}\right)^2 - \bar{c}_{V}\left(\frac{V}{T_{\rm K}}\right)^2 - \bar{c}_{T}\pi^2\left(\frac{T}{T_{\rm K}}\right)^2\nonumber\\+ &\alpha a_4\frac{V}{T_{\rm K}}+2\alpha a_5\frac{V\omega}{T_{\rm K}^2},\label{eq:Spec-splitting}   
 \end{align}
where the curvature coefficients $\bar{c}_{\omega}, \bar{c}_{V}$ and $\bar{c}_T$ are given by,
\begin{align}
  \bar{c}_{\omega} = & -a_{2},\label{eq:cbar_Omega}\\ 
  \bar{c}_{V} = & -(\beta_0+3\alpha^2)a_{3},\label{eq:cbar_V}\\
  \bar{c}_T = & -a_{3}.\label{eq:cbar_T}
\end{align}
As with the splitting fields $B_V$ and $B_T$ for $dI/dV$, we similarly define 
analogous fields $\bar{B}_V, \bar{B}_T$ and $\bar{B}_{\omega}$
for the spectral function (\ref{eq:Spec-splitting}), as the fields 
where the curvature coefficients $\bar{c}_V$, $\bar{c}_T$ and $\bar{c}_\omega$ change sign \cite{Filippone2018}.

We now estimate the splitting fields via the curvatures defined above. 
From Eqs.~(\ref{eq:cbar_V})-(\ref{eq:cbar_T}), the curvatures $\bar{c}_V$ and $\bar{c}_T$ are proportional, implying that $\bar{B}_V=\bar{B}_T$ \cite{Filippone2018}.
The four independent curvatures $c_V,c_T$ for $dI/dV$ and $\bar{c}_V,\bar{c}_{\omega}$ for the spectral function,
are calculated by evaluating the coefficients $a_{i}(B) =\sum_{\sigma}a_{i\sigma}(B)$ within the
NRG approach for a typical parameter set in the Kondo regime.  The results are shown in Fig.~\ref{fig:acoeff+ccoeff}(a) for the coefficients $a_i(B)$ and in Fig.~\ref{fig:acoeff+ccoeff}(b) for the curvatures. We see that all four curvatures change sign at approximately the same field  $B_{V}\approx B_{T}\approx \bar{B}_{V}(= \bar{B}_{T})\approx \bar{B}_{\omega}$ \cite{Oguri2018a,Filippone2018}, with the difference between these values lying at the one $\%$ level for $U/\Gamma=8$ and decreasing for larger values of $U/\Gamma$. Hence, the splitting of the Kondo resonance (``zero-bias peak'') in $dI/dV$ (vs $V$ or $T$) and in the spectral function $A(\omega,V,T,B)$ (vs $V,T$ or $\omega$), occurs at essentially the same magnetic field. We denote this field by $B_c$. In terms of the definition of $T_{\rm K}$ in Eq.~(\ref{eq:tk-spin}), this field is estimated as $B_c\approx 0.75T_{\rm K}$ [see Fig.~\ref{fig:acoeff+ccoeff}(b)], consistent with estimates from Refs.~\onlinecite{Oguri2018a,Filippone2018} [the former found $B_c\approx 0.76 T_{\rm K}$ while the latter found $B_c\approx 0.75T_{\rm K}$], and also with that found for the spectral function of the Kondo model $B_{c}/T_{\rm K}^{\mathrm{HWHM}}\approx 0.5$ \cite{Costi2000}, upon using $T_{\rm K}^{\mathrm{HWHM}}\approx 1.5T_{\rm K}$ (see Sec.~\ref{subsec:Kondo-scales}).

\begin{figure}[thb!]
  \centering 
\includegraphics[width=0.8\textwidth]{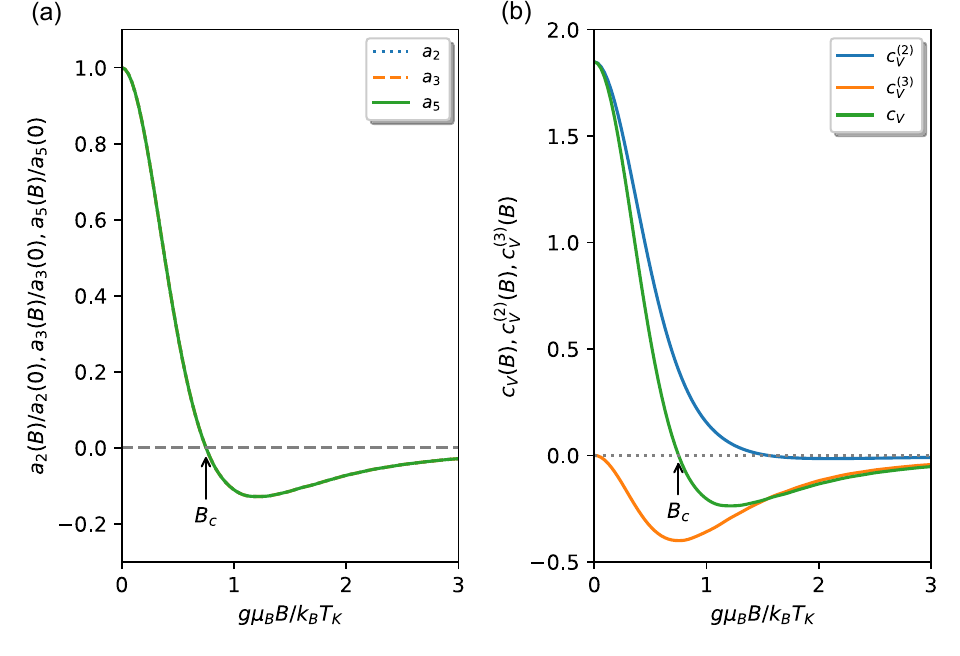}
\caption 
{ (a) The $B$-dependence of the normalized coefficients $a_2(B)/a_2(0),a_3(B)/a_3(0)$ and $a_5(B)/a_5(0)$ appearing in the spectral function (\ref{eq:Spec-fliq}). (b) The $B$-dependence of the curvature coefficient $c_V$ and its two- and three-body contributions,
$c_{V}^{(2)}$ and $c_{V}^{(3)}$, respectively. The parameters used ($U=16\Gamma$, $\varepsilon_0=-7\Gamma,V_g=-1$) 
correspond to typical values for the asymmetric Kondo regime. NRG parameters as in Fig.~\ref{fig:acoeff+ccoeff}.
}
\label{fig:a-scaled}
\end{figure}

The near exact coincidence of the splitting field for the various curvatures is ultimately due to the universality of the coefficients $a_2,a_3$ (and  $a_5$) in the Kondo regime, where one can observe, to within around one $\%$ accuracy for $U/\Gamma=8$, that 
\begin{equation}
        a_2\approx 3a_3\approx -a_5\;\; (\mbox{Kondo regime}),\label{eq:acoeff-prop}
\end{equation} 
with equality setting in for $U/\Gamma\to\infty$. Thus, any linear combination of these (such as the curvatures), scaled by their $B=0$ value, results, to within the same accuracy, in a single universal function of $B/T_{\rm K}$  which changes sign at the universal field $B_c$. This is illustrated in Fig.~\ref{fig:a-scaled}(a).  The two- and three-body contributions to the curvature
$c_V$ are also shown in Fig.~\ref{fig:a-scaled}(b) as a function of magnetic field, demonstrating the statement made at the end of Sec.~\ref{subsec:higher-order+NRG}, that the competition between these contributions drives the splitting of the Kondo resonance in $dI/dV$. Indeed, the three-body contributions contribute maximally at  $B=B_\mathrm{c}$.
\subsubsection{Experimental estimates of the splitting field $B_c$}
\label{subsubsec:Bc-estimates}
Experiments which attempt to verify the result $g\mu_{\mathrm{B}}B_c/k_{\mathrm{B}}T_{\rm K}^{\mathrm{HWHM}}\approx 0.5$ (or, equivalently, $g\mu_{\mathrm{B}}B_c/k_{\mathrm{B}}T_{\rm K}\approx 0.75$) by analyzing $dI/dV$ data
\cite{Kogan2004,Amasha2005,Houck2005,Quay2007,Liu2009,Kretinin2011} have so far been hampered by the intrinsic problem 
of extracting $B_c$ from  $dI/dV$ data [discussed in the main text in connection with Figs.~2(e)-2(f)], and by  the absence of good estimates for $T_{\rm K}^{\mathrm{HWHM}}$ in terms of the more usual scale used in quantum dots, that from the conductance $T_{\rm K}^{\mathrm{G}}\approx T_{\rm K}$. This has resulted in estimates for $g\mu_{\mathrm{B}}B_\mathrm{c}/k_{\mathrm{B}}T_{\mathrm{K}}$ varying from $0.5$ to $1.5$ \cite{Kretinin2011}. With the precise relation between these different scales, as discussed in Sec.~\ref{subsec:Kondo-scales}, it may be possible to obtain
more precise estimates of $B_c$.  Nevertheless, recent high precision measurements analyzing $dI/dV$ data and working with a certain Kondo scale $T_{\rm K}'$, find $g\mu_{\mathrm{B}}B_{\mathrm{c}}/k_{\mathrm{B}}T_{\rm K}'=0.23$ whereas that extracted from their $c_V(B_c)=0$ yields instead $g\mu_{\mathrm{B}}B_{\mathrm{c}}/k_{\mathrm{B}}T_{\rm K}'=0.38$, values which are inconsistent by $65\%$ \cite{Hata2021}. While these measurements were precise, the carbon nanotube quantum dot investigated in this study might not be the best system to access Kondo universality, since level spacings are small and may invalidate the use of a single level Anderson model. Clearly, one needs both an appropriate system that is in the strongly correlated Kondo regime with $U/\Gamma \gg 1$ (compared to the typical value $3-4$ of carbon nanotube and semiconductor quantum dots) and an alternative to $dI/dV$ measurements in order to reliably detect the onset of the splitting of the Kondo resonance. We show in the following section, that this alternative is provided by thermocurrent spectroscopy for molecular quantum dots. Since the latter generally have very large charging energies, with $U/\Gamma$ of order $20-100$, accessing the universal splitting on the basis
of a single level Anderson model is also expected to be more reliable than for semiconductor or carbon nanotube quantum dots.

\subsubsection{Thermocurrent \texorpdfstring{$I_{\rm th}(V)$}{} versus {\it{V}} in the Fermi-liquid regime}
\label{subsubsec:probe-splitting-Ith}
We now consider the thermocurrent within the above higher-order Fermi-liquid theory. This yields results valid for $V\ll T_{\rm K}$ and $\Delta T \ll T\ll T_{\rm K}$.
Substituting the derivative of the spectral function at $\omega=\mu_L=\alpha_LV$ from Eq.~(\ref{eq:Spec-fliq})
\begin{align}
\left(\frac{\partial A(\omega,V,T=0,B,\Delta T=0)}{\partial\omega}\right)_{\omega=\mu_L} = & \frac{1}{\pi\Gamma T_{\rm K}}a_{1} + \frac{1}{\pi \Gamma T_{\rm K}^2} \left(2a_{2}\alpha_L V + 2a_{5}\alpha V\right)\label{eq:Spec-derivative}
\end{align}
into the fist term of the expression for the thermocurrent in Eq.~(\ref{eq:Ith-linearDeltaT}) yields, assuming a symmetric voltage drop across the leads ($\alpha_L=\alpha_R=-1/2$) and a lead coupling asymmetry $\Gamma_L/\Gamma_R=\lambda$, noting that $2\alpha = (1-\lambda)/(1+\lambda)$,
\begin{align}
  I_{\rm th}^a(V)= & \gamma\frac{\pi^2k_{B}^2T}{3}\frac{\Delta T}{\pi\Gamma T_{\rm K}}a_{1}
+ \gamma\frac{\pi^2k_{B}^2TV}{3}\frac{\Delta T}{\pi\Gamma T_{\rm K}^2}[\frac{1-\lambda}{1+\lambda} a_{5}-a_{2}].\label{eq:Ith-a}
\end{align}
In order to evaluate the second term in Eq.~(\ref{eq:Ith-linearDeltaT}), i.e., $I_{\rm th}^b$ in Eq.~(\ref{eq:Sommerfeld3a}), we require the $\Delta T$-dependence of the spectral function $A(\omega,T,V,B,\Delta T)$. For the molecular quantum dot in the experiment, a highly asymmetric coupling to the leads is found. Thus we only need to consider the two cases $\Gamma_L\ll \Gamma_R$ and $\Gamma_L\gg \Gamma_R$. In the former, the molecular quantum dot couples strongly to the right (cold) lead which is at temperature $T_R=T$ and the Kondo resonance is pinned to the chemical potential of the right lead. Hence, its temperature dependence is determined by $T_R=T$. In this case, the spectral function has a negligible dependence on $\Delta T$ for $\omega$ in the relevant transport window $[\mu_L,\mu_R]=[-V/2,+V/2]$, i.e. $(\partial A/\partial \Delta T)=0$, so that $I_{\rm th}^b\approx 0$. In the second case when the molecular quantum dot couples strongly to the left (hot) lead, which is at temperature $T_L=T+\Delta T$,  the Kondo resonance is pinned to the chemical potential of the left lead and its temperature dependence is determined by $T_L=T+\Delta T$ and will therefore be dependent on $\Delta T$. In the Fermi liquid regime, this dependence will be given by Eq.~(\ref{eq:Spec-splitting}) with $T$ replaced by $T_L=T+\Delta T$. We find using Eq.~(\ref{eq:Spec-splitting}), 
\begin{equation}
(\partial A/\partial \Delta T)_{\Delta T=0}= (\partial A/\partial T_{L})_{T_{L}= T}=-(1/\pi\Gamma)\bar{c}_T\pi^2 2T/T_{\rm K}^2. 
\end{equation}
Substituting this into Eq.~(\ref{eq:Sommerfeld3a}), using $\bar{c}_T=-a_3$ [Eq.~(\ref{eq:cbar_T})], gives for $\Gamma_L\gg \Gamma_R$, 
\begin{align}
  I_{\rm th}^b(V) = &
\gamma\frac{\pi^2k_{B}^2TV}{3}\frac{\Delta T}{\pi\Gamma T_{\rm K}^2}[-6a_3].\label{eq:Ith-b}
\end{align}
Thus, the total thermocurrent $I_{\rm th}=I_{\rm th}^a+I_{\rm th}^b$ for $V\ll T_{\rm K}$ and $\Delta T \ll T\ll T_{\rm K}$ is given by,
\begin{equation}
  I_{\rm th}(V) =  \gamma\frac{\pi^2k_{B}^2T}{3}\frac{\Delta T}{\pi\Gamma T_{\rm K}}a_{1}
+ \begin{cases}\gamma\frac{\pi^2k_{B}^2TV}{3}\frac{\Delta T}{\pi\Gamma T_{\rm K}^2}[\frac{1-\lambda}{1+\lambda} a_{5}-a_{2}], &\text{if $\lambda=\Gamma_L/\Gamma_R\ll 1$,}\\
\gamma\frac{\pi^2k_{B}^2TV}{3}\frac{\Delta T}{\pi\Gamma T_{\rm K}^2}[\frac{1-\lambda}{1+\lambda}a_{5}-a_{2}-6a_3], &\text{if $\lambda=\Gamma_L/\Gamma_R\gg 1$,}
\end{cases}\label{eq:Ith0}
\end{equation}
i.e.,
\begin{align}
  I_{\rm th}(V)  \approx & \gamma\frac{\pi^2k_{B}^2}{3}T\Delta T\left[s_0(B)+ s_1(B)V\right],\label{eq:Ith2}
  \end{align}
  where $s_0(B)$ and $s_1(B)$ are given by,
  \begin{equation}
    s_0(B) =  \frac{1}{\pi\Gamma T_{\rm K}}a_{1},\label{eq:s0}
    \end{equation}
    \begin{equation}
    s_1(B) =  \frac{1}{\pi\Gamma T_{\rm K}^2}\begin{cases}\left[\frac{1-\lambda}{1+\lambda}a_{5}-a_{2}\right], &\text{if $\lambda=\Gamma_L/\Gamma_R\ll 1$,}\\
    \left[\frac{1-\lambda}{1+\lambda}a_{5}-a_{2}-6a_3\right], &\text{if $\lambda=\Gamma_L/\Gamma_R\gg 1$.}
    \end{cases}
    \label{eq:s1}
  \end{equation}
  Using the aforementioned proportionalities $a_2\approx 3a_3\approx -a_5$ (valid deep in the Kondo regime), we find for the zero bias thermocurrent slope  
  $(\partial I_{\rm th}/\partial V)_{V=0}\propto s_1(B)$, for highly asymmetric lead couplings $\lambda \to 0$ and $\lambda \to \infty$,
 \begin{equation}
    s_1(B) =\begin{cases} 
    \frac{1}{\pi\Gamma T_{\rm K}^2}\left[+a_5-a_2\right]\approx -\frac{2a_2}{\pi\Gamma T_{\rm K}^2}, &\text{if $\lambda =\Gamma_L/\Gamma_R \ll 1$,}\\
    \frac{1}{\pi\Gamma T_{\rm K}^2}\left[-a_5-a_2-6a_3\right]\approx -\frac{2a_2}{\pi\Gamma T_{\rm K}^2}, &\text{if $\lambda =\Gamma_L/\Gamma_R \gg 1$.}\end{cases}\label{eq:Ith-explicit}
    \end{equation}
i.e., for highly asymmetric lead couplings, the thermocurrent slope in the Kondo regime is independent of whether the molecule couples strongly to the left (hot) or right (cold) lead. The dependence of $s_1(B)$ on the coupling asymmetry $\lambda$ is weak and enters through the prefactor of the $a_5$ coefficient in Eq.~(\ref{eq:s1}). 

\begin{figure}[thb!]
\centering 
\includegraphics[width=0.7\textwidth]{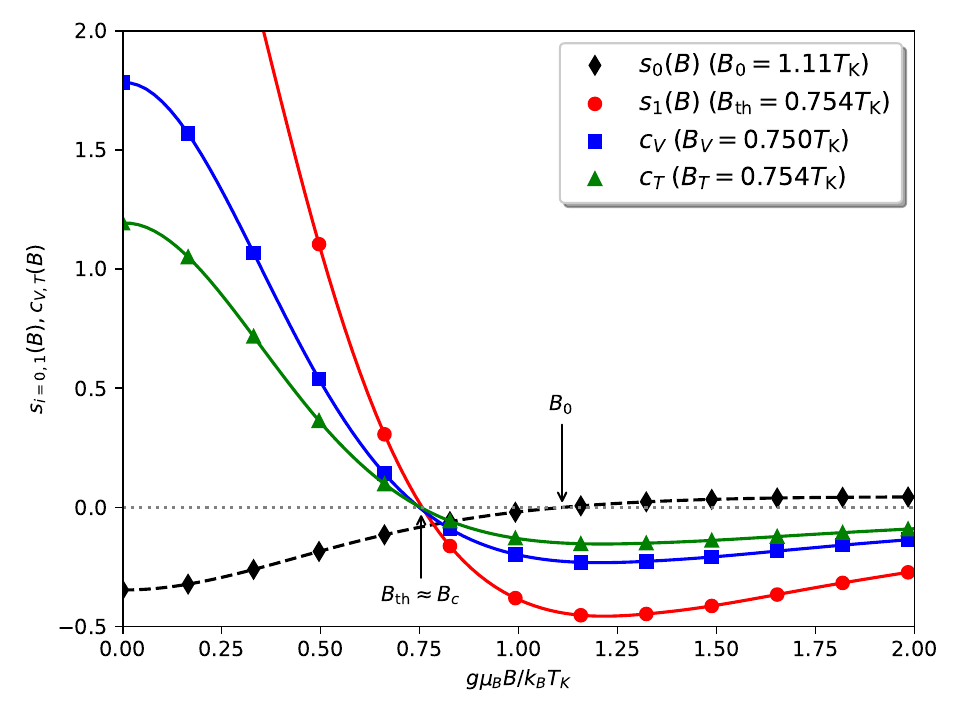}
\caption 
{NRG calculations for the $B$-dependence of the coefficients $s_{0}$ and $s_1$ appearing in the low temperature thermocurrent Eq~(\ref{eq:Ith2}) [for $\lambda=\Gamma_L/\Gamma_R \to 0$ and dropping the 
  prefactors $1/\pi\Gamma T_{\rm K}$ and $1/\pi\Gamma T_{\rm K}^2$], and compared to the $B$-dependence of the coefficents $c_V$ and $c_T$ appearing in the low temperature differential conductance.  The coefficient $s_1(B)$, and hence the slope of $I_{\rm th}(V)$ at $V=0$, changes sign at $B=B_{\rm th}\approx 0.754T_{\rm K}$ close to the value at which $c_V$ and $c_T$ in the differential conductance change sign. In contrast, the sign of the $V=0$ thermocurrent, determined by $s_0(B)$, changes 
  sign at a higher field $B_0(V_g=-1)\approx 1.11 T_{\rm K}>B_{\rm th}\approx B_{c}$ (see  Sec.~\ref{subsubsec:contrast-offset+slope} and Refs.~\onlinecite{Costi2019a}). Parameters typical for the asymmetric Kondo regime are used: $U/\Gamma=8, \varepsilon_0=-5\Gamma,V_g=-1$. NRG parameters as in Fig.~\ref{fig:acoeff+ccoeff}.
  }
\label{fig:scoeff+cvct}
\end{figure}
Equation~(\ref{eq:Ith2}) describes the leading temperature and voltage corrections to the nonlinear thermocurrent valid for $T,V\ll T_{\rm K}$, asymmetric lead couplings, and arbitrary magnetic field $B$. It allows us to address the experimentally observed sign change of   $\left(\partial I_{\rm th}(V)/\partial V\right)_{V=0} \propto  s_1(B) $ upon increasing $B$ above $B_{\rm th}$ and also to determine the value of $B_{\rm th}$. 
The field-dependence of the coefficients $a_{i},i=0,\dots,5$, evaluated within the NRG for a typical parameter set in the Kondo regime, were shown above in Fig.~\ref{fig:acoeff+ccoeff}. In Fig.~\ref{fig:scoeff+cvct} we show the $B$-dependence of the coefficients  $s_0(B)$ and $s_1(B)$. For comparison, the field-dependence
of the curvature coefficients $c_T$ and $c_V$ of $dI/dV$ are also shown. Remarkably, we find that $s_1(B)$ changes sign at
  essentially the same field at which the Kondo resonance splits in a magnetic field, i.e., $B_{\rm th}=B_c$ (the deviation of $B_{\rm th}$ from $B_c$ decreases, with increasing $U$, see next section). Moreover, deep in the Kondo regime, this value is largely independent of the lead coupling asymmetry $\lambda$ due to the aforementioned proportionality of the coefficients $a_5, a_2$ and $a_3$ which all change sign at the same (universal) field [$B=B_c$, see Fig.~\ref{fig:acoeff+ccoeff}(a)]. Thus, the sign change of the slope of the thermocurrent $\left(\partial I_{\rm th}(V)/\partial V\right)_{V=0} \propto  s_1(B) $ occurs at the splitting field $B_c$ independent on the precise value of the lead coupling asymmetry.
Thus, thermocurrent measurements of Kondo correlated quantum dots at finite bias voltage provide a new way to determine
the splitting of the Kondo resonance, independent of the usual way via the differential conductance. 

Finally, we note that while the first term in Eq.~(\ref{eq:Ith2}), proportional to $s_0$, is a linear response quantity [see Eqs.~(S12), (S39) and (S31)], the second term, proportional to $s_1(B)$, and linear in $V$, is a nonequilibrium quantity, as it arises from contributions of $O(V\omega)$ to the spectral function [Eqs.~(\ref{eq:Spec-splitting}) and (\ref{eq:Spec-derivative})].

\subsubsection{Universal scaling functions for \texorpdfstring{$s_{1}(B)$}{} and \texorpdfstring{$c_{V}(B)$}{}}
\label{subsubsec:universal-scaling-functions}
\begin{figure}[thb!]
\centering 
\includegraphics[width=0.9\textwidth]{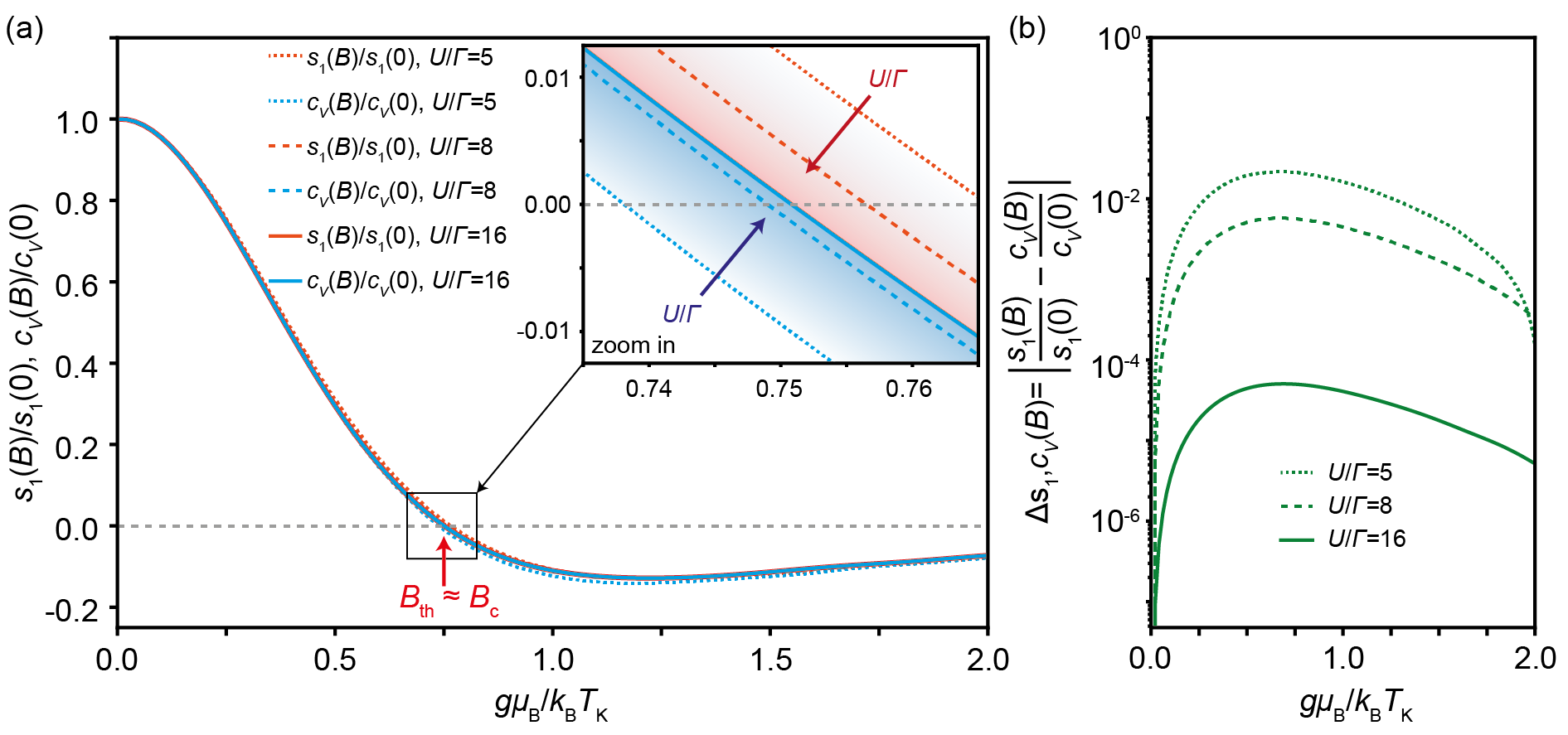}
\caption 
{(a) Scaling of the normalized zero-bias thermocurrent slope $\propto s_1(B)/s_1(0)$ (blue) and the normalized curvature coefficient $c_V(B)/c_V(0)$ (red) with increasing $U/\Gamma$ for $V_g$ in the Kondo regime ($V_g=1$ for $U/\Gamma=16$, $V_g=1.5$ for $U/\Gamma=8$ and $V_g=0.5$ for $U/\Gamma=5$). The inset shows the behaviour near $B_{\rm th}\approx B_c$.
(b) The absolute difference $\Delta_{s_1,c_V}(B)=|\frac{s_1(B)}{s_1(0)}-\frac{c_V(B)}{c_V(0)}|$ between the normalized $s_1$ and $c_V$ versus $B$ for $U/\Gamma=5,8$ and $16$. NRG parameters as in Fig.~\ref{fig:acoeff+ccoeff}.
}
\label{fig:s1-cv-scaling}
\end{figure}

The main text [Fig~4(a)] showed that for $U/\Gamma\gg 1$, $s_1(B)/s_1(0)$ and $c_V(B)/c_V(0)$ are universal scaling functions of their argument ($g\mu_B B/k_BT_{\rm K}$) in the Kondo regime of gate voltages $V_g$.  The dependence of these functions on $U/\Gamma \gg 1$
for gate voltages close to mid-valley (i.e., in the Kondo regime) is shown in Fig.~\ref{fig:s1-cv-scaling} for three different values of $U/\Gamma=5,8$ and 16. One observes that the difference between these two functions decreases with increasing $U/\Gamma$ (inset to Fig.~\ref{fig:s1-cv-scaling}), with the maximum absolute difference being below $2\%$ for $U/\Gamma=5$, below $0.6\%$ for $U/\Gamma=8$ and below $0.01\%$ for $U/\Gamma=16$. Hence, in the Kondo limit and for $U/\Gamma \gg 1$, the normalized scaling functions for the slope of the thermocurrent and the curvature coefficient of the differential conductance, while strictly different, can nevertheless be well approximated by a single scaling function. An approximate interpolation formula for this function, valid for a large range of $b=g\mu_BB/k_BT_{\rm K}$ from $b\ll 1$ to $b\gg 1$, is given by
\begin{equation}\label{eqn:slopefit}
    f_1(b)= \frac{(1-\alpha_0 b^2)}{(1+\beta_0 b^2)^{z}},
\end{equation}
where $\alpha_0\approx 1.77, \beta_0\approx 1.12$ and $z\approx 2.6$. This was used in the main text to fit the experimental data for the zero-bias thermocurrent slope to the above universal curve (see \ref{subsec:fitting-to-s1}).

The above universality in $c_V$ and $s_1$ follows, from the observation made previously that in the Kondo regime $a_2,a_3$ and $a_5$ are proportional to each other [Eq.~(\ref{eq:acoeff-prop})], and when normalized by their $B=0$ values, collapse onto the same universal curve for $U/\Gamma\gg 1$ [Fig.~\ref{fig:a-scaled}(a)]. Hence, any linear combination of these, such as $c_V$ or $s_1$, will also follow this curve and will also change sign at the universal field $B_c$. The same competition between two- and three-body fluctuations that drives the splitting of the Kondo resonance in $dI/dV$ via the vanishing of $c_V$ at $B=B_c$ [see Fig.~\ref{fig:a-scaled}(b)], is also responsible for the sign change of the zero-bias thermocurrent slope $s_1$ at $B=B_{\rm th}\approx B_c$, i.e.,
the three-body fluctuations also play an important role for the thermocurrent. 

Finally, one can summarize the universal aspects of magnetic field effects on the spectral function (through the coefficients $a_2,a_3$ and $a_5$), differential conductance (through the curvatures, themselves functions of $a_2$ and $a_3$) and the
thermocurrent (through $s_1$, a function of $a_2$ and $a_5$) in the Kondo limit for $U/\Gamma\gg 1$ by a single function, approximately given by Eq.~(\ref{eqn:slopefit}).
Deviations from this universality arise for weak correlations or outside the Kondo regime (e.g. on approaching the mixed-valence regime). In the latter case, the deviations result from particle-hole asymmetry and are described by the coefficients $a_1$ and $a_4$, where $a_1$ ($a_4$) shifts the zero-energy (zero-bias) peak in $A(\omega)$ ($dI/dV$) to finite energy (voltage) [see Eqs.~(\ref{eq:Spec-splitting}) and (\ref{eq:dIdV-fliq})], with $a_1$ also being responsible for the finite thermocurrent offset at $V=0$ through the coefficient $s_0$ [see Eq.~(\ref{eq:s0})]. 

\subsubsection{Zero-bias thermocurrent and zero-bias thermocurrent slope}
\label{subsubsec:contrast-offset+slope}
We contrast the above findings to the experiment of Svilans et al. in Ref~\onlinecite{Svilans2018}. In this experiment, the
gate-voltage ($V_g$) dependence of the zero-bias thermocurrent $I_{\rm th}(V=0,V_g)\propto s_0$ was measured,  as a function of
of temperature and as a function of magnetic field for a range of gate voltages around mid-valley ($V_g=0$). A sign change
of $I_{\rm th}(V=0,V_g)$ and its $V_g=0$ slope $(\frac{\partial I_{\rm th}(V=0,V_g)}{\partial V_g})_{V_g=0}$ was observed upon increasing
the magnetic field at low temperatures $T\lesssim T_{\rm K}$. 

The above observations can be understood from Eq.~(\ref{eq:linear-transport-thermocurrent}), which shows that $I_{\rm th}(V=0,V_g)=G(T,B)S(T,B)\Delta T$ is a linear-response quantity, whose sign changes, as a function of temperature or magnetic field, are determined by sign changes of the linear Seebeck coefficient $S(T,B)$. Ref.~\onlinecite{Costi2019a}, showed that the sign change
in  the zero-bias thermocurrent occurs at a gate-voltage dependent field $B_0(V_g)>B_c$. The sign change in the slope $(\frac{\partial I_{\rm th}(V=0,V_g)}{\partial V_g})_{V_g=0}$ also occurs at $B_0(V_g=0)$, as can be seen by expanding $I_{\rm th}(V=0,V_g)$ about $V_g=0$, 
\begin{align}
I_{\rm th}(V=0,|V_g|\ll 1) = & I_{\rm th}(V=0,V_g=0) + V_g \left.\frac{\partial I_{\rm th}(V,V_g)}{\partial V_g}\right\rvert_{V_g=0} + \dots \label{eq:zero-bias-Ith}
\end{align}
and noting that, since   $I_{\rm th}(V=0,V_g=0) =0$ by particle-hole symmetry, $I_{\rm th}(V=0,|V_g|\ll 1)$, 
and its derivative with respect to $V_g$, both change sign at the same field, i.e., at $B=B_0(|V_g|\ll 1)$. Indeed, Svilans et al. \cite{Svilans2018} commented on the fact that the experimental field at which the low temperature value of $(\frac{\partial I_{\rm th}(V=0,V_g)}{\partial V_g})_{V_g=0}$ changes sign exceeds the expected field $B=B_c$ for the splitting. Equation~(\ref{eq:zero-bias-Ith}), together with
the result $B_0(V_g\to 0)>B_c$ \cite{Costi2019a} explains this observation. For typical values of $U/\Gamma \gg 1$, $B_0(0)$ can be a factor two larger than $B_c$ \cite{Costi2019a}, hence, $B_0$ cannot be used to estimate the splitting field $B_c$. 

In contrast to the non-universal zero-bias thermocurrent $s_0$ [see Fig.~4(a) of the main text] and its slope with respect to $V_g$ at $V_g=0$,
the zero-bias slope of the thermocurrent $\propto s_1(B)$ measured in the present experiment is universal and changes sign at the universal field $B_{\rm th}\approx B_c$.
Thus thermocurrent spectroscopy, via the slope of the zero-bias thermocurrent, is able to access universal aspects of Kondo physics.
Non-universal aspects, such as the small finite offset of the zero bias thermocurrent $I_{\rm th}(V=0,V_g)\propto T\Delta T s_0(B)$ can, however, also be of some practical interest. In particular, the sign of this offset at $T\ll T_{\rm K}$ and $B=0$ measures the deviation from particle-hole symmetry for a given gate voltage $V_g$, and indicates whether the Kondo resonance lies above (for $V_g>0$) or below (for $V_g<0$) the Fermi level.

\subsubsection{Temperature and thermal bias effects}
\label{subsubsec:temperature-effects}
 
The higher-order Fermi-liquid theory suffices to capture the main observation of the experiment, i.e., the kink (slope change) in $I_{\rm th}$ vs $V$ upon increasing $B$ above $B_c$
in the low-temperature strong coupling regime $\Delta T \ll T \ll T_{\rm K}$. 
We comment briefly on the effect of increasing either temperature $T$ or the thermal bias $\Delta T$ to values above $T_{\rm K}$ starting from the above regime, i.e., $B>B_c$ and $\Delta T \ll T \ll T_{\rm K}$. Since increasing temperature increases the effective splitting field $B_c(T)>B_c(T=0)$ \cite{Costi2000}, for sufficiently high temperature one will eventually be in the regime where $B<B_c(T)$ and the kink in $I_{\rm th}$ vs $V$ at $V=0$ will vanish. Similarly, increasing the thermal bias acts like increasing the effective base temperature $T\to T_{eff}(\Delta T)>T$, as noted in the
experiment in Fig.~S7, so again the kink will vanish for sufficiently large thermal bias.
The vanishing of the kink in $I_{\rm th}$ vs $V$ at $V=0$ with increasing $T$ or $\Delta T$ is, indeed, consistent
with the measurements [see Figs.~\ref{fig:T_sweep_8T}(a)-\ref{fig:T_sweep_8T}(b) and Figs.~\ref{fig:Ith_vs_dT}(a)-\ref{fig:Ith_vs_dT}(b)].
In the next section, we quantify the above expectations for the temperature and thermal bias dependence of the nonlinear thermocurrent, by using an approximate method, which applies to a wider range of bias voltages, thermal biases and temperatures, than those accessible within the higher-order Fermi-liquid theory.

\subsection{Equation of motion method for nonlinear thermoelectric transport}
\label{subsec:EOM-method}
In the strongly nonlinear regime, where $V, T$ or $\Delta T$ become comparable to or larger than the low temperature scale $T_{\rm K}$,  we require a more generally applicable, albeit less exact, theory of nonequilibrium thermoelectric transport through a Kondo-correlated quantum dot as compared to that offered
by the NRG for linear transport \cite{Costi1994,Yoshida2009,Costi2010,Merker2013,Costi2019a}, or the higher-order Fermi-liquid theory described above, which is valid for finite small $V, \Delta T \ll T\ll T_{\rm K}$ \cite{Oguri2018a,Oguri2018b}. 
%While there are many such approximate theories \cite{Wingreen1994,Boese2001,Kim2002,Dong2002,Rosch2003a,Roermund2010,Lavagna2015,Khedri2018,Schinabeck2018},
Here, we shall use the recently developed  Green function equation of motion decoupling technique \cite{Roermund2010,Lavagna2015}, which generalizes an earlier approach of Lacroix \cite{Lacroix1981} for the finite $U$
Anderson impurity model in equilibrium to the two-lead Anderson model (\ref{eq:ham}) out of equilibrium.  Within this approach, electrical transport out of equilibrium was investigated at $B=0$ and at finite $B$ for the symmetric case \cite{Roermund2010}. A first application of this technique to thermal transport, at zero magnetic field,
investigated efficiency, power and generalized Seebeck coefficients at large thermal ($\Delta T$) and voltage ($V$) biases of Kondo-correlated
quantum dots \cite{Eckern2020}\footnote{A related approach has been used to investigate the thermocurrent of a spin-$1/2$ quantum dot in the Kondo regime at zero bias voltage and zero magnetic field as a function of the thermal bias \cite{Sierra2017}.}. Here, we go one step further and include a finite magnetic field $B$ in order to investigate the field, temperature, voltage and thermal bias dependence of the thermocurrent $I_{\rm th}$. We compare the results obtained with the corresponding measurements in the experiment.

\subsubsection{Equation of motion method and numerical caclulations}
\label{subsubsec:EOM-method+numerics}
Full details of the approach may be found elsewhere \cite{Roermund2010,Lavagna2015}, so here we shall only give a brief account. Starting with the local retarded Green function of the dot $G_{0\sigma}(\omega)=\langle\langle d_{\sigma};d_{\sigma}^{\dagger} \rangle\rangle$, one considers its equation of motion (EOM), %\cite{Zubarev1960},
\begin{align}
 (\omega - \varepsilon_{0\sigma}) G_{0\sigma}(\omega) = &  1 + \sum_{k\alpha}t_{\alpha}\langle\langle c_{k\alpha\sigma};d^{\dagger}_{\sigma} \rangle \rangle + U \langle\langle n_{0,-\sigma}d_{\sigma};d^{\dagger}_{\sigma}\rangle \rangle,\label{eq:one}
\end{align} 
which results in two new Green functions on the RHS. The EOM for the first Green function  $\langle\langle c_{k\alpha\sigma};d^{\dagger}_{\sigma} \rangle \rangle$,
\begin{align}
(\omega - \epsilon_{k\alpha\sigma}) \langle\langle c_{k\alpha\sigma};d^{\dagger}_{\sigma} \rangle \rangle  = &  t_{\alpha} \langle\langle d_{\sigma};d^{\dagger}_{\sigma}\rangle \rangle,\label{eq:two}
\end{align}
returns $G_{0\sigma}$, while the EOM of the second Green function,
$\langle\langle n_{0,-\sigma}d_{\sigma};d^{\dagger}_{\sigma}\rangle \rangle$,
results in three new Green functions (see Refs.~\onlinecite{Lacroix1981,Roermund2010,Eckern2020}).
A closed set of equations can be obtained by writing
down the EOM of the latter three Green functions and decoupling the resulting higher-order Green functions with two lead operators
(e.g., of the kind $\langle\langle c_{k\alpha\,-\sigma}^{\dagger}c_{k'\alpha',-\sigma}d_{\sigma};d^{\dagger}_{\sigma}\rangle \rangle$) in terms of mean fields $\langle d^{\dagger}_{\sigma}c_{k\alpha\sigma}\rangle$, $\langle c^{\dagger}_{k\alpha\sigma}c_{k'\alpha\sigma}\rangle$ and lower order Green functions, e.g.,
\begin{align}
  \langle\langle c_{k\alpha,-\sigma}^{\dagger}c_{k'\alpha',-\sigma}d_{\sigma};d^{\dagger}_{\sigma}\rangle \rangle \approx & \langle c_{k\alpha,-\sigma}^{\dagger}c_{k'\alpha',-\sigma}\rangle\langle\langle d_{\sigma};d_{\sigma}^{\dagger} \rangle\rangle,\label{eq:approx1}\\
    \langle\langle d_{-\sigma}^{\dagger}c_{k'\alpha',-\sigma}c_{k\alpha,\sigma};d^{\dagger}_{\sigma}\rangle \rangle \approx & \langle d_{-\sigma}^{\dagger}c_{k'\alpha',-\sigma}\rangle\langle\langle c_{k\alpha\sigma};d_{\sigma}^{\dagger} \rangle\rangle.\label{eq:approx2}
  \end{align}
  This then allows one to obtain a closed expression for the dot Green function \cite{Lacroix1981,Lavagna2015}: 
\begin{align}
 G_{0\sigma}(\omega) = &  \frac{1 + K_{\sigma}(\omega)\left[\langle n_{0,-\sigma}\rangle +\Sigma_{\sigma,4}(\omega)\right]}
     {\omega-\varepsilon_{0\sigma}-\Sigma_{\sigma,0}(\omega) + K_{\sigma}(\omega)\left[\Sigma_{\sigma,1}(\omega)-\Sigma_{\sigma,0}(\omega)\Sigma_{\sigma,4}(\omega)\right]},\label{eq:three}
\end{align}
where the self-energies $\Sigma_{\sigma,i=0,\dots,4}(\omega)$ and $K_{\sigma}(\omega)$ (denoted by $I_{\sigma}(\omega)$ in \cite{Lavagna2015})
are given by\cite{Lavagna2015}
\begin{align}
   \Sigma_{\sigma,0}(\omega) =&   \sum_{k\alpha}\frac{t_{\alpha}^2}{\omega - \epsilon_{k\alpha}},\label{eq:sigma0}\\
  \Sigma_{\sigma,1}(\omega) =&   \sum_{k\alpha}{t_{\alpha}^2}\left[\frac{\sum_{k'}\langle c^{\dagger}_{k\alpha,-\sigma}c_{k'\alpha,-\sigma}\rangle}{\omega - (\varepsilon_{0\sigma}-\varepsilon_{0,-\sigma})-\epsilon_{k\alpha} + i\gamma_{\sigma}} + \frac{\sum_{k'}\langle c^{\dagger}_{k'\alpha,-\sigma}c_{k\alpha,-\sigma}\rangle}{\omega - (\varepsilon_{0\sigma}+\varepsilon_{0,-\sigma}+U)+\epsilon_{k\alpha} + i\gamma_{D}}\right],\label{eq:sigma1}\\
    \Sigma_{\sigma,2}(\omega) =&   \sum_{k\alpha}{t_{\alpha}^2}\left[\frac{1-\sum_{k'}\langle c^{\dagger}_{k\alpha,-\sigma}c_{k'\alpha,-\sigma}\rangle}{\omega - (\varepsilon_{0\sigma}-\varepsilon_{0,-\sigma})-\epsilon_{k\alpha} + i\gamma_{\sigma}} + \frac{1-\sum_{k'}\langle c^{\dagger}_{k'\alpha,-\sigma}c_{k\alpha,-\sigma}\rangle}{\omega - (\varepsilon_{0\sigma}+\varepsilon_{0,-\sigma}+U)+\epsilon_{k\alpha} + i\gamma_{D}}\right],\label{eq:sigma2}\\
    \Sigma_{\sigma,3}(\omega) =&   \sum_{k\alpha}{t_{\alpha}^2}\left[\frac{1}{\omega - (\varepsilon_{0\sigma}-\varepsilon_{0,-\sigma})-\epsilon_{k\alpha} + i\gamma_{\sigma}} + \frac{1}{\omega - (\varepsilon_{0\sigma}+\varepsilon_{0,-\sigma} +U)+\epsilon_{k\alpha} + i\gamma_{D}}\right],\label{eq:sigma3}\\
   \Sigma_{\sigma,4}(\omega) =&   \sum_{k\alpha}{t_{\alpha}}\left[\frac{\langle d^{\dagger}_{-\sigma}c_{k\alpha,-\sigma}\rangle}{\omega - (\varepsilon_{0\sigma}-\varepsilon_{0,-\sigma})-\epsilon_{k\alpha} + i\gamma_{\sigma}} - \frac{\langle c^{\dagger}_{k\alpha,-\sigma}d_{-\sigma}\rangle}{\omega - (\varepsilon_{0\sigma}+\varepsilon_{0,-\sigma}+U)+\epsilon_{k\alpha} + i\gamma_{D}}\right],\label{eq:sigma4}\\
 K_{\sigma}(\omega) =&
     \frac{U}
     {\omega-\varepsilon_{0\sigma}-\Sigma_{\sigma,0}(\omega)  - \Sigma_{\sigma,3}(\omega)}.\label{eq:Isigma}
\end{align}
Within the Lacroix theory, the lifetime of the  single particle ($\gamma_{\sigma}$) and doubly occupied ($\gamma_D$) states of the dot appearing in the denominators of the self-energies $\Sigma_{\sigma,1},\dots,\Sigma_{\sigma,4}$ vanish, being given by the usual $i\delta=i0^{+}$ of retarded Green functions.  Additional considerations, however, show that these have finite values \cite{Roermund2010,Lavagna2015}.  They are evaluated in perturbation theory for the finite $U$ Anderson model to fourth order in $t_\alpha$ for $\gamma_{\sigma}$ and to second order in $t_\alpha$ for $\gamma_D$ and describe lifetime broadening due to finite $B, T$ or $V$ and generalize the results for the infinite $U$ Anderson model \cite{Wingreen1994}.  The effect of
$\gamma_{\sigma}$ and $\gamma_D$ is to cut off the logarithmic singularities in the self-energies at $\mu_{L,R}\pm g\mu_BB/2$
at low $T$. This in turn leads to several improvements over the Lacroix theory in which $i\gamma_\sigma = i\gamma_D = i\delta=i0^{+}$. For example, 
the Kondo resonance at the particle-hole symmetric point ($2\varepsilon_0+U=0$), absent in the Lacroix theory, is recovered within this approach\cite{Roermund2010,Lavagna2015}. Another improvement over the Lacroix theory concerns the exponent in the Kondo scale. Within this theory, $T_{\rm K}\sim e^{\pi \varepsilon_0(\varepsilon_0+U)/(1.5\Gamma U)}$, which comes closer to the exact one,  $T_{\rm K}\sim e^{\pi \varepsilon_0(\varepsilon_0+U)/(2\Gamma U)}$, as compared to the Lacroix theory which gives $T_{\rm K}\sim e^{\pi\varepsilon_0 (\varepsilon_0+U)/(\Gamma U)}$\cite{Roermund2010}. Moreover, the expected
features of the nonequilibrium Kondo resonance, such as the peaks in the spectral function at the two chemical potentials $\mu_L=+eV/2$ and $\mu_R=-eV/2$ for $B=0$, and the peaks at  $\mu_{L,R}\pm g\mu_B B/2$ on applying a magnetic field, are also recovered \cite{Roermund2010,Eckern2020}. Hence, the approach captures the qualitative features of the nonequilibrium Kondo effect in a magnetic field and we therefore expect that the same will hold for the field and voltage dependent thermocurrent of interest to us here.

The expectation values $n_{0,-\sigma}=\langle d^{\dagger}_{-\sigma}d_{-\sigma}\rangle$, $\langle d^{\dagger}_{-\sigma}c_{k-\alpha\sigma}\rangle$ and $\langle c^{\dagger}_{k\alpha,-\sigma}c_{k'\alpha,-\sigma}\rangle$
appearing in the above equations can be related to the dot Green function $G_{0,-\sigma}$ in the wide band limit\cite{Roermund2010}. Hence, Eq.(\ref{eq:three}) represents a set of two coupled integral equations relating  $G_{0\uparrow}$ and $G_{0\downarrow}$. They reduce to a single integral equation for
$G_{0\uparrow}(\omega)=G_{0,\downarrow}(\omega)$ at zero magnetic field $|\varepsilon_{0\sigma}-\varepsilon_{0,-\sigma}|=|g\mu_B B| = 0$. The integral equations are
solved self-consistently by numerical iteration subject to the constraint on the occupation numbers \cite{Lavagna2015}
\begin{align}
n_{0\sigma}(\omega) =& \int d\omega f_{neq}(\omega,V,T) A_{\sigma}(\omega,V,T,B), \label{eq:constraint}
\end{align}
where $A_{\sigma}(\omega,V,T,B)=-{\rm Im}[G_{0\sigma}(\omega+i\delta,V,T,B)]/\pi$ is the nonequilibrium spectral function of the dot\footnote{We omit the dependence of the spectral function (and Green functions) on the thermal bias $\Delta T$ for simplicity of notation, but the dependence on this is implied} and $f_{\rm neq}(\omega,V,T)$ is a nonequilibrium distribution function given by, 
\begin{align}
f_{neq}(\omega,V,T) =& \frac{\Gamma_{L}f_{L}(\omega)+\Gamma_{R}f_{R}(\omega)}{\Gamma_L+\Gamma_R}. \label{eq:effectiveFermi}
\end{align}

  In practice, the above theory suffers from a small violation of particle-hole symmetry \cite{Roermund2010}. This is negligible for
  $B=0$, but needs to be corrected for finite field calculations. This is done by implementing particle-hole symmetry 
  $H(\varepsilon_{0},U,B,\mu_L,\mu_R)\leftrightarrow H(-(\varepsilon_0+U),U,-B,\mu_R,\mu_L)$ explicitly.
  This requires solving two sets of coupled integral equations for
  $G_{0\uparrow}^{\varepsilon_0},G_{0\downarrow}^{\varepsilon_0}$ and their particle-hole transformed pairs $G_{0\uparrow}^{-(\varepsilon_0+U)},G_{0\downarrow}^{-(\varepsilon_0+U)}$ and symmetrizing at each iteration to maintain particle-hole symmetry
  \footnote{Note that by ``maintaining particle-hole symmetry'' we do not imply that we are working at the special particle-hole symmetric point $\varepsilon_0 = -(\varepsilon_0+U)$, i.e., at $\varepsilon_0=-U/2$ (mid-valley). Instead, we are
    working at an arbitrary $\varepsilon_0$ and we symmetrize the results to ensure that the particle-hole transformation (symmetry) for the given $\varepsilon_0$ is satisfied. }.
 Stable results are obtained at the cost of a slower convergence. We used adaptive integration routines with addititional intervals for the regions around the points $\mu_{L,R}\pm g\mu_B B/2$ where the integrands of the self-energies (\ref{eq:sigma1})-(\ref{eq:sigma4}) have a rapid variation. The calculations become feasible by distributing the integrations over 256 OpenMP threads on the JURECA supercomputer of Forschungszentrum J\"ulich (each
  node on this supercomputer has 2 AMD EPYC 7742 CPUS, and each CPU has 64 cores running at 2.25 GHz. With hyperthreading this gives 256 threads/node; 2 nodes were used in the calculations, one for $\Delta T=0$ and one for $\Delta T>0$). Our convergence criterion was that the Euclidean norm of the difference of successive iterates be less than $10^{-3}$.
  
  \subsubsection{Model parameters and ranges for \texorpdfstring{$V,B,T$}{} and \texorpdfstring{$\Delta T$}{}}
  \label{subsubsec:model-parameters}
 Since all the interesting effects in Kondo-correlated quantum dots occur on a scale of order $T_{\rm K}$, the theoretical results are shown as functions
 of the dimensionless ratios $eV/k_BT_{\rm K}$, $g\mu_{B}B/k_{B}T_{\rm K}$, $\Delta T/T_{\rm K}$. From experiment
 $T_{\rm K}\approx 12.8 K$ is extracted from the universal Kondo conductance curve via $G(T=T_{\rm K})=G(T=0)/2$  so an experimental $V$-range from $-10mV$ to $+10mV$ would correspond to $eV/k_BT_{\rm K}$ values up to $10 \times 0.001 \times 1.6\times 10^{-19}/(1.38\times 10^{-23}\times 12.8)=9.055\approx 10$, so using $-10 \leq eV/k_BT_{\rm K}\leq +10$ corresponds
 closely to an experimental range $-10mV \leq V \leq +10mV$. The splitting of the Kondo resonance occurs at a field $g\mu_BB_c=0.5k_{B}T^{\mathrm{HWHM}}_{\rm K}= 0.75k_{B} T_{\rm K}$ where $T_{\rm K}^{\mathrm{HWHM}}$ is the Kondo scale defined as the HWHM of the $T=0$ Kondo resonance \cite{Costi2000} and $T_{\rm K}$, defined in Eq.~(\ref{eq:tk-spin}), is close the Kondo scale extracted from the conductance\cite{Merker2013}. Hence, from $g\mu_BB_c=0.75k_{B}T_{\rm K}$ we find
 $B_c=7.15$~T using $T_{\rm K}=12.8 K$ and the measured $g=2$ from EPR in Sec.~\ref{subsec:EPR}. The sign change of the slope of $I_{\rm th}$ for $V\to 0$ 
in the main text occurs at  $B\approx  6.6 T$, which is within 10 per cent of  the predicted $B_{\rm th}\approx B_c$ within the higher-order Fermi-liquid theory.
 
The asymmetry ratio, $\Gamma_L/\Gamma_R=\lambda =0.017$, extracted from the measured conductance in the main text was used for the calculations. 
The gate coupling for the device in the main text was weak (Fig.~\ref{fig:stability}) so estimates for the charging energy and local level position $\varepsilon_0$ were not possible.  However, the Kondo resonance in $dI/dV$ has a marked asymmetry about $V=0$, suggesting a level energy $\varepsilon_0$ in the asymmetric Kondo regime a few $\Gamma$ away from the mixed valence regime, since in this case the spectral function shows the above kind of asymmetry. Taking  $U/\Gamma=8$, for example, one may reproduce the observed asymmetry in $dI/dV$ by using $\varepsilon_0/\Gamma=-5$ or $-5.5$. The precise values, are, however, not important for a description of the main trends in the thermocurrent that we wish to describe. The field range used for the device in the main text is $0 \leq g\mu_{B}B/k_BT_{\rm K}\lesssim 0.85$, in the calculations we show a wider range $0\leq  g\mu_{B}B/k_BT_{\rm K}\lesssim 4$ to illustrate more clearly the differences between the low and high field behavior of $I_{\rm th}$. In the experiments, temperature and thermal bias ranges were $0.16 \lesssim T/T_{\rm K}\lesssim 1.2$ and  $0.05 \lesssim \Delta T/T_{\rm K}\lesssim 0.4$, respectively.  We shall use $0.5 \leq T/T_{\rm K}\leq 4$ and $0.2 \leq \Delta T/T_{\rm K}\leq 3$ extending the higher end for these quantities, while ensuring that we also access values for these quanties in the strong coupling regime. 

\subsubsection{Magnetic field dependence}
\label{subsubsec:b-dependence}
\begin{figure}[htb!]
  \centering 
\includegraphics[width=0.9\textwidth]{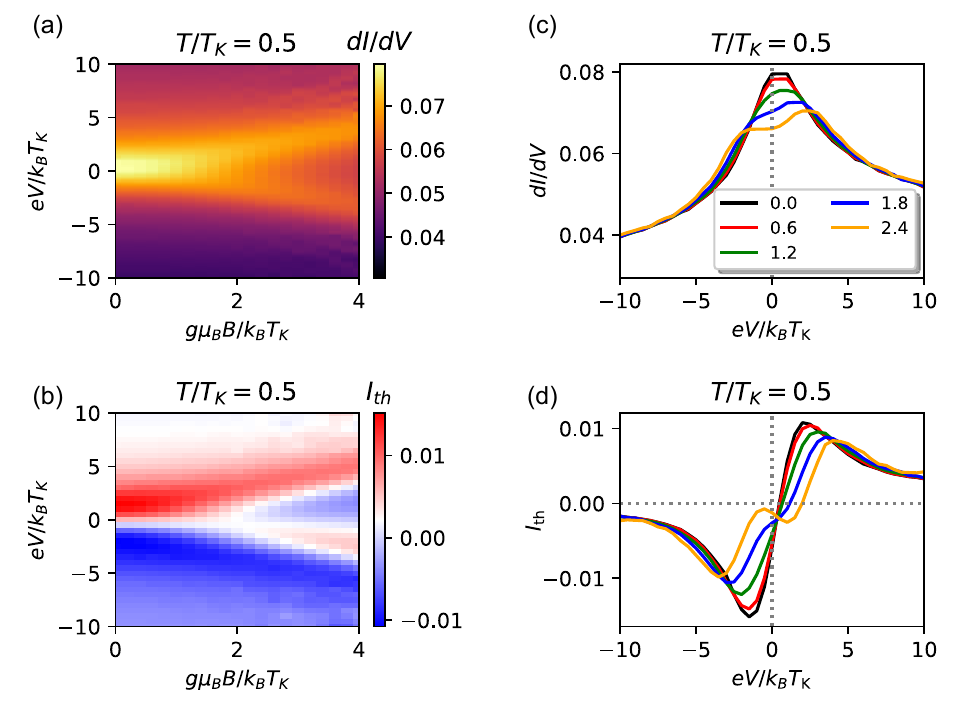}
\caption 
{Magnetic field dependence at low base temperature $T/T_{\rm K}=0.5$ and $\Delta T/T_{\rm K}=0.2$. (a),(c) Differential conductance vs bias voltage at differenet magnetic fields. (b),(d), Thermocurrent vs bias voltage at different magnetic fields. The linecuts in (c),(d) are for magnetic fields $g\mu_B B/k_BT_{\rm K}$ indicated in the legend. 
}
\label{fig:B-dep-lowT}
\end{figure}

\begin{figure}[htb!]
  \centering 
\includegraphics[width=0.9\textwidth]{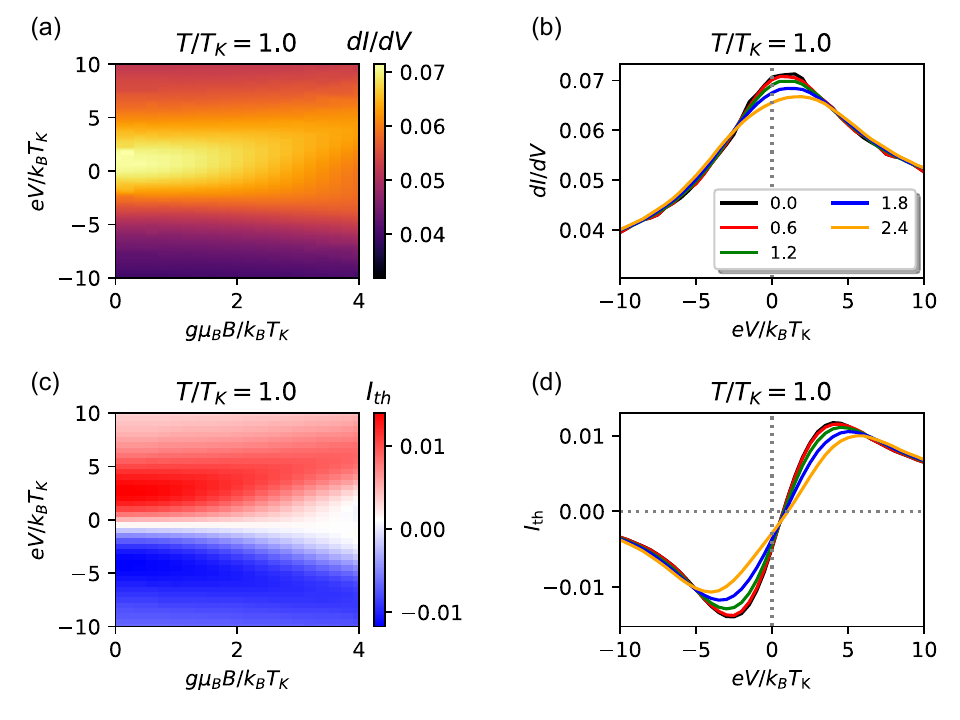}
\caption 
{Magnetic field dependence at higher base temperature $T/T_{\rm K}=1.0$ and $\Delta T/T_{\rm K}=0.2$. (a),(c) Differential conductance vs bias voltage at differenet magnetic fields. (b),(d), Thermocurrent vs bias voltage at different magnetic fields. The linecuts in (c),(d) are for magnetic fields $g\mu_B B/k_BT_{\rm K}$ indicated in the legend. 
}
\label{fig:B-dep-highT}
\end{figure}

Figs.~3(a) and 3(b) (main text) show the measured $I_{\rm th}$ vs $B$ and $V$ at a small thermal bias $\Delta T/T_{\rm K}=0.6/12.8\approx 0.05$
and at a low base temperature $T/T_{\rm K}=2/12.8\approx 0.16$,  while Figs.~2(c) and 2(d) (main text) show the corresponding $dI/dV$. Calculations within the EOM approach described above, for finite $B$, become numerically intractable for $T\ll T_{\rm K}$, and for too small a thermal bias $\Delta T\ll T_{\rm K}$. For temperatures much below $T/T_{\rm K}=0.5$, the equations are difficult to converge, while a thermal bias much below $\Delta T/T_{\rm K}=0.2$ results in a large signal to noise ratio in the procedure used to obtain $I_{\rm th}$ [see Eq.(\ref{eq:Ith-subtraction})].
We therefore set $\Delta T/T_{\rm K}=0.2$. In Fig.~\ref{fig:B-dep-lowT} we show results for
$T/T_{\rm K}=0.5$ (as in the main text) while in  Fig.~\ref{fig:B-dep-highT} we show the effect of using a higher temperature $T/T_{\rm K}=1.0$. One sees, that at the lower temperature, the same qualitative features are found as in the experiment, namely a kink in $I_{\rm th}$ vs $V$ at high fields $B>T_{\rm K}$ which is absent at low fields $B\ll T_{\rm K}$. Upon increasing temperature, see Fig.~\ref{fig:B-dep-highT}, the kink is absent at high fields.

\subsubsection{Thermal bias dependence}
\label{subsubsec:deltaT-dependence}
\begin{figure}[htb!]
  \centering 
  \includegraphics[width=0.9\textwidth]{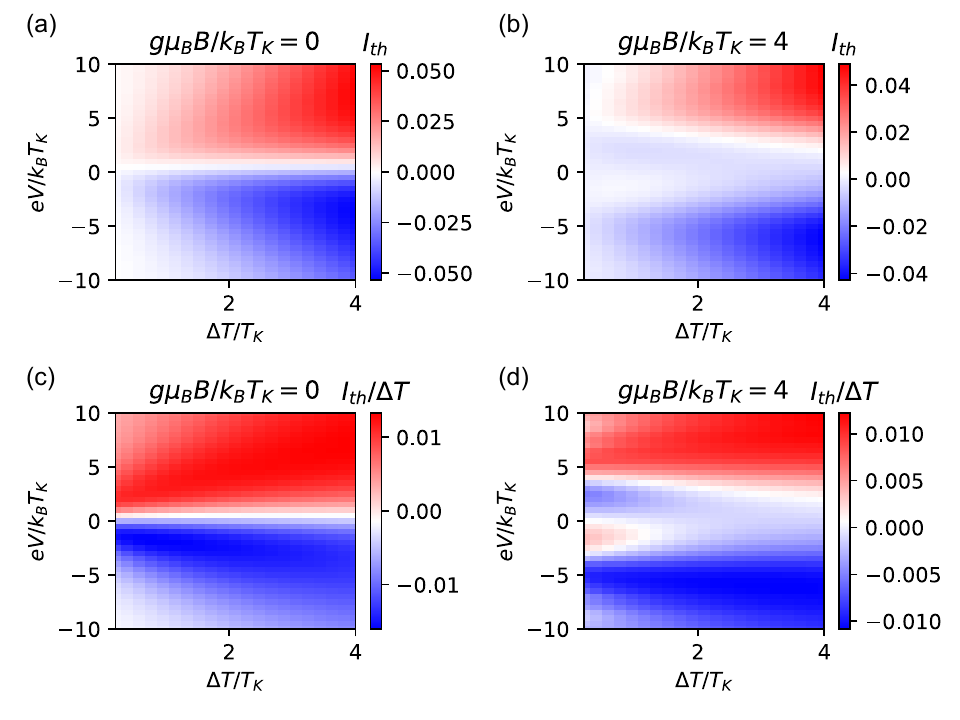}
\caption 
{Thermal bias dependence of the thermocurrent at low base temperature $T/T_{\rm K}=0.5$.
  Absolute thermocurrent  $I_{\rm th}$ (top panels) and normalized thermocurrent $I_{\rm th}/\Delta T$ (bottom panels) vs
  thermal bias $\Delta T/T_{\rm K}$  and 
  $eV/k_{B}T_{\rm K}$ for $g\mu_B B/k_{B}T_{\rm K}=0$ (left panels) and  $g\mu_B B/k_{B}T_{\rm K}=4$ (right panels). 
}
\label{fig:dT-dep-lowT}
\end{figure}

\begin{figure}[htb!]
  \centering 
  \includegraphics[width=0.9\textwidth]{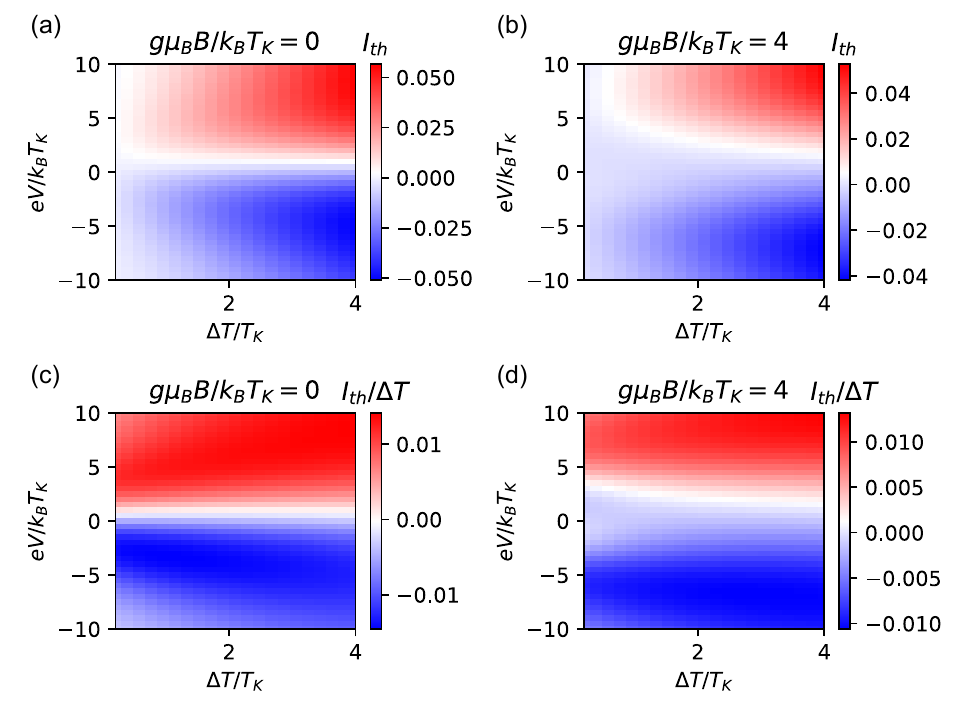}
\caption 
{Thermal bias dependence of the thermocurrent at a higher base temperature $T/T_{\rm K}=1.0$.
Absolute thermocurrent $I_{\rm th}$ (top panels) and normalized thermocurrent $I_{\rm th}/\Delta T$ (bottom panels) vs thermal bias $\Delta T/T_{\rm K}$  and 
  $eV/k_{B}T_{\rm K}$ for $g\mu_B B/k_{B}T_{\rm K}=0$ (left panels) and  $g\mu_B B/k_{B}T_{\rm K}=4$ (right panels). 
}
\label{fig:dT-dep-highT}
\end{figure}

The thermal bias dependence of $I_{\rm th}$ at two characteristic magnetic fields $g\mu_B B/k_BT_{\rm K}=0$ (i.e. at $B<B_{c}$)
and $g\mu_B B/k_BT_{\rm K}=4$ (i.e., for $B>B_{c}$) is shown in Fig.~\ref{fig:dT-dep-lowT} at a low base temperature $T/T_{\rm K}=0.5$ and
in Fig.~\ref{fig:dT-dep-highT} at a higher base temperature $T/T_{\rm K}=1$. The top panels in each figure show $I_{\rm th}$, while the bottom panels show $I_{\rm th}/\Delta T$, where the latter representation brings out more clearly the evolution of the kink at $V=0$ with
thermal bias. The results at the lower base temperature and finite field, show more clearly the sign changes of the thermocurrent at low temperature and how these evolve (vanish) with increasing thermal bias. These results capture the main trends in the thermal-bias dependence of the experiments in Figs.~S5a-b.

\clearpage
\subsubsection{Temperature dependence}
\label{subsubsec:T-dependence}
\begin{figure}[htb!]
  \centering 
\includegraphics[width=0.9\textwidth]{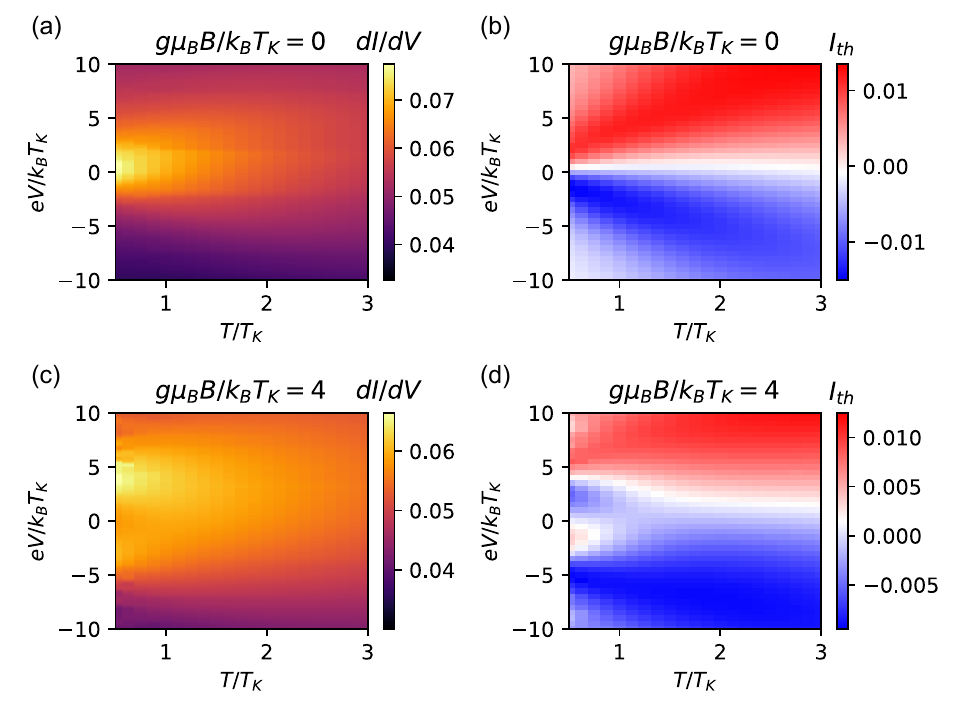}
\caption 
{Temperature and bias-voltage dependence of $dI/dV$ (left panels) and $I_{\rm th}$ (right panels) at zero magnetic field  $g\mu_B B/k_{B}T_{\rm K}=0$ (top panels) and at a high magnetic field $g\mu_B B/k_{B}T_{\rm K}=4$ (lower panels).
}
\label{fig:T-dep}
\end{figure}

We show the temperature dependence of $dI/dV$ and $I_{\rm th}$ in Fig.~\ref{fig:T-dep} for  $g\mu_B B/k_{B}T_{\rm K}=0$ and 
$g\mu_B B/k_{B}T_{\rm K}=4$. A comparison for $B>B_c$ with the experimental
data showed good qualitative agreement (see  Fig.S4). In particular, the finite field thermocurrent for $B>T_{\rm K}$ captures the feature seen in the experiment of an enhanced positive thermocurrent at low temperatures and $V<0$.

\clearpage
\subsubsection{Proportionality between \texorpdfstring{$I_{\rm th}$}{} and \texorpdfstring{$d^2I/dV^2$}{} at \texorpdfstring{$\Delta T\ll T\ll T_{\rm K}$}{}}
\label{sec:proportionality}
\label{subsubsec:prop-Ith}

  Finally, we comment briefly on the observed proportionality between $I_{\rm th}$ and $d^{2}I/dV^2$ seen experimentally for small $\Delta T\ll T\ll T_{\rm K}$ in Fig.~\ref{fig:dIdVIth} at both zero and finite magnetic fields. For small bias voltage $V\ll T_{\rm K}$,  one can understand this 
within the higher-order Fermi-liquid theory of Sec.~\ref{subsec:higher-order+NRG} as follows. From Eq.~(\ref{eq:dIdV-fliq}), we have for $\Delta T\ll T\ll T_{\rm K}$ and $V \ll T_{\rm K}$,  $d^{2}I/dV^2\propto  -\frac{c}{T_{\rm K}} -2c_V \frac{V}{T_{\rm K}^2}$, 
while from Eq.~(\ref{eq:Sommerfeld4}) we have that $I_{\rm th}/\Delta T\propto T[s_0(B)+s_{1}(B)V]$, i.e., both $I_{\rm th}$ and $d^2I/dV^2$ are linear in $V$ at low bias voltages, as also seen in the experiment in  Fig.~\ref{fig:dIdVIth}. The slopes with respect to $V$ of $I_{\rm th}$ and $d^2I/dV^2$ are determined by the coefficients $s_1$ and $c_V$, respectively. These correlate as a function of field $B$ as shown in Fig.~\ref{fig:scoeff+cvct} and  Fig.~\ref{fig:s1-cv-scaling}(a). Moreover, these slopes both change sign at the same field $B\approx B_c$. Hence, the presence (for $B>B_c$) or absence (for $B<B_c$) of a kink in $I_{\rm th}$ at $V=0$ correlates with the presence or absence of a similar kink in  $d^2I/dV^2$, as observed in Fig.~\ref{fig:dIdVIth}.
The above proportionality may be of practical use for detecting possible sign reversals of the zero-bias thermocurrent slope through purely electrical current measurements.

\begin{figure}[htb!]
  \centering 
\includegraphics[width=0.8\textwidth]{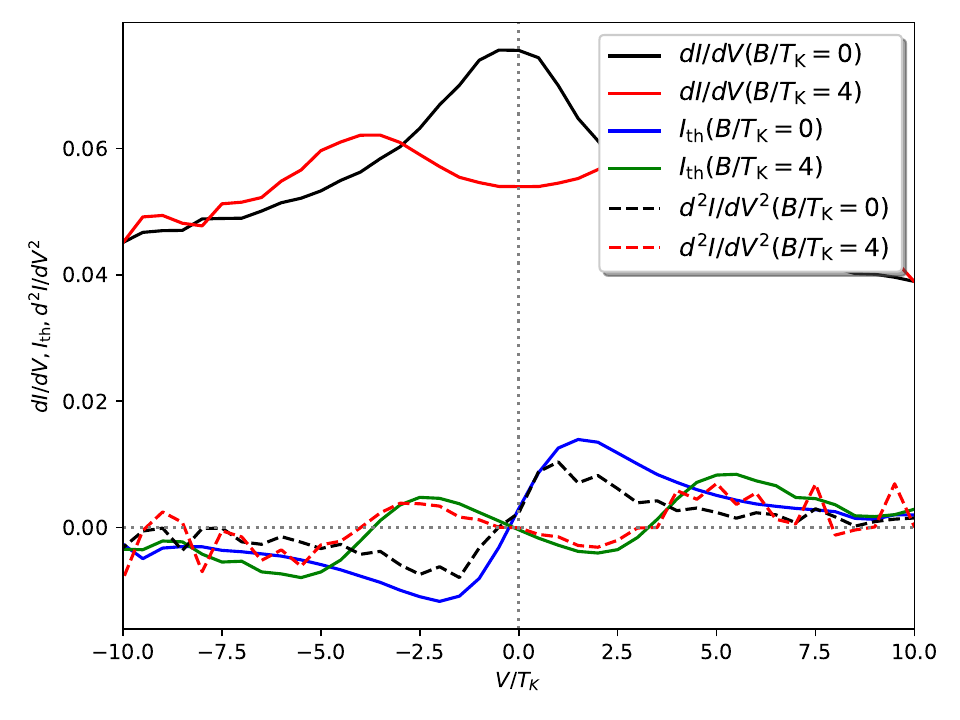}
\caption 
{Equation of motion method results showing the approximate proportionality between $d^2I/dV^2$ and $I_{\rm th}$ at $\Delta T\ll T\ll T_{\rm K}$  for low and high magnetic fields and  over a wide range of voltages. $dI/dV$ is also shown. Parameters: $U=8\Gamma$, $\varepsilon_0=-3\Gamma$, $\Delta T/T_{\rm K}=0.2$, $T/T_{\rm K}=0.5$. Note the significant noise in the second numerical derivative of the current ($d^2I/dV^2$).
}
\label{fig:Ith-prop-dGdV}
\end{figure}

The proportionality between $I_{\rm th}$ and $d^2I/dV^2$ at $\Delta T\ll T \ll T_{\rm K}$ extends, qualitatively, to larger voltages $V\gg T_{\rm K}$, outside the regime of validity of the Fermi-liquid calculations. To access these voltages, we use the above equation of motion method (EOM). The results, shown in Fig.~\ref{fig:Ith-prop-dGdV}, show that a qualitative proportionality between $I_{\rm th}$ and $d^2I/dV^2$ extends to larger bias voltages, as observed experimentally in Fig.~\ref{fig:dIdVIth}. 

\pagebreak
\clearpage
\section{Synthesis \& characterization of the organic radical}
\begin{figure}[htb!]
    \centering
    \includegraphics[width=0.3\textwidth]{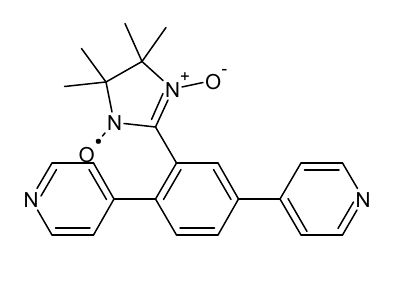}
    \caption{Organic radical molecule with a backbone and a nitronyl-nitroxide side group where an unpaired electron resides.}
    \label{fig:radical}
\end{figure}
In the experiment described in the main text and also here in the Supplemental information, an organic radical is used.
The chemical structure of the organic radical is shown in Fig.~\ref{fig:radical}, where an unpaired electron resides on a nitronyle-nitroxide side group. The structure of the molecule is very close to the previously reported Kondo-correlated system in STM~ \cite{Zhang2013}, where a spin-1/2 Kondo effect was observed in the weak coupling regime. The radical in this letter has shorter pyridine anchoring sites, where the coupling between the electrodes and the unpaired electron is expected to be stronger. Confirmation is made based on the higher Kondo temperature $T_\mathrm{K}$ in our system, indicating our system is much stronger coupled.
\pagebreak
\subsection{Synthesis}
\label{subsec:synthesis}
All chemicals and anhydrous solvents were used as purchased without further purification, unless stated otherwise. Deuterated solvents were obtained from Cambridge Isotope Laboratories, Inc. (Andover, MA, USA). All other commercial available starting materials were purchased from Sigma-Aldrich, Acros or Fluorochem. NMR experiments were acquired on a 400 or 500 MHz Bruker Avance III spectrometer equipped with a QNP or BBFO probe head respectively. The chemical shifts ($\delta$) are reported in parts per million (ppm) relative to tetramethylsilane or referenced to residual solvent peaks and the \textit{J} values are given in Hz ($\pm 0.1$ Hz). For high-resolution mass spectrometry (HRMS) a HR-ESI-ToF-MS measurement on a maXisTM 4G instrument from Bruker was performed. Column chromatography was performed on SiliaFlash$^\text{\textregistered}$ P60 from SILICYCLE with a particle size of 40-63 $\upmu$m (230-400 mesh). Thin layer chromatography (TLC) was performed on Silica gel 60 F254 glass plates with a thickness of 0.25 mm from Merck using fluorescent quenching under UV light at 254 nm for the localization of sample spots. EPR spectra (X-band) were recorded on a Bruker ELEXSYS-II E500 CW-EPR spectrometer with attachment of a Bruker N\textsubscript{2} temperature controller. Samples for EPR measurements were prepared in capillary tubes.
The software packages eview4wr and esimX were used for simulation~\cite{software_packages}.UV–vis absorption spectra were recorded at 20 $^{\circ}$C on a Jasco V-770 spectrophotometer. IR spectra were recorded with a Shimadzu IRTracer-100.

\begin{figure}[htb!]
    \centering
    \includegraphics[width=\columnwidth]{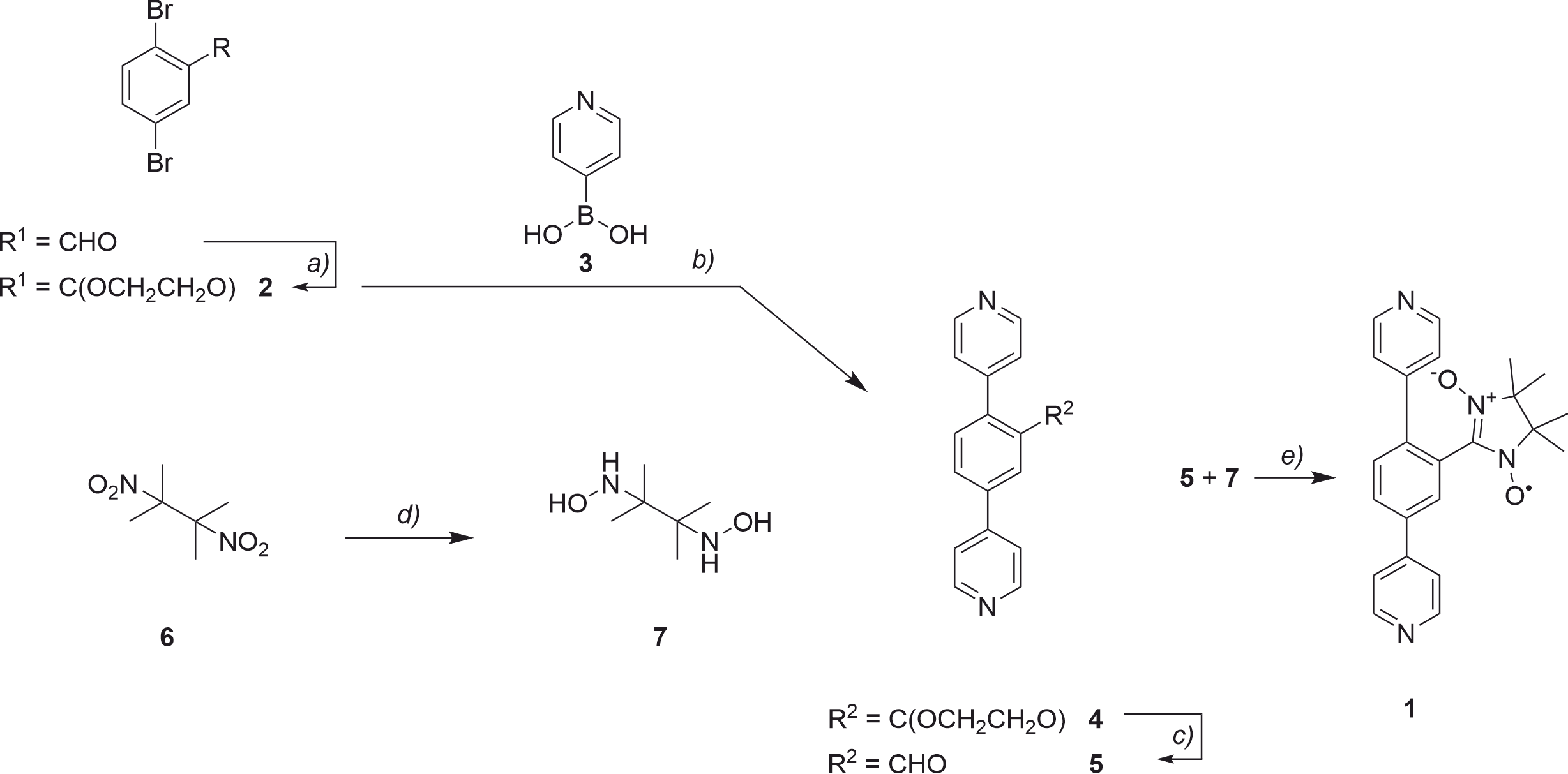}
    \caption{Scheme 1: (a) TsOH, Toluene, dean-stark, reflux, quant. (b) Pd(PPh\textsubscript{3})\textsubscript{2}Cl\textsubscript{2}, K\textsubscript{2}CO\textsubscript{3}, MeOH, Dioxane, 80 $^{\circ}$C, 36 h, 90\%, (c) 1 M HCl, THF, 1 h, r.t., 91\%, (d), Zn, NH\textsubscript{4}Cl, THF, H\textsubscript{2}O, 0 - 10 $^{\circ}$C, 48 h, 57\%, (e) MeOH, PbO\textsubscript{2}, r.t., in the dark, 37 h, 36\%}
    \label{fig:syn}
\end{figure}
The synthesis of the nitronyl nitroxide radical (NNR) \textbf{1} was already reported elsewhere~\cite{Pyurbeeva2021}, and is summarized in Fig.~\ref{fig:syn}. Starting from the commercially available 2.5-dibromobenz-alehyde, which was first protected as acetal \textbf{2} following a reported protocol~\cite{turnbull_phosphonofluoresceins_2021}. A twofold \textit{Suzuki} coupling with two equivalents of 4-pyridineboronic acid \textbf{3} followed by acidic deprotection yielded 2,5-di(pyridin-4-yl)benzaldehyde \textbf{5} in a good yield of 82\% over three steps. The nitro compound \textbf{6} was converted to the hydroxyl amine \textbf{7} following a literature known procedure~\cite{hirel_nitronyl_2001}. Condensation of the aldehyde \textbf{5} with the hydroxylamine \textbf{7} and successive in-situ oxidation with lead dioxide yielded NNR \textbf{1} in 36 \% yield. The lead dioxide was freshly prepared form lead(IV) acetate as described by Wilmarth~\cite{wilmarth_application_1955}. While scrutinizing the transport properties of NNR, G. V. Romanenko reported the synthesis of NNR, although by a different approach. The analytical data obtained by us matches the reported data of G.V.Romanenko ~\cite{romanenko_spin-labeled_2020}.  
Detailed experimental description and analytical details of compounds \textbf{4}, \textbf{5} and \textbf{1} as well as cyclic voltammetry of \textbf{1} are published elsewhere~\cite{Pyurbeeva2021}.

%The synthesis of the target structure starts from the commercially available 2,5-dibromobenzaldehyde 1. The aldehyde of 1 was masked as an acetal following a reported procedure.\cite{turnbull_phosphonofluoresceins_2021} The 2,5-di(pyridin-4-yl)benzaldehyde 5 was obtained by a twofold Suzuki cross-coupling reaction with one equivalent of the acetal 2 and two equivalents of the 4-pyridineboronic acid 3 followed by acidic liberation of the aldehyde functionality with HCl in THF/water. The aldehyde 5 was obtained in an excellent yield of 82\% over three steps. The hydroxylamine 7 was prepared from the corresponding nitro compound 6 following a reported protocol.\cite{hirel_nitronyl_2001} \linebreak Condensation of the hydroxylamine 7 and the aldehyde 5 followed by in-situ oxidation yielded the nitronyl nitroxide radical (NNR) 8 in 36\%. The lead dioxide used for this reaction was freshly precipitated as described by Wilmarth.\cite{wilmarth_application_1955} During the determination and theoretical investigation of the transport properties of NNR, G. V. Romanenko published its structure, although synthesized by a different approach. The obtained analytical data of our molecule matches the published molecule.
\pagebreak
\newpage
\clearpage

\subsection{Electron paramagnetic resonance spectroscopic analysis of the nitronyl nitroxide Radical (NNR) 8}
\label{subsec:EPR}
\begin{figure}[htb!]
    \centering
    \includegraphics[width=0.7\textwidth]{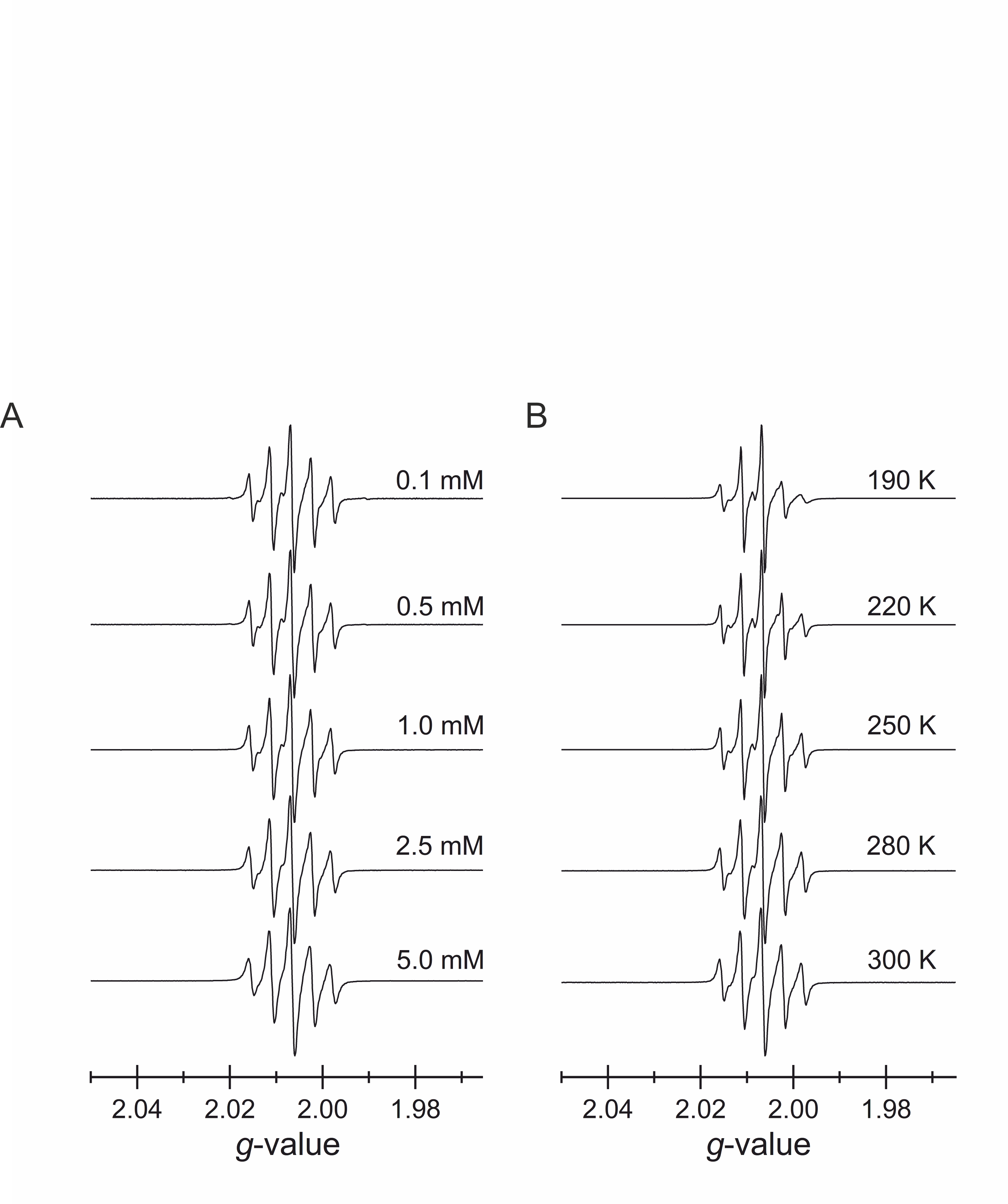}
    \caption{(A) Concentration dependent EPR (X-band) spectra of NNR at 300 K, (B) Temperature dependent EPR (X-band) spectra of NNR (2.5 mM). All data was measured in DCM. Power attenuation: 25 dB. Microwave frequency: 9.45 GHz.}
    \label{fig:EPR}
\end{figure}
The EPR spectrum of NNR at 300 K (Fig.~\ref{fig:EPR}(A)) shows five lines of relative intensity 1:2:3:2:1 due to the coupling of the unpaired electron with two equivalent \textsuperscript{14}N atoms (\textit{I} = 1). The \textit{g}-value is 2.0065 with a spacing of \textit{a}\textsubscript{N} = 7.10 G, which are normal parameters for nitronyl nitroxide monoradicals~\cite{wang_temperature-dependent_2018,zhivetyeva_interaction_2018}. There is no indication of a coupling to a third inequivalent \textsuperscript{14}N atom demonstrating that the N atom of the \textit{o}-pyridyl is too far away for interaction with the radical. With increasing concentration, the line shaping gets as expected broader, but the overall shape is relatively concentration independent. 
A splitting of the five lines in the EPR spectrum of NNR appears, when cooling the sample from 300 K to 190 K (Fig.~\ref{fig:EPR}(B)) along with line broadening and line asymmicity. At lower temperature the rotation around the phenyl-nitronyl nitroxide axis will be slowed down, hence resulting in inequality of the two N atoms in the nitronyl nitroxide unit. In agreement with our observations, the change from so-called “fast rotation” regime to “slow rotation” regime has previously been used to explain line broadening and asymmetric lines for nitronyl nitroxide monoradicals~\cite{stone_spin-labeled_1965,chechik_spin-labelled_2004}.
\pagebreak
\clearpage

\bibliographystyle{apsrev4-2}
%\bibliography{ref_SI.bib}% P
\providecommand{\noopsort}[1]{}\providecommand{\singleletter}[1]{#1}%

\pagebreak